\documentclass[aps,prd,10pt,twocolumn,showpacs,showkeys,preprintnumbers,floatfix,nofootinbib,superscriptaddress,longbibliography]{revtex4-2}
\usepackage{amsfonts} 
\usepackage{amssymb} 
\usepackage{amsmath} 
\usepackage{graphicx} 
\usepackage{subfigure} 
\usepackage{array} 
\usepackage{dcolumn} 
\usepackage{latexsym} 
\usepackage{longtable} 
\usepackage{bbold}
\usepackage{color}
\usepackage{multirow}
\usepackage{rotating}
\usepackage{comment}
\usepackage{dsfont}
\usepackage{slashed}

\providecommand \doibase {http://dx.doi.org/}%

\graphicspath{{./figs/}{./figs_gS/}}

\newcommand{\qslash}[1]{\text{$\not \! #1$}}

\newcommand{\tskip}{\mathop{\tau_{\rm skip}}\nolimits}
\newcommand{\MeV}{\mathop{\rm MeV}\nolimits}
\newcommand{\GeV}{\mathop{\rm GeV}\nolimits}

\newcommand{\fm}{\mathop{\rm fm}\nolimits}

\newcommand{\CO}{\mathcal{O}}

\newcommand{\mpi}{M_\pi}
\newcommand{\mN}{M_N}
\newcommand{\spiN}{\sigma_{\pi N}}

\newcommand{\MSb}{{\overline{\text{MS}}}}
\newcommand{\SMOM}{\text{SMOM}}

\newcommand{\changed}[1]{{{#1}}}

\usepackage{textgreek}
\usepackage[unicode]{hyperref} 

\DeclareMathOperator{\Tr}{Tr}

\definecolor{green}{rgb}{0.1, 0.8, 0.1}

\usepackage{textgreek}
\usepackage[unicode]{hyperref} 

\begin{document}


\title{Flavor diagonal nucleon charges using clover fermions on MILC HISQ ensembles}

\author{Sungwoo~Park}
\email{park49@llnl.gov}
\affiliation{Physical and Life Sciences Division, Lawrence Livermore National Laboratory, Livermore, CA 94550, USA}
\affiliation{Nuclear Science Division, Lawrence Berkeley National Laboratory, Berkeley, CA 94720, USA}
\affiliation{Thomas Jefferson National Accelerator Facility, 12000 Jefferson Avenue, Newport News, VA 23606, USA}

\author{Rajan~Gupta}
\email{rajan@lanl.gov}
\affiliation{Los Alamos National Laboratory, Theoretical Division T-2, Los Alamos, NM 87545, USA}

\author{Tanmoy~Bhattacharya}
\email{tanmoy@lanl.gov}
\affiliation{Los Alamos National Laboratory, Theoretical Division T-2, Los Alamos, NM 87545, USA}

\author{Fangcheng~He}
\email{fangchenghe@lanl.gov}
\affiliation{Los Alamos National Laboratory, Theoretical Division T-2, Los Alamos, NM 87545, USA}

\author{Santanu~Mondal}
\email{santanu.sinp@gmail.com}
\affiliation{Ibsyn Scientific, 75C Park St, Kolkata, India 700016}

\author{Huey-Wen Lin}
\email{hwlin@pa.msu.edu}
\affiliation{Department of Physics and Astronomy, Michigan State University, MI, 48824, USA}

\author{Boram~Yoon}
\email{byoon@nvidia.com}
\affiliation{NVIDIA Corporation, Santa Clara, CA 95051, USA}

\preprint{LLNL-JRNL-858665, LA-UR-22053}

\begin{abstract}
We present lattice results for the flavor diagonal charges of the
proton from the analysis of eight ensembles generated using
2+1+1-flavors of highly improved staggered quarks (HISQ) by the MILC
collaboration. The calculation includes all the needed connected and
disconnected contributions to nucleon three-point function.  For
extracting matrix elements using fits to the spectral decomposition of
these correlation functions, two strategies to remove excited state
contributions are employed and compared. To renormalize these charges,
the 2+1-flavor mixing matrix is calculated in the RI-sMOM intermediate
scheme on the lattice.  The final results are presented in the $\MSb$
scheme at scale 2~GeV.
The axial charges for the proton are 
$g_A^u = 0.781(25)$,  $g_A^d = -0.440(39)$, and $g_A^s = -0.055(9)$;  
the tensor charges are 
$g_T^u = 0.782(28)$, $g_T^d = -0.195(16)$, and $g_T^s = -0.0016(12)$; 
and the scalar charges are 
$g_S^u = 9.39(88)$,  $g_S^d = 8.84(93)$,   and $g_S^s = 0.37(14)$. 
Results for the neutron are given by the $u \leftrightarrow d$ interchange. 
Results for the sigma terms are 
$\sigma_{\pi N}|_{\rm standard} = 42(6)~{\rm MeV}$ from a ``standard'' analysis and  
$\sigma_{\pi N}|_{N \pi}        = 61(6)~{\rm MeV}$ from a ``$N\pi$'' analysis that  
includes the contributions of multihadron $N\pi $ excited states 
as motivated by chiral perturbation theory. Our preferred value 
$\sigma_{\pi N}|_{N \pi}$ is consistent with the 
phenomenological extraction from $\pi- N$ scattering data. The strangeness content 
of the proton, for which the ``standard'' analysis is appropriate, 
is $\sigma_{s}|_{\rm standard} = 35(13)~{\rm MeV}$.

\end{abstract}
\maketitle

\section{Introduction}
\label{sec:into}

Results for the flavor diagonal nucleon charges, $g_{A,S,T}^{u,d,s}$,
are extracted from the matrix elements of axial, scalar, and tensor
quark bilinear operators, ${\overline q}\Gamma q$ with the Dirac
matrix $\Gamma = \gamma_\mu \gamma_5, I, \sigma_{\mu\nu}$,
respectively, within ground state nucleons.  The lattice QCD
calculations were done using Wilson-clover fermions on eight ensembles
generated by the MILC collaboration with 2+1+1-flavors of highly
improved staggered quarks (HISQ) \cite{Bazavov:2012xda}. The lattice
parameters used in this clover-on-HISQ calculation are given in
Tables~\ref{tab:ens} and~\ref{tab:ens_params}.

The motivation for these calculations and much of the methodology used
has already been published for $g_{A}^{q}$ in Ref.~\cite{Lin:2018obj},
$g_T^{q}$ in \cite{Gupta:2018lvp} and for the pion-nucleon sigma term,
$\sigma_{\pi N}=m_{u,d}\times g_S^{u+d}$, in~\cite{Gupta:2021ahb}.
The charges $g_A^{u,d,s}$ give the contributions of the intrinsic spin
of the quarks to the nucleon spin; the $g_T^{u,d,s}$ give the
contribution of the quark electric dipole moment (EDM) operator to the
nucleon EDM; and all three, $g_{A,S,T}^{u,d,s}$, give the coupling of
dark matter or Higgs-like interactions with nucleons in the respective Lorentz 
channels.  In addition, with $g_{S}^{u,d,s}$ in hand, we calculate the
pion-nucleon sigma term, $\sigma_{\pi N}=m_{u,d} g_S^{u+d}$ and the
strangeness content of the nucleon, $\sigma_{s}=m_{s} g_S^{s}$. We
compare our results with those given in the latest FLAG report
2024~\cite{FlavourLatticeAveragingGroupFLAG:2024oxs}, a community
review of calculations done until May 2024 by various lattice
collaborations, and new results since then in
Section~\ref{sec:conclusion}.

This work supersedes our earlier
publications~\cite{Lin:2018obj,Gupta:2018lvp,Gupta:2021ahb} and
updates in conference proceedings~\cite{Park:2023tsj,Park:2024vjp}. It
includes a number of improvements:
\begin{itemize}
\item 
The calculation has been extended to eight ensembles described in Tables~\ref{tab:ens}  
and~\ref{tab:ens_params}. The point $a06m220$ is new. 
\item
The disconnected contributions on all ensembles except $a12m220$ have
now been calculated with operator insertion at all intermediate points
$t$ between the nucleon source and the sink separated by
Euclidean time $\tau$.
\item 
The statistics on most of the ensembles have been increased.
\item 
Results for $g_S^{u,d,s}$ are new. In Ref.~\cite{Gupta:2021ahb}, the
result for only the renormalization group invariant pion-nucleon sigma
term, $\sigma_{\pi N}=m_{u,d}\times g_S^{u+d}$, was presented. Using the fact that $m \overline \psi\psi$ is the mass term in the action,
implies $Z_m \times Z_S = 1$. \looseness-1
\item 
Complete results for the renormalization of quark bilinears for the
clover-on-HISQ formulation, \emph{i.e.}, the 2+1-flavor mixing matrix,
$Z_{A,S,T}$, are obtained nonperturbatively on the lattice using the
regularization independent symmetric momentum subtraction (RI-sMOM)
scheme~\cite{Martinelli:1994ty,Sturm:2009kb}. These $Z$s are then
matched to the $\overline {MS}$ scheme and run to $2$~GeV using
results from perturbation theory. Results in the basis $(u-d,u+d,s)$,
most relevant to this work, are given in Appendix~\ref{sec:NPR}. Our
previous calculations used $Z_\text{isoscalar}\approx
Z_\text{isovector}$, i.e., $Z_\text{u+d}\approx Z_\text{u-d}$ for the
renormalization constants of the axial and tensor operators. Having
completed the full calculation, we find removing this approximation
has not changed the results significantly because the corrections are
small as can be inferred from the numbers in
Appendix~\ref{sec:NPR}.\looseness-1
\item 
Resolving and removing the contributions of excited states to nucleon
correlation function continues to be a leading systematic.  These
artifacts have to be removed to get ground state matrix elements.  In
our recent lattice QCD calculation of the nucleon axial vector form
factor $G_A(Q^2)$ \cite{Jang:2019vkm,Park:2021ypf,Jang:2023zts} and of
the pion-nucleon sigma term $\sigma_{\pi N}$ \cite{Gupta:2021ahb}, we
presented evidence of larger-than-expected excited-state contributions
(ESC) from $N \pi$ and $N\pi\pi$ multihadron excited states in these
nucleon 3-point correlation functions. Motivated by these works, we
study the impact of including $N\pi/N\pi\pi$ states in the analysis of
all the flavor diagonal nucleon matrix elements.

For correlation functions with resolvable signal of ESC, fits to get
ground state matrix elements (GSME) now include two excited states
in the spectral decomposition, and in each case compare fits with and
without including the $N\pi$ excited state. These fits have been made
using the full covariance matrix.  To further control ESC, we vary
$t_{\rm skip}$, the number of points skipped adjacent to the source
and the sink, and the range of $\tau$ values included in the fits.
For final results we take the average of these various fits, weighted
by the Akaika Information Criteria (AIC) score~\cite{1100705}.
\item The chiral-continuum-finite-volume (CCFV) extrapolation to the physical point, 
defined by $M_\pi \to 135$~MeV, $a \to 0$,  and $M_\pi L \to \infty$, is done keeping 
the leading correction in each of the three variables. 
 \end{itemize}

This paper is organized as follows. A brief introduction to the
lattice methods is given in Section~\ref{sec:Methodology}, and the
strategies (models) used to remove excited state contributions are
described in Section~\ref{sec:ESC}. The averaging of models to get
results for the charges using the AIC score is discussed in
Section~\ref{sec:models}.  The framework for obtaining the
renormalized charges in the $\MSb$ scheme at 2~GeV is described in
Appendix~\ref{sec:NPR}.  The lattice calculation of the
renormalization factors for the 2+1-flavor theory is done using the
RI-sMOM scheme. The extrapolation of the renormalized data to the
physical point using the CCFV fits is discussed in
Section~\ref{sec:CCFV}. A comparison with previous results and our
conclusions are given in Section~\ref{sec:conclusion}.

\begin{table*}[tbp]    
\begin{center}
\renewcommand{\arraystretch}{1.2} 
\begin{ruledtabular}
\begin{tabular}{l|cccccccc}
Ensemble ID    & $a$ (fm) & $M_\pi$ (MeV) & $\beta$ & $C_{\rm SW}$ & $ am_{ud}$ & $ am_{s}$  & $L^3\times T$   & $M_\pi L$  \\ \hline
$a15m310$      & 0.1510(20) & 320.6(4.3)   & 5.8      & 1.05094    & -0.0893    & -0.0210    & $16^3\times 48$ & 3.93    \\
$a12m310$      & 0.1207(11) & 310.2(2.8)   & 6.0      & 1.05094    & -0.0695    & -0.018718  & $24^3\times 64$ & 4.55    \\
$a12m220$      & 0.1184(10) & 227.9(1.9)   & 6.0      & 1.05091    & -0.075     & -0.02118   & $32^3\times 64$ & 4.38    \\
$a09m310$      & 0.0888(8)  & 313.0(2.8)   & 6.3      & 1.04243    & -0.05138   & -0.016075  & $32^3\times 96$ & 4.51    \\
$a09m220$      & 0.0872(7)  & 225.9(1.8)   & 6.3      & 1.04239    & -0.0554    & -0.01761   & $48^3\times 96$ & 4.79    \\
$a09m130$      & 0.0871(6)  & 138.1(1.0)   & 6.3      & 1.04239    & -0.058     & -0.0174    & $64^3\times 96$ & 3.90    \\
$a06m310$      & 0.0582(4)  & 319.6(2.2)   & 6.72     & 1.03493    & -0.0398    & -0.01841   & $48^3\times 144$& 4.52    \\
$a06m220$      & 0.0578(4)  & 235.2(1.7)   & 6.72     & 1.03493    & -0.04222   & -0.01801   & $64^3\times 144$& 4.41    \\
\end{tabular}
\end{ruledtabular}
\caption{The lattice spacing $a$, the valence pion mass $M_\pi$, gauge
  coupling $\beta$, the Sheikholeslami-Wohlert coefficient $C_{SW}$
  defining the clover term in the Wilson action, the light quark mass
  $am_{ud} = 1/2\kappa_{ud} -4$, the strange quark mass $am_{s} =
  1/2\kappa_{s} -4$, the lattice volume $L^3 \times T$, and the
  lattice size in units of $M_\pi$ for the eight ensembles analyzed
  for flavor diagonal charges $g_\Gamma^{q}$.}
\label{tab:ens}
\end{center}
\end{table*}

\begin{table*}[tbp]    
\begin{center}
\renewcommand{\arraystretch}{1.2} 
\begin{ruledtabular}
\begin{tabular}{l|cccccc|ccc|ccc}
Ensemble ID & $\sigma$ &  $\tau/a$ & $N_\text{conf}^{\rm 2pt}$ & $N_\text{conf}^{\rm conn}$ & $N_\text{LP}$  & $N_\text{HP}$   & $N_\text{conf}^{l}$ & $N_\text{src}^l$   &  $\frac{N_{\rm LP}^l}{N_{\rm HP}^l}$   & $N_\text{conf}^{s}$ & $N_\text{src}^s$   &  $\frac{N_{\rm LP}^s}{N_{\rm HP}^s}$ \\ \hline
$a15m310$   & 4.2 & \{6,7,8,9\}    & 1917 & 1917        & 64      &  8  & 1917  & 2000   &  50 & 1917  & 2000   &  50   \\
$a12m310$   & 5.5 & \{8,10,12,14\} & 1013 & 1013        & 64      &  8  & 1013  & 10000  &  50 & 1013  & 8000   &  50   \\
$a12m220$   & 5.5 & \{8,10,12,14\} & 959  &  744        & 64      &  4  &958(\#)& 11000  &  30 &  870  & 5000   &  50   \\
$a09m310$   & 7.0 & \{10,12,14,16\}& 2263 & 2263        & 64      &  4  & 1017  & 10000  &  50 & 1024  & 6000   &  50   \\
$a09m220$   & 7.0 & \{10,12,14,16\}& 964  &  964        & 128     &  8  & 712   & 8000   &  30 & 847   & 10000  &  50   \\
$a09m130$   & 7.0 & \{10,12,14,16\}& 1274 & 1290        & 128     &  4  & 1270  & 10000  &  50 & 994   & 10000/4000$(^\dagger)$  &  50 \\
$a06m310$   & 12.0& \{18,20,22,24\}& 977  &  500        & 128     &  4  & 808   & 12000  &  50 & 976   & 10000/4000$(^\dagger)$  &  50 \\
$a06m220$   & 11.0/9.0 (*) & \{18,20,22,24\} & 1010 & 649& 64      &  4  & 1001  & 10000  &  50 & 1002  & 10000  &  50   \\
\end{tabular}
\end{ruledtabular}
\caption{Parameters used in the calculation of the charges
  $g_\Gamma^{q}$ on the eight ensembles. The second column gives the
  smearing parameter $\sigma$ used in the creation of the covariant
  Gaussian smeared source for quark propagators. For ensemble $a06m220
  $, marked with (*), the connected contribution was done with
  $\sigma=11$ while the disconnected was done with $\sigma=9$. The
  list of source-sink time separations $\tau$ analyzed for the
  connected 3-point functions is given in column three. The
  disconnected contributions were calculated for all $\tau$ and all intermediate $t$ except on
  $a12m220$ for light quarks ($N^l_{\rm conf}$ marked with (\#)). On this ensemble, 
  data were collected with fixed $t_{\rm skip}=3$ and the four 
  source time slices were fixed at $t_{src}=2,18,34,50$ rather than being chosen randomly. 
  $N_\text{conf}^{\rm 2pt}$ and $N_\text{conf}^{\rm conn}$ give the number of configurations
  analyzed for 2-point and connected 3-point correlation functions,
  respectively. In the truncated solver with bias correction method,
  we used $N_\text{LP}$ low and $N_\text{HP}$ high precision
  measurements per configuration for the connected contributions. Columns 8--10 (11--13), specify the number of   configurations, $N_\text{conf}^{l}$ ($N_\text{conf}^{s}$),
 analyzed for the light (strange) quark
  disconnected contributions with $N_{\rm src}^l$ ($N_{\rm src}^s$)
  $Z_4$ random noise sources and $N_{\rm LP}/N_{\rm HP}$ low to high
  precision measurements.  In cases marked with $(^\dagger)$, the
  $N^s_\text{conf}$ configurations were split into two sets and
  analyzed with different number of sources $N_{\rm src}^s$, and the
  results averaged. Results for isovector charges on full
  set of thirteen configurations including these has been published in   Ref.~\protect\cite{Gupta:2018qil}.  }
\label{tab:ens_params}
\end{center}
\end{table*}

\section{Lattice Methodology}
\label{sec:Methodology}

\begin{figure}[thb]   
  \subfigure{ \includegraphics[height=0.72in]{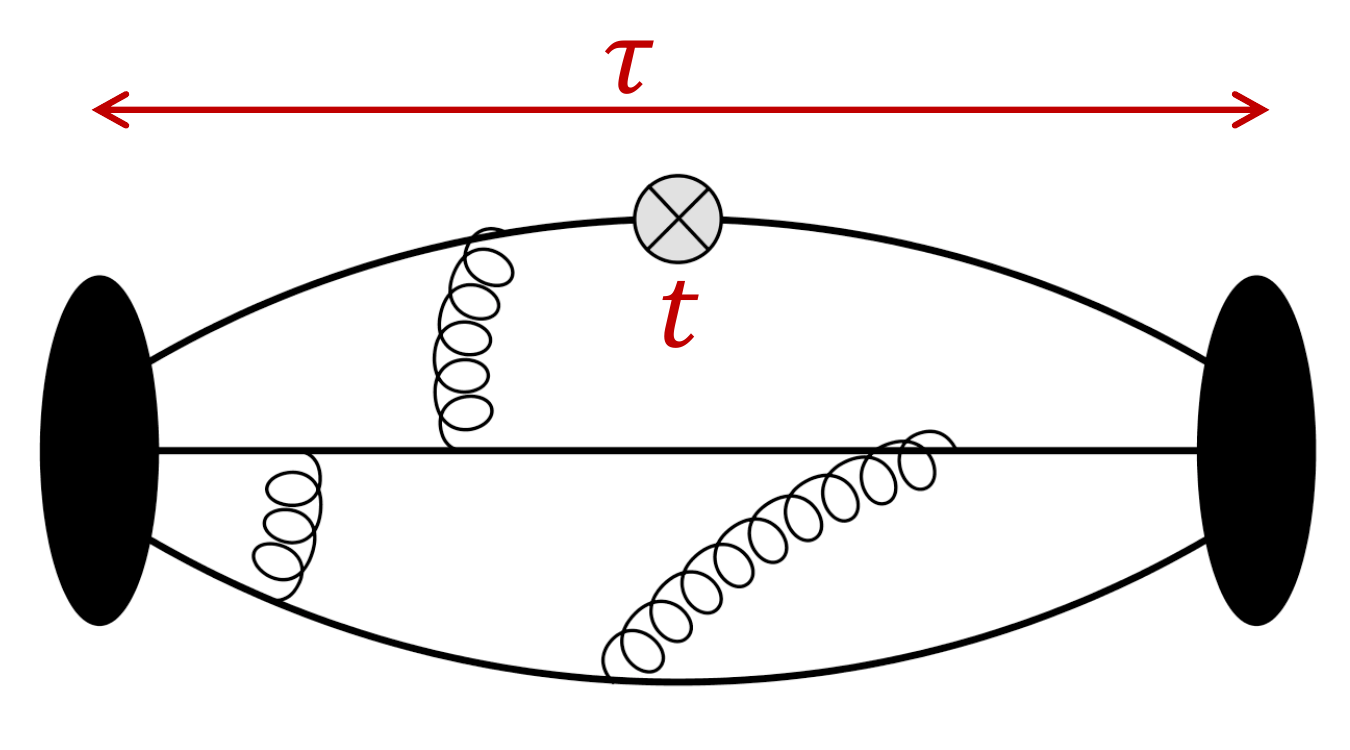}  \hspace{0.09\linewidth}
    \includegraphics[height=0.9in]{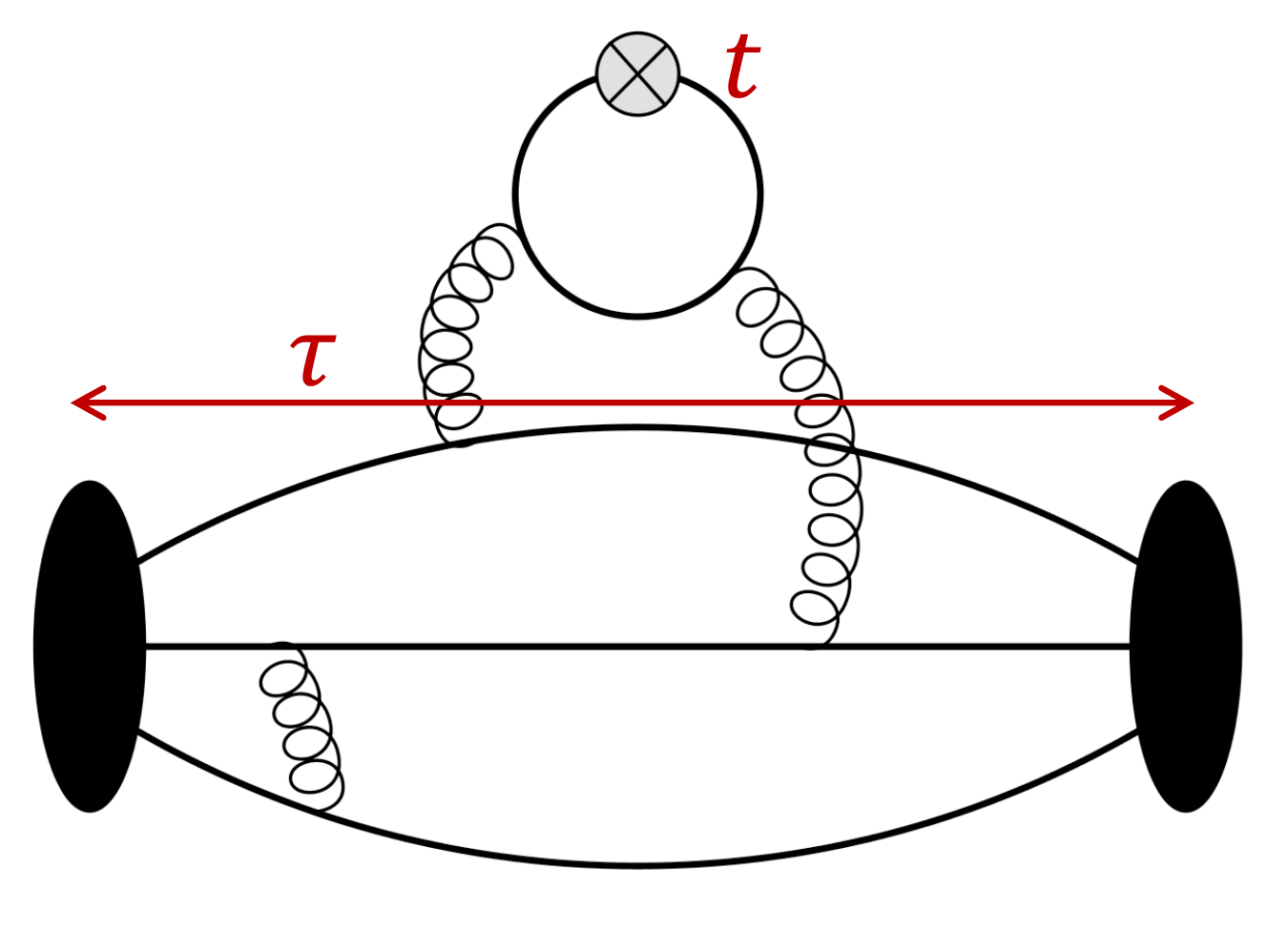}\hspace{0.09\linewidth}
  }
 \changed{\includegraphics[height=0.72in]{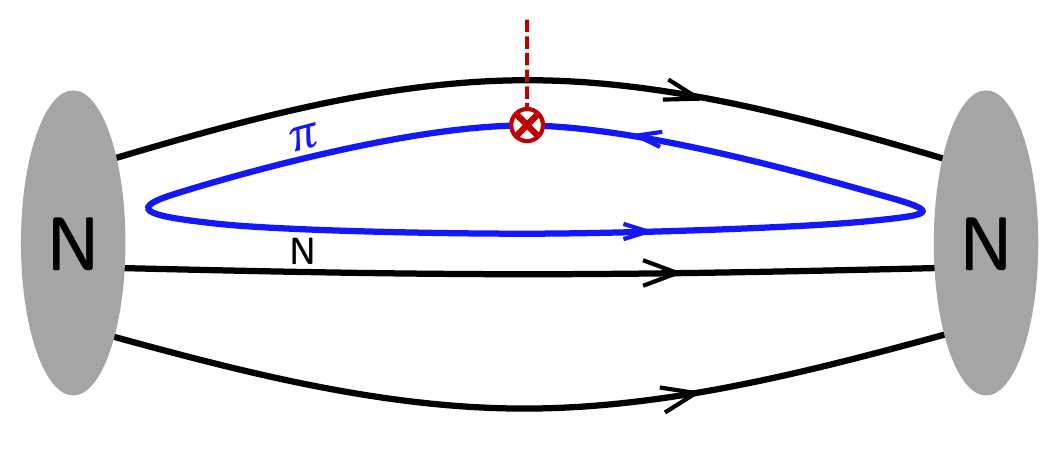}}
\vspace{-0.1in}
\caption{ \changed{The top row shows the connected (left) and
    disconnected (right) diagrams that contribute to the 3-point
    functions from which the matrix element of flavor-diagonal
    operators are extracted. The black \changed{and gray} blobs denote
    the nucleon source and sink separated by Euclidean time
    $\tau$. The operator, shown by the symbol $\otimes$,
    is inserted at all intermediate times $t$ between the
    nucleon source and sink points. The bottom diagram (redraw of the
    top right) illustrates why the disconnected contribution for the
    scalar operator with $u$ and $d$ flavors can have an enhanced
    contribution due to the $N\pi$-intermediate
    state.}\looseness-1 \label{fig:Npisketch} \label{fig:conn_disc}}
\vspace{-0.1in}
\end{figure}

The charges $g_{A,S,T}^{u,d,s}$ are extracted from the forward matrix
elements of the flavor diagonal quark bilinear operator,
$\mathcal{O}_\Gamma^q(x)=\bar{q}(x)\Gamma q(x)$ with $q\in \{u,d,s\}$,
inserted with zero momentum at time $t$ as illustrated in
Fig.~\ref{fig:conn_disc} using the relation
\begin{align}
   \langle n(s_f,{\mathbf{p}}=0) | {\bar q} \Gamma q | n(s_i,{\mathbf{p}}=0) \rangle = g_\Gamma^q {\bar u}(s_f) \Gamma u(s_i) \,,
\label{eq:ME}
\end{align}
with $u(s)$ the nucleon spinor. These ME are obtained from the 3-point functions 
\begin{align}
  C_\Gamma^\text{3pt} (t;\tau) = \Tr[ \mathcal{P} \langle 0 |
    \mathcal{N}(\tau,\mathbf{p}=0) \mathcal{O}_\Gamma^q (t,\mathbf{q}=0)
    \bar{\mathcal{N}}(0)| 0 \rangle]
\label{eq:C3pt}
\end{align}
where $\mathcal{P}$ is the spin projection defined in
Ref.~\cite{Bhattacharya:2015wna}. 
The interpolating operator ${\mathcal{N}}$ used to create/annihilate the nucleon 
at the source/sink time slices $0$ and $\tau$ is
\begin{align}
\mathcal{N}(x)=\epsilon^{abc}\left[ u^{aT}(x) C\gamma_5 \frac{1\pm \gamma_4}{2}
                                    d^{b}(x)\right] u^{c}(x) \,.
\end{align}
Here, $\{a,b,c\}$ are color indices and $C=\gamma_4\gamma_2$ is the
charge conjugation matrix. The ${\mathcal{N}}$ is projected to zero
momentum at the sink to get the forward matrix elements.  Henceforth,
we will drop the momentum indices as $\mathbf{p}=\mathbf{q}=0$ for all
the calculations presented here.  The goal is to design better
${\mathcal{N}}$, and increase the statistical precision so that there
is good signal at large enough $t$ and $\tau-t$ such that the
extracted matrix elements are within the nucleon ground state.
Otherwise ESC have to be removed using fits to the spectral
decomposition of the correlation functions as discussed in
Sec.~\ref{sec:ESC}.

The nucleon 2- and 3-point correlation functions are calculated using
the Wilson-clover fermion action with the Sheikholeslami-Wohlert
coefficient $c_{SW}$ fixed to its tree-level tadpole improved value,
$c_{SW}=1/u_0^3$ where $u_0$ is the tadpole factor given by the fourth
root of the plaquette expectation value calculated on HISQ lattices
that have been hypercubic (HYP) smeared~\cite{Hasenfratz:2001hp}.

Our calculation assumes isospin symmetry, i.e., $m_u=m_d\equiv m_l$,
and the valence light quark mass $m_l$ is tuned so that clover-on-HISQ
pion reproduces the HISQ-on-HISQ Goldstone (sea) pion mass
$M_\pi^\text{sea}$. The valence strange quark mass is tuned so that
the clover-on-HISQ pseudoscalar meson mass reproduces
$M_{s\bar{s}}=\sqrt{m_s^\text{sea}/m_l^\text{sea}}M_\pi^\text{ sea}$
\cite{Bhattacharya:2015wna} where $m_s^\text{sea}/m_l^\text{ sea}$ is
the ratio of HISQ (sea) quark masses and $M_\pi^\text{sea}$ is fixed
at the physical value 135~MeV \cite{Bazavov:2012xda}. While this
clover-on-HISQ mixed action formulation is not unitary and could have
an addition systematic uncertainty, no obvious effects have been
observed in our calculations, presumably because taking the continuum
limit matching the Goldstone pseudoscalar spectra of the clover and
HISQ formulations is sufficient.

The quark propagators used to construct the quark line diagrams shown
in Fig.~\ref{fig:conn_disc} are the inverse of the clover Dirac
operator with a smeared (Wuppertal gauge covariant
Gaussian~\cite{Gusken:1989ad}) source. The smearing parameter $\sigma$
is given in Table~\ref{tab:ens_params} and the same smearing is
applied at the source and sink points. Efficacy of this smeared source
in suppressing ESC was explored in
Refs.~\cite{Yoon:2016dij,Yoon:2016jzj}.

As illustrated in Fig.~\ref{fig:conn_disc}, the correlation function,
$C_\Gamma^\text{3pt}(t;\tau)$ defined in Eq.~\eqref{eq:C3pt}, after
Wick-contractions, is the sum of two types of quark lines diagrams
called ``connected" and ``disconnected":
\begin{align}
  C_\Gamma^\text{3pt}(t;\tau) = C_\Gamma^\text{conn}(t;\tau)
+ C_\Gamma^\text{disc}(t;\tau).
\end{align}
To reduce the computational cost, the connected diagrams are
constructed using a coherent sequential source \cite{LHPC:2010jcs,
  Yoon:2016dij} and the truncated solver with bias correction
\cite{Bali:2009hu,Blum:2012uh,Yoon:2016dij} methods.  The quark loop
with zero-momentum operator insertion, i.e., $\mathbf{q=0}$, for the
disconnected contribution, $C_\Gamma^\text{disc}$, is estimated
stochastically using $Z_4$ random noise sources. This calculation was
accelerated with a combination of the truncated solver with bias
correction method \cite{Bhattacharya:2015wna}, and the hopping
parameter expansion used as a pre-conditioner
\cite{Thron:1997iy,Michael:1999rs,Bhattacharya:2015wna}. For the
scalar case, the disconnected contribution is calculated using the
vacuum subtracted operator $\mathcal{O}_S^q - \langle \mathcal{O}_S^q
\rangle$~\cite{Bhattacharya:2005rb}. Additional details are given in
Ref.~\cite{Bhattacharya:2015wna}.
The statistics for the connected and disconnected calculations are
given in Table~\ref{tab:ens_params}.

The calculation of the quark propagators and their contractions to
construct correlation functions was done using the Chroma software
suite~\cite{Edwards:2004sx}. On GPU nodes, we use the QUDA
library~\cite{Clark:2009wm,Babich:2010mu,Babich:2011np} with the
multigrid invertor~\cite{Brannick:2007ue,Babich:2010qb,Osborn:2010mb}
built into it~\cite{Clark:2016rdz}.

In our previous works \cite{Lin:2018obj,Gupta:2018lvp}, the removal of
excited-state contamination (ESC) was carried out by making separate
fits to $C^{\text{conn}}_\Gamma(t;\tau)$ and
$C^{\text{disc}}_\Gamma(t;\tau)$.  Similarly, the chiral-continuum
(CC) extrapolations of $g_\Gamma^{q,\text{disc}}$ and
$g_\Gamma^{q,\text{conn}}$ to $a \to 0$ and $M_\pi \to 135$~MeV were
performed independently. This procedure  may have introduced an
unquantified systematic bias as discussed in Ref.~\cite{Lin:2018obj}.
This systematic has been removed in this work by making ESC fits to
the sum and the subsequent CCFV extrapolation of the renormalized
charges $g_\Gamma^{q}$.  To understand the size of possible
systematics in our previous work, we obtained results (ESC fits and
CCFV extrapolation) both ways, and find the differences to be smaller
than the statistical errors.\looseness-1

\section{Strategies to Remove Excited-State Contributions}
\label{sec:ESC}

The GSME $\langle 0 | O_\Gamma^q | 0 \rangle$, and from them
$g_{\Gamma}^{q;\text{bare}}$ using Eq.~\eqref{eq:ME}, are extracted
from fits to the spectral decomposition of the spin-projected
$C^{\textrm{3pt}}_\Gamma (t;\tau)$:
\begin{align}
    C^{\textrm{3pt}}_\Gamma (t;\tau) &=\sum_{i,j=0} \mathcal{A}_i^\ast
    \mathcal{A}_j \langle i |O_\Gamma^q| j \rangle e^{-M_i t - M_j (t- \tau)}
\label{eq:3pt-sd}
\end{align}
The challenge lies in removing all ESC since, even in our highest
statistics ensembles, the right hand side of Eq.~\eqref{eq:3pt-sd} is
truncated at 3-states. While the ESC decrease exponentially with the mass gap $\Delta M$ and 
source-sink separation $\tau$ as $e^{-\Delta M \tau}$, the signal-to-noise ratio for 
$C^{\textrm{3pt}}_\Gamma $ degrades even faster   exponentially as
$e^{-(M_N-3/2M_\pi)\tau}$~\cite{Hamber:1983vu,Lepage:1989hd}. The
$C^{\textrm{3pt}}_\Gamma $ are well-measured only up to source-sink
separation $\tau \approx 1.4\fm$ and for $\tau \lesssim 1.4\fm$, ESC
are found to be significant. Three- state fits to remove these using
Eq.~\eqref{eq:3pt-sd} with $\mathcal{A}_0$, $M_i$ and the $A_i^\ast A_j
\langle i | O_\Gamma^q | j \rangle$ left as free parameters are not
stable with current statistics. Additional information on these
parameters is needed.

To make robust fits to Eq.~\eqref{eq:3pt-sd}, the key parameters are
the ground state amplitude $\mathcal{A}_0$ and the finite volume
spectrum $M_i$. If these are known {\it a priori}, then to extract
$\langle 0 | O_\Gamma^q | 0 \rangle$ only the
combinations $A_i^\ast A_j \langle i | O_\Gamma^q | j \rangle$
involving the excited states remain as free parameters. This then becomes a tractable
problem. Note, these $A_i^\ast A_j \langle i | O_\Gamma^q | j \rangle$
are not used in any subsequent analysis.
 
In principle, the $\mathcal{A}_0$ and $M_i$  can be obtained from fits to 
the spectral decomposition of the 2-point function:
\begin{align}
  C^\textrm{2pt}(\tau)=\sum_{i=0}^3 | \mathcal{A}_i|^2 e^{-M_i \tau}  \,.
  \label{eq:2pt-sd}
\end{align}
To make n-state fits to $C^{\text{3pt}}_\Gamma(t;\tau)$ with input
from $C^\textrm{2pt}(\tau)$, we need $C^\textrm{2pt}(\tau)$ truncated
to $n+1$ (or higher) number of states presuming that the effects of
all the neglected states are subsumed in the parameters of the $n+1$
(and higher) state. Of these, $M_1$ is the most important for
controlling ESC in fits to $C^{\text{3pt}}_\Gamma(t;\tau)$ with the
contamination falling  exponentially with the mass gap. Unfortunately,
with current statistics, 4-state fits to Eq.~\eqref{eq:2pt-sd} give
large regions of parameter space with roughly the same
$\chi^2/dof$. We, therefore, resort to using physics motivated priors to determine the $M_i$.

The second challenge is that fits to $C^\textrm{2pt}(\tau)$ do not
expose multihadron (such as $N \pi$, $N \pi \pi, \ldots$) excited
states.  The reason is that the coupling of multihadron states with
nucleon quantum numbers to $\cal{N}$ is suppressed by factors of
$1/L^3$ for each extra state~\cite{Bar:2015zwa}. On the other hand, as discussed
later, their contribution to the matrix elements of certain operators
is enhanced and compensates for the volume suppression. Therefore, in
cases with enhanced contributions, these states need to be included in
fits to $C^{\text{3pt}}_\Gamma(t;\tau)$ because the mass of $N(1)
\pi(-1)$ (or $N(0) \pi(0) \pi(0) $) state is about 1230~MeV, which is
significantly smaller than the first radial excitation $N(1440)$.  In
short, our first goal is to determine all the excited states, i.e.,
their $M_i$, that make significant contributions.

Our overall approach consists of making simultaneous fits to
$C^\textrm{2pt}(\tau)$ (truncated at four states) and
$C^{\textrm{3pt}}_\Gamma (t;\tau)$ (truncated at 2 or 3 states) within
a single elimination jackknife process using binned data.\footnote{
The binning of the data is done in two steps. First, the $O(100)$
measurements made on each configuration are averaged. Second, these
configuration averages are further binned using bin sizes ranging
between 2--9 resulting in 250--325 data points for the various
ensembles. The auto-correlation function calculated using these binned
data, which enter in the single elimination jackknife process, show no
measurable correlations. Consequently, no augmentation of the single
elimination jackknife errors is deemed necessary.}  All fits are made
using the full covariance matrix with respect to the time slices
included in the fits.  It is, however, restricted to being block
diagonal for simultaneous fits to the two-and three-point correlation
functions as the off diagonal elements between them are not
well-determined with the current statistics.

We also carried out separate fits to the two- and three-point
functions and the two sets of results overlap.  No anomalous case
showing a bias in the simultaneous fits was found. Since the
simultaneous fits give more conservative errors and are better
motivated as a method, the results are presented using them.

In fits without priors, or wide priors, the $A_i$ and
$M_i$ are predominantly fixed by $C^\textrm{2pt}(\tau)$, while the
$C^{\textrm{3pt}}_\Gamma (t;\tau)$ give $A_i^\ast A_j \langle i |
O_\Gamma^q | j \rangle$ including the GSME $\langle 0 | O_\Gamma^q | 0
\rangle$. In this case the priors are used mainly to stabilize the
fits. The drawback, with current statistics, is that there is a whole
region in the $\{A_i,M_i\}$ space that gives fits to
$C^\textrm{2pt}(\tau)$ with similar $\chi^2/dof$.  Thus, there is
considerable range of plausible values for $M_1$ with the lowest,
theoretically, being the $N(1) \pi(-1)$ or $N(0) \pi(0) \pi(0) $.  One way to reduce this uncertainty is to input physically motivated values of
the $M_i$ through narrow priors.

Ideally, one wants a data driven analysis procedure, i.e., the data and the fit
should fix the $M_i$. With the current statistics, we are not able to
achieve this desired situation. We, therefore, carry out the full
analysis using two physically motivated but different values for $M_1$:
\begin{enumerate}
\item The ``standard'' fit to $C^\textrm{2pt}(\tau)$ and
  $C^\textrm{3pt}(t;\tau)$ employs wide priors for all excited-state
  amplitudes, $\mathcal{A}_i$, and masses, $M_i$. In these simultaneous fits, the
  priors are used only to stabilize the fits.  The $A_0$ and $M_i$,
  are mainly constrained by $C^\textrm{2pt}(\tau)$.  The value of $M_1
  $ found in this case is $\gtrsim 1.5$~GeV.  This simultaneous fit is
  labeled ``$\{4\}$" and $M_1$ is loosely identified with the radial
  excitation $N(1440)$.
\item The ``$N \pi$'' fit uses a narrow prior for $M_1$, centered on
  the non-interacting energy of the lowest allowed, $N(1) \pi(-1)$ or $N(0) \pi(0)
  \pi(0)$, lattice state on that ensemble.  In practice, the energies of
  these two states are very similar, so they can be considered as one
  in Eq.~\eqref{eq:2pt-sd} with a combined amplitude given by the
  sum. Throughout this work, for brevity we use the label $N \pi$ for
  their combined contribution. The priors for the other states are the same as in the ``standard'' fit.  This fit is labeled
  ``$\{4^{N \pi}\}$".
\end{enumerate}
Since the two fit strategies are not distinguished by the $\chi^2$, we
explain, when quoting the final results, the physics motivations for
the choices made. Also, where appropriate, we quote the difference as
an estimate of an associated systematic uncertainty. Further details of our implementation of these two fits strategies are
given in Refs.~\cite{Park:2021ypf,Gupta:2021ahb}.

\begin{figure*}[t]    
  \center
  \includegraphics[width=0.27\linewidth]{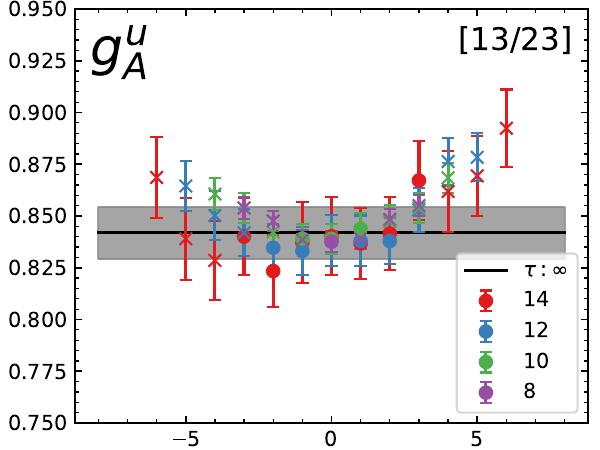} \hspace{0.6cm}
  \includegraphics[width=0.27\linewidth]{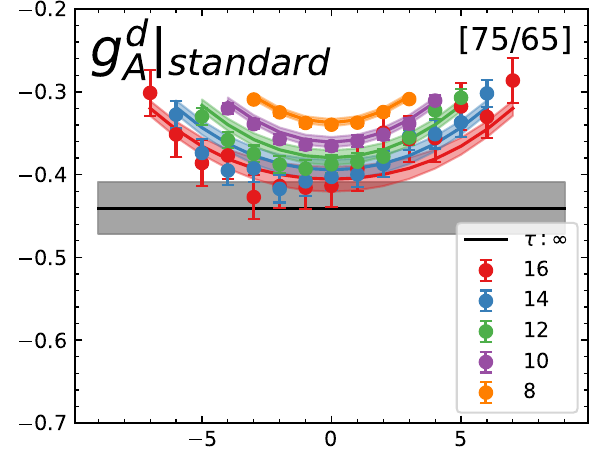}\hspace{0.6cm}
  \includegraphics[width=0.27\linewidth]{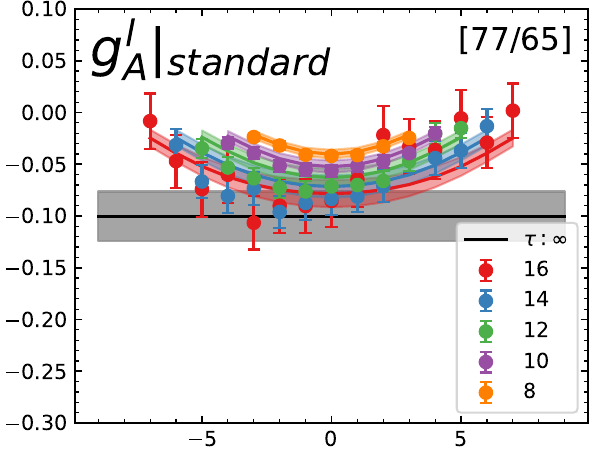}\\
  
  \phantom{\includegraphics[width=0.27\linewidth]{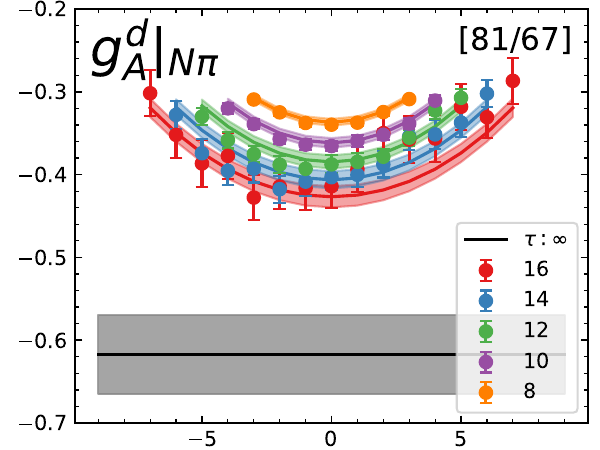}} \hspace{0.6cm} 
  \includegraphics[width=0.27\linewidth]{figs/3pt_AIC/gAd_4Npi_a09m130.pdf} \hspace{0.6cm}
  \includegraphics[width=0.27\linewidth]{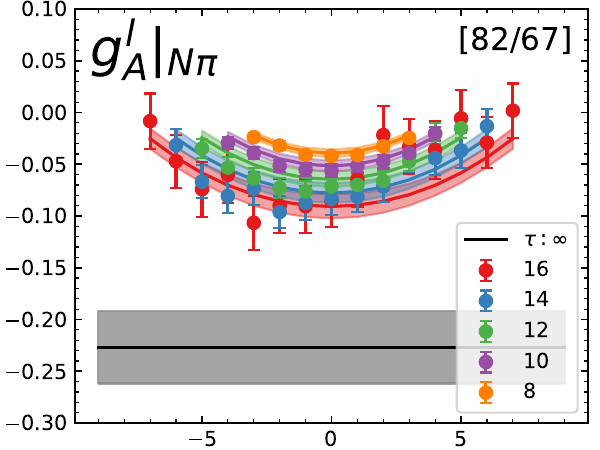}\\
  
  \includegraphics[width=0.27\linewidth]{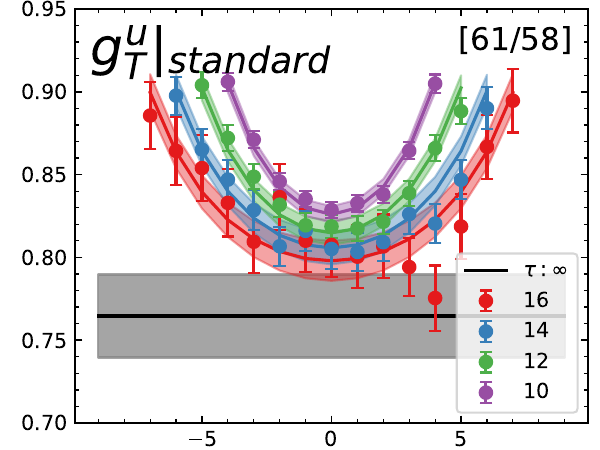}  \hspace{0.6cm}
  \includegraphics[width=0.27\linewidth]{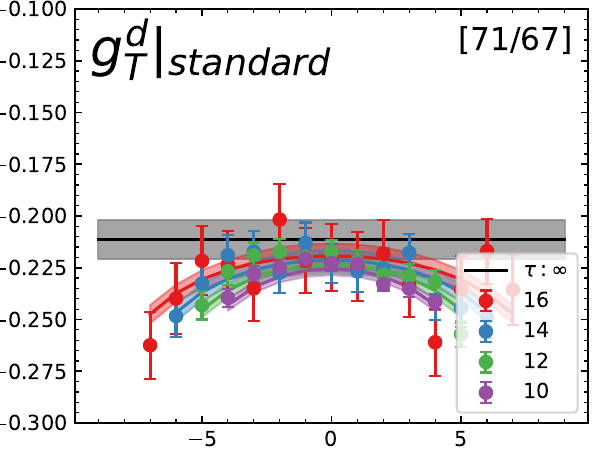}\hspace{0.6cm}
  \includegraphics[width=0.27\linewidth]{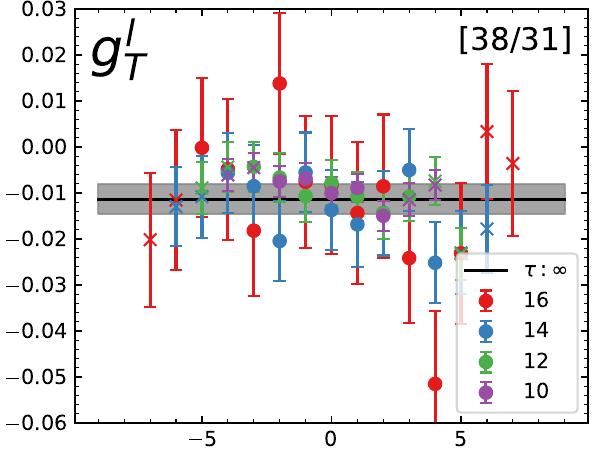}\\

  \includegraphics[width=0.27\linewidth]{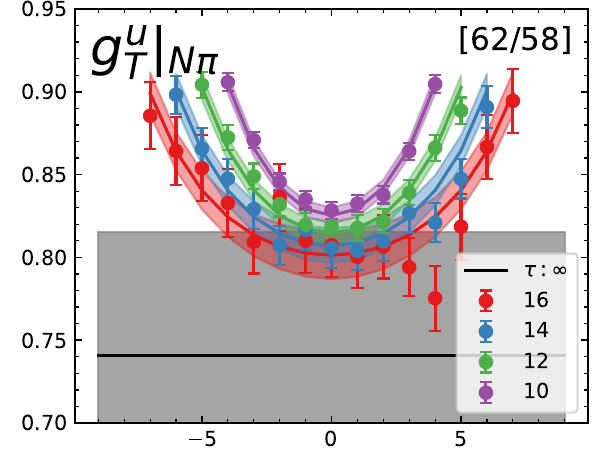}\hspace{0.6cm}
  \includegraphics[width=0.27\linewidth]{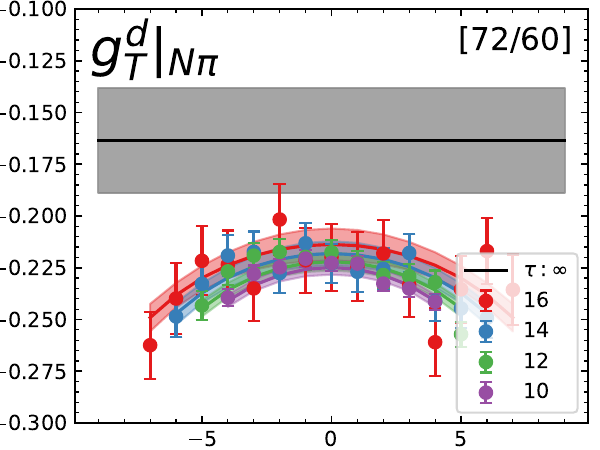}\hspace{0.6cm}
  \phantom{\includegraphics[width=0.27\linewidth]{figs/3pt_AIC/gTd_4Npi_a09m130.pdf}} 

  \includegraphics[width=0.27\linewidth]{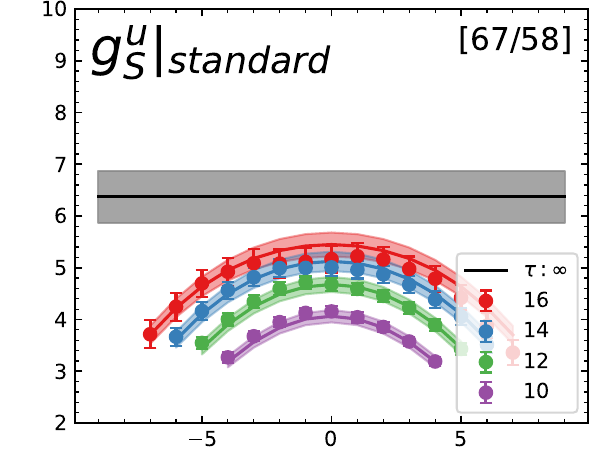}\hspace{0.6cm}
  \includegraphics[width=0.27\linewidth]{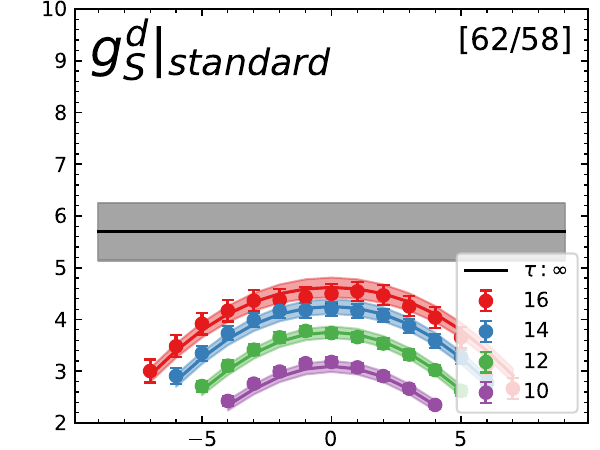}\hspace{0.6cm}
  \includegraphics[width=0.27\linewidth]{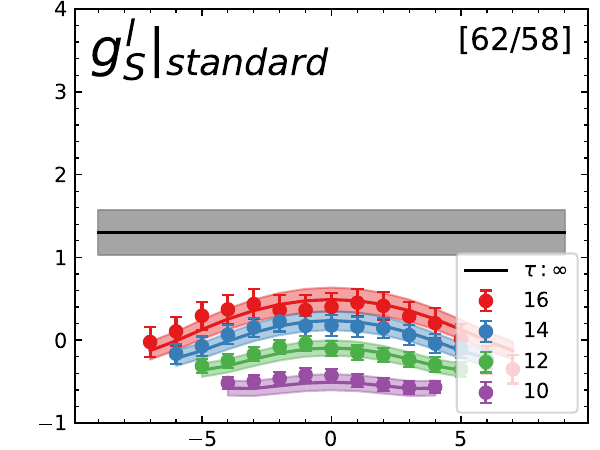}\\
  
  \includegraphics[width=0.27\linewidth]{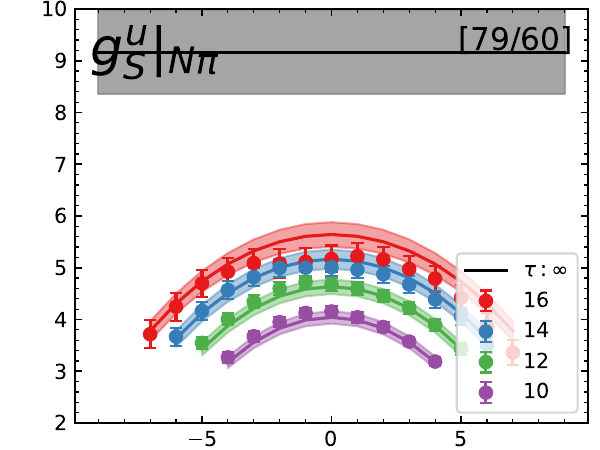}\hspace{0.6cm}
  \includegraphics[width=0.27\linewidth]{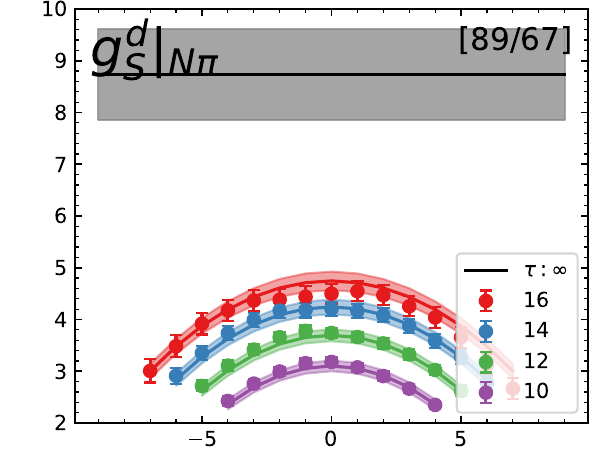}\hspace{0.6cm}
  \includegraphics[width=0.27\linewidth]{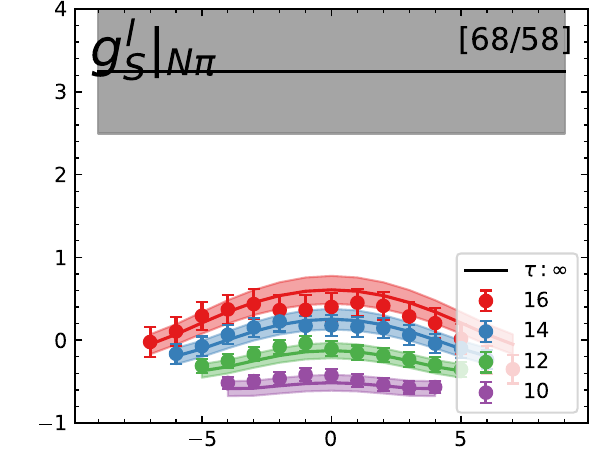}
  
  \caption{Data for the bare charges $g_{A,T,S}^{u,d}$ (sum of the
    connected and disconnected contributions) obtained at various
    separations $\{\tau,t\}$ are plotted versus $(t-\tau/2)/a$ for the
    physical $\mpi$ ensemble $a09m130$. We show both the ``standard''
    and ``$N \pi$'' fits, defined in the text, when ESC are
    manifest. In each case, the result of the fit is shown by lines of
    the same color as the data for the various $\tau/a$ listed in the
    label, and the $\tau\to\infty$ value is given by the gray band. We
    also show, for comparison, the light quark disconnected
    contributions, $g_{A,S,T}^{l}$, in the panels on the right and the
    ES fits to them. The $\chi^2/dof$ of the fit with the largest AIC
    score is given within square parenthesis in the top right corner.}
  \label{fig:gAgTgS_ESC}
\end{figure*}

\begin{figure*}[t]      
  \center
  \includegraphics[width=0.32\linewidth]{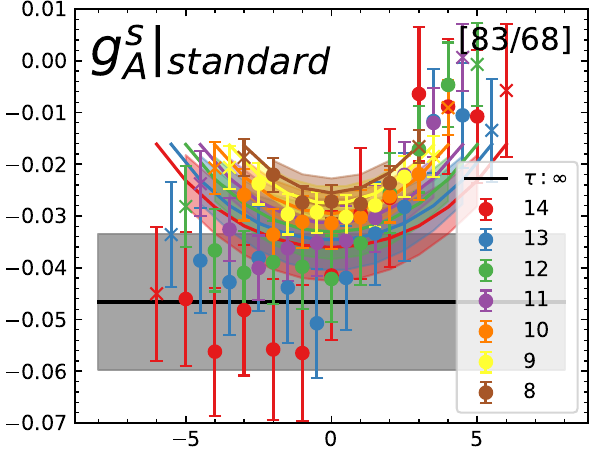}
  \includegraphics[width=0.32\linewidth]{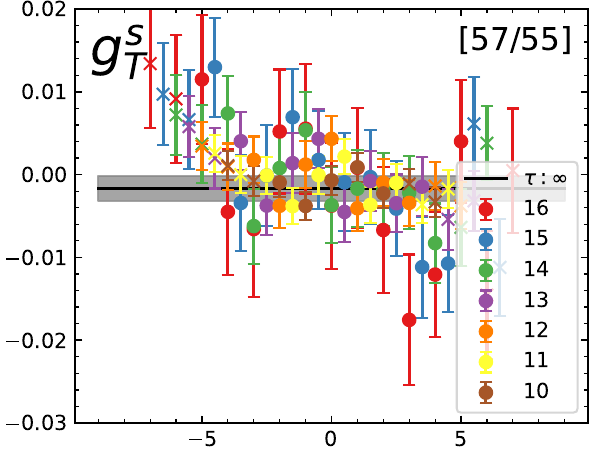}
  \includegraphics[width=0.32\linewidth]{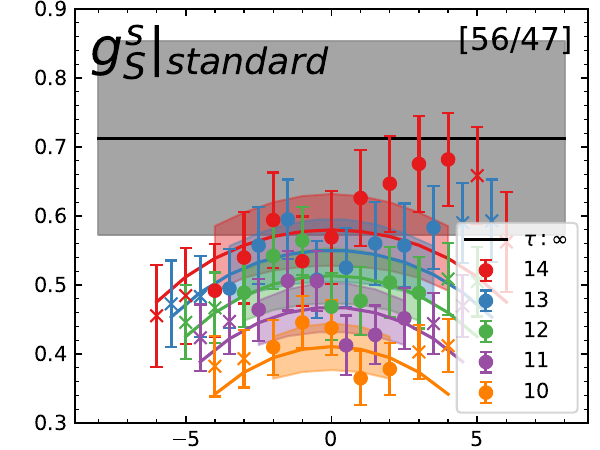}     
  \caption{Data for the bare strange charges,  $g_{A,T,S}^{s}$, obtained
    at various separations $\{\tau,t\}$, are plotted versus
    $(t-\tau/2)/a$ for the physical $\mpi$ ensemble $a09m130$.  These
    have only disconnected contributions. Only the ``standard'' fits
    to remove ESC are shown and considered since the expected lowest
    multihadron state, $\Sigma K$, is heavier than the nucleon's
    radial excitation.  Rest is the same as in
    Fig.~\protect\ref{fig:gAgTgS_ESC}.  }
  \label{fig:gAgTgS_s_ESC}
\end{figure*}

It is important to note that the mass of the $N \pi$ or the $N \pi
\pi$ state decreases below that for the first radial excitation
$N(1440)$ only for ensembles with $M_\pi \lesssim  200$~MeV, reaching $\sim
1230$~MeV for physical pion mass ensembles. Thus, in cases with
enhanced matrix elements, one expects to observe a strong dependence
on $M_\pi$ only for $M_\pi < 200$~MeV, which in our calculation leaves only
the $a09m130$ ensemble. Fits to these data are shown in
Figs.~\ref{fig:gAgTgS_ESC} and~\ref{fig:gAgTgS_s_ESC}.

Furthermore, previous studies indicate that the difference in nucleon
isovector axial and tensor charges, $g_{A,T}^{u-d}$, from the two
strategies is small \cite{Park:2021ypf}. This is also observed for
$g_{A,T}^{u,d,s}$ in this study, and the errors in results with the
``$N \pi$'' strategy are comparable to the differences. Only for
$g_{S}^{u,d}$ do the differences stand out.  Thus, to nail down the
best analysis strategy with up to 3-state fits requires future higher
statistics studies on more ensembles with $M_\pi \lesssim 200$ MeV.

To illustrate the difference in the values of $M_1$ coming out of the
``standard'' and ``$N\pi$'' analyses, we quote the
$a09m130$ values, which have the maximum difference. The
Bayesian prior and width used for the mass gap $a\Delta
M_1=a(M_1-M_0)$ is $0.35(20)$ for the ``standard'' and $0.12(2)$ for
the ``$N\pi$'' analysis with the central values corresponding to 0.79
GeV and 0.27 GeV in physical units, respectively.  The output
$a\Delta M_1$ from the simultaneous fits to $C^\text{2pt}$ and
$C^\text{3pt}_\Gamma$  are consistent with the input priors,
i.e., they lie in the range $0.24\sim 0.30$ for the ``standard'' and $0.11\sim
0.14$ for the ``$N\pi$'' analysis. We note little dependence of output values on the quark flavor and/or the operator. 

\section{Model average of various fits made to remove ESC}
\label{sec:models}

\begin{figure}[t]      
  \center
      \includegraphics[width=1.0\linewidth]{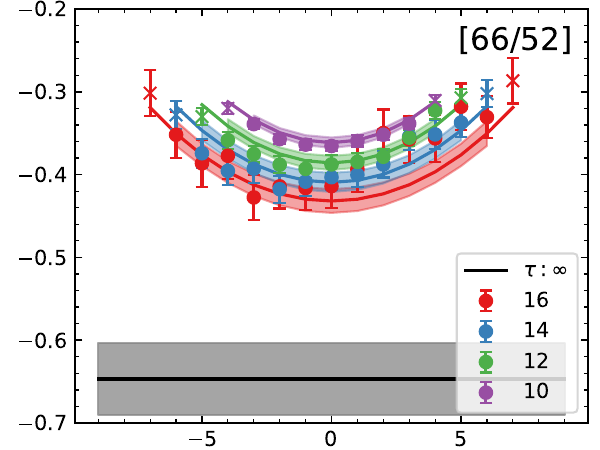}
      \includegraphics[width=1.0\linewidth]{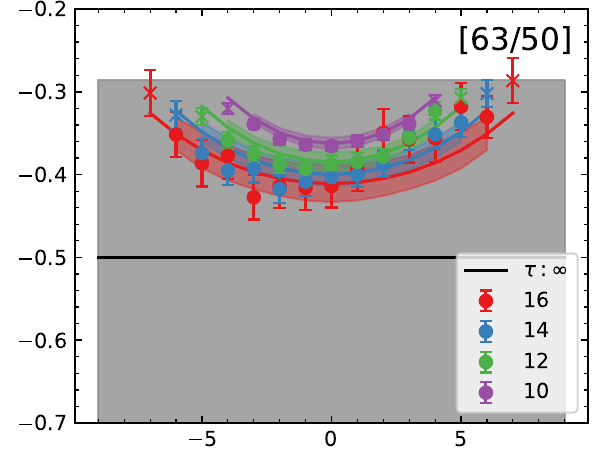}
  \caption{The two panels shows the 2-state (top) and 3-state (bottom)
    ``$N\pi$'' fit to the same data for $g_A^d$ from the $a09m130$
    ensemble.  Points with $\tau=\{10,12,14,16\}$ and
    $t_\text{skip}=2$ are included in the fits and are shown using
    filled circles. These two fits have the largest and similar AIC
    scores and dominate the model average, but differ
    qualitatively---the 2-state fit shows a large extrapolation, while
    the 3-state fit has a large error---and give different central
    values. }
  \label{fig:gAD+l_ESC}
\end{figure}

We have carried out the following twelve ESC fits, i.e., the 12 ``models",  for both the
``standard'' and the ``$N\pi$'' strategies to quantify change under
variations of the fit parameters.
\begin{itemize}
\item The full list of $\tau$ values analyzed (covering 0.9--1.5fm) is
  given in Table~\ref{tab:ens_params}. The ESC analysis is done (i)
  with and (ii) without using the data with the largest $\tau$. These largest $\tau$ data have the largest statistical but smallest systematic errors.
\item For each $\tau$, we neglect data on $\tskip$ time slices
  next to the source and the sink that have the largest ESC. Fits with
  three different $t_\text{skip}$ values in the range of 0.2--0.4~fm
  were made.
\item The spectral decomposition of the 3-point function,
  Eq.~\eqref{eq:3pt-sd}, is truncated at (i) 2-states and (ii)
  3-states. In all 3-state fits, the $\langle2|\CO|2\rangle$ term is
  set to zero as it is not resolved by our data.
\end{itemize}
For the final results, we take the average weighted by the AIC score
times the inverse of the variance, $\propto
\text{exp}[-(\chi^2-2N_\text{dof})/2]/\sigma^2$
\cite{1100705,Jay:2020jkz}.
In most cases, one set of values of $\tau$ and $t_\text{skip}$ gives
scores that are much higher than all others for both the 2- and
3-state fit.  Thus, in practice, the model average is mainly over the
variation of the result with the order of the truncation of
Eq.~\eqref{eq:3pt-sd}. While most 2- and 3-state fits are similar, an
example in the ``$N \pi$'' analysis where we note a qualitative
difference, even though the AIC scores were similar, is shown in
Fig.~\ref{fig:gAD+l_ESC}. Since we do not expect an enhanced $N\pi$
state contribution to $g_{A,T}^q$, the large extrapolation in the
2-state fit indicates that this is an artifact due to inadequate
statistics. The large error and smaller extrapolation in the 3-state
fit is further indication that the data, in this and such cases, are
statistics limited. More on this point in Sec.~\ref{sec:ESCpatterns}.

Results from a 2- versus 3-state truncation of Eq.~\eqref{eq:3pt-sd}
highlight the tension between balancing systematic and statistical
errors, which is accentuated in the ``$N\pi$'' analysis.  The 2-state
fit results may have significant systematic error from residual ESC
whereas the statistical precision of the data limits the robustness of
the 3-state fits. The same tension is also present between the fits
with the two sets of $\tau$ values.  The $\chi^2/dof$ or the AIC score
or the $p$ value of the fit, in most cases, does not distinguish
between them. We regard cases with large extrapolation in the ESC fits
or a large growth in errors or a large difference between 2- and
3-state fits as outliers due to inadequate statistics rather than a
physical effect.  The differences are, nevertheless, used to assign an
appropriate systematic uncertainty in these cases.

\section{Pattern of Excited-State Contributions to \texorpdfstring{$g_{A,S,T}^{u,d,s}$}{gA,S,T(u,d,s)}}
\label{sec:ESCpatterns}

The $a09m130$ ensemble data for the ratio
$C^\text{3pt}(t;\tau)/C^\text{2pt}(\tau)$ are shown in
Figs.~\ref{fig:gAgTgS_ESC} and ~\ref{fig:gAgTgS_s_ESC} as a function
of the values of separations $\tau$ and current insertion time $t$
simulated. As already stated above, we choose this physical pion mass
ensemble to point out the notable features since the differences in
$\Delta M_1$ between the ``standard'' and the ``$N \pi$'' analyses on
it are the largest.  In each panel, the final result, the AIC weighted
average of the twelve ESC fits, the models, is shown by the horizontal grey
band. The uncertainty bands on data points with a given color are
obtained as follows: The value at each $\{\tau,t\}$ point is the
weighted average of the twelve model values using the individual AIC
weights, and the "error" is the spread. These points are then
simply joined to form the bands. Note that these bands are not related
in any simple way to the final grey band, but are presented to show
the variation in the twelve values, which is given by the thickness of
the bands.

Theoretically, the data for all $\tau$ must become symmetric about
$t-\tau/2$ for the charges as the statistical precision improves, and
the convergence should become monotonic for large enough $\tau$. Also,
when choosing the $\tau_{\rm max}$ to use in the fits, the errors at
that $\tau$ should ideally be small enough to resolve the differences
in data at a given $t$ but different $\tau$.  Clearly, for the
physical mass ensemble $a09m130$, the data are already marginal by
these criteria at $\tau/a=14$ ($\approx 1.2$ fm).

In Refs.~\cite{Park:2021ypf,Lin:2018obj,Gupta:2018lvp,Gupta:2021ahb},
we had studied the pattern of ESC to $g_{A,S,T}^{u,d}$ separately for
the connected and disconnected
contributions. Fig.~\ref{fig:gAgTgS_ESC} here shows fits to the sum,
$C^{\text{conn}}_\Gamma(t;\tau) + C^{\text{disc}}_\Gamma(t;\tau)$.
For completeness, we have compared these results for charges versus
those obtained by doing separate fits to the connected and
disconnected contributions, and find the differences are smaller than
the errors.  A summary of our observations is as follows:
\begin{itemize}
\item $g_A^{u}$: The convergence of the connected contribution is from
  below. The disconnected contribution is negative ($ \sim -0.06$) and
  converges to a more negative value as illustrated by $g_A^{l,s}$ in
  Figs.~\ref{fig:gAgTgS_ESC} and~\ref{fig:gAgTgS_s_ESC}. A notable
  consequence is that the ESC mostly cancel in the sum and the data at
  different $\tau$ overlap within the current statistical precision as
  illustrated for $g_A^u$ in the top left panel in
  Fig.~\ref{fig:gAgTgS_ESC}. We, therefore, estimate the $\tau \to
  \infty$ value by the weighted average of the data for a range of
  $\tau$ and $t$ values, i.e., the ``plateau method''. Thus, the ``standard'' and the ``$N \pi$'' results are the same.
\item $g_A^{d}$: The connected and disconnected contributions are both
  negative and their ESC add in contrast to what is observed for
  $g_A^{u}$.  We, therefore, perform 2- and 3-state fits to obtain the
  $\tau \to \infty$ value and the extrapolation is significantly
  larger in the ``$N \pi$'' analysis.
\item $g_A^{l}$ and $g_A^{s}$: The disconnected contribution is small
  and negative. The convergence is towards a more negative value for
  both as evident from Figs.~\ref{fig:gAgTgS_ESC}
  and~\ref{fig:gAgTgS_s_ESC}.
\item $g_T^{u}$ and $g_T^{d}$: The ESC contribution to $g_T^{u}$
  ($g_T^{d}$) is significant (small). The disconnected contribution is
  small ($<0.01$) and negative.  Thus, it only marginally reduces
  $g_T^{u}$ and increases $|g_T^{d}|$. More important, the fits to
  $g_T^{u}$ and $g_T^{d}$ are good and the size of the
  extrapolation to the $\tau \to \infty$ value is small.
\item $g_T^{l}$ and $g_T^{s}$: The values are much smaller than the
  errors and there is no clear pattern of convergence.  We, therefore,
  assume that there are no significant ESC and make two fits to a
  constant ansatz---with and without using the full covariance
  matrix. The results are consistent, however, the fit with the
  covariance matrix gives a larger error, and is chosen for 
  the final result.
\item $g_S^{u}$ and $g_S^{d}$: The connected and disconnected
  contributions to {\it both} are positive, the ESC are large, and the
  convergence is from below. The size of the extrapolation in the ``$N
  \pi$'' fits is large, giving larger values for $g_S^{u,d}$, which
  implies a larger pion-nucleon sigma term~\cite{Gupta:2021ahb}.
\item $g_S^{l}$ and $g_S^{s}$: Their values are also positive and
  significant, however, the errors in the current data are large. In
  one case, $g_S^s$ from the $a09m310$ ensemble, only the diagonal
  elements of the covariance matrix were used since we did not find
  reasonable ranges of $\{\tau,t,t_{\rm skip}\}$ for which the
  off-diagonal elements were stable.
\end{itemize}

Results for the bare charges extracted using the two strategies to
remove ESC, and after model averaging, are summarized in
Tables~\ref{tab:gAbare_final},~\ref{tab:gSbare_final}
and~\ref{tab:gTbare_final}. Their renormalization and conversion to the
${\MSb}$ scheme at the scale $2$~GeV is done  using the analysis
for the 2+1-flavor theory described in Appendix~\ref{sec:NPR}.

The main question, in addition to the truncation of
Eq.~\eqref{eq:3pt-sd} to 2- and 3-state fits, is which analysis should
be selected for the final values. In the ``$N \pi$'' analysis, $M_1^{N
  \pi}$ is set to the non-interacting energy of the $N \pi$ state
through a narrow prior and is not the output of a data driven fit.
While the $N \pi$ state is expected to contribute in principle to all
the charges (there is a contribution at one-loop in $\chi$PT
~\cite{Bar:2016uoj}), no enhanced contribution is expected except to
the scalar charges~\cite{Gupta:2021ahb} as evident from
Fig.~\ref{fig:gAgTgS_ESC}.  Furthermore, ESC from multihadron states,
relative to those from the radial excitations, are expected to become
significant only for $M_\pi \lesssim 200$~MeV, i.e., below which the $M_1^{N
  \pi}$ starts to be smaller than $M_1^{\rm radial}$. Thus, to
distinguish between the ``standard'' and ``$N\pi$'' fits, we must rely
on data close to the physical pion mass, i.e., in this work on only
the $a09m130$ data shown in Figs.~\ref{fig:gAgTgS_ESC} and
~\ref{fig:gAgTgS_s_ESC}. With these caveats, our reasons for choosing
the analysis strategy for presenting the final results for each of the
charges are given below.

The error in the $\tau \to \infty$ extrapolated value for
$g_{A}^{u,d}$ in the ``$N \pi$'' analysis is large. For $g_A^u$, we
see no significant ESC, so the ``standard'' fit value is taken. In
$g_A^d$, the ESC to the connected and disconnected terms add and the
``$N \pi$'' fit value is about $30\%$ larger. We do not have a
theoretical reason for such a large effect, nevertheless we take the
average of the two estimates. Note, results for the isovector axial
charge $g_{A}^{u-d}$, which has no disconnected contributions, are
consistent with our final values already given in
Ref.~\cite{Jang:2023zts} as shown in Table~\ref{tab:isovector}.

In the case of $g_{T}^{u,d}$, the ``standard'' and ``$N \pi$''
analyses give essentially the same central value and the extrapolation
is small. Also, the ESC to the connected and disconnected parts mostly
cancel in $g_T^d$.  The errors, however, are much larger, beyond what
data would indicate, with the ``$N \pi$'' fit. Since there is no
phenomenological or $\chi$PT motivation for a large contribution from
multihadron states in the tensor channel, we choose the ``standard''
analysis.

The $g_{A,T}^s$, illustrated in Fig.~\ref{fig:gAgTgS_s_ESC}, get only
disconnected contributions and the lowest multihadron state is likely
to be $\Sigma K$ as illustrated in Fig.~\ref{fig:conn_disc}). This is
heavier than the radial excitation of the nucleon. Also, the data are
noisy and the values are small. For these reasons we choose the
``standard'' analysis. It also gives an error estimate that we
consider conservative.

The data for $g_{S}^{u,d}$ are the cleanest and the extrapolated
values have reasonable errors. The two strategies, however, give
different values but are not distinguished by the $\chi^2/dof$.
The size of the extrapolation is large in the ``$N \pi$'' analysis
and, on its own, calls into question even a 3-state fit.  However,  $\chi$PT suggests such an  enhanced contribution from the $N \pi$
and $N \pi \pi $ states as discussed in
Ref.~\cite{Gupta:2021ahb}. There we presented results for the
pion-nucleon sigma term, $\sigma_{\pi N}$, and highlighted the
$\approx 50\%$ difference in $g_S^{u,d}$ from the two analysis
strategies. The same reasoning leads us to continue to choose the ``$N\pi$''
analysis in this work. Overall, our conclusion is that $g_S^{u,d}$
show large enhanced contributions from the $N\pi$ excited states
unlike $g_{A,T}^{u,d}$. For $g_S^{s}$, the lowest multihadron state
would again be the heavier $\Sigma K $, so no enhancement is expected
with respect to the ``standard'' analysis.

Lastly, even though we have chosen the ``standard" analysis values for
$g_{A,S}^s$, we do show the CCFV extrapolation of the data from the ``$N\pi$''
analysis in Figs.~\ref{fig:CCFV_gAs} and \ref{fig:CCFV_gSs}.  The
procedure for the renormalization of these charges is discussed in
Appendix~\ref{sec:NPR}.

\begin{table*}[!p]

\begin{ruledtabular}
\footnotesize
  \begin{tabular}{l l lllllllllll}
  Ensemble & & $g_A^{u-d}$ & $g_A^{u}$ & $g_A^{d}$ & $g_A^{u+d}$ & $g_A^{l}$ & $g_A^{s}$  \\
\hline

\multirow{2}{*}{a15m310} & $\{3\}$ & 1.2416(57) & 0.892(12) & -0.3538(57) & 0.535(13) & -0.0431(28) & -0.0277(16)\\

 & $\{3^{N\pi}\}$ & 1.2507(82) & 0.892(12) & -0.3646(76) & 0.516(16) & -0.0510(39) & -0.0315(22)\\

\hline

\multirow{2}{*}{a12m310} & $\{4\}$ & 1.267(11) & 0.8896(97) & -0.372(24) & 0.469(29) & -0.0696(77) & -0.0401(48)\\

 & $\{4^{N\pi}\}$ & 1.2741(79) & 0.8896(97) & -0.372(21) & 0.479(20) & -0.0699(64) & -0.0420(27)\\

\hline

\multirow{2}{*}{a12m220} & $\{4\}$ & 1.285(30) & 0.886(17) & -0.415(28) & 0.457(43) & -0.083(16) & -0.049(13)\\

 & $\{4^{N\pi}\}$ & 1.304(23) & 0.886(17) & -0.441(28) & 0.425(55) & -0.098(21) & -0.0595(72)\\

\hline

\multirow{2}{*}{a09m310} & $\{4\}$ & 1.256(45) & 0.8695(87) & -0.3614(78) & 0.506(14) & -0.0532(57) & -0.0304(25)\\

 & $\{4^{N\pi}\}$ & 1.251(11) & 0.8695(87) & -0.3688(85) & 0.497(16) & -0.0578(66) & -0.0336(35)\\

\hline

\multirow{2}{*}{a09m220} & $\{4\}$ & 1.307(80) & 0.8558(93) & -0.414(17) & 0.441(24) & -0.084(15) & -0.0448(60)\\

 & $\{4^{N\pi}\}$ & 1.318(33) & 0.8558(93) & -0.436(25) & 0.382(30) & -0.115(17) & -0.0620(85)\\

\hline

\multirow{2}{*}{a09m130} & $\{4\}$ & 1.281(32) & 0.842(13) & -0.441(32) & 0.386(46) & -0.100(24) & -0.047(13)\\

 & $\{4^{N\pi}\}$ & 1.406(42) & 0.842(13) & -0.618(48) & 0.104(77) & -0.227(35) & -0.118(34)\\

\hline

\multirow{2}{*}{a06m310} & $\{4\}$ & 1.196(45) & 0.870(31) & -0.372(23) & 0.488(47) & -0.067(20) & -0.0396(49)\\

 & $\{4^{N\pi}\}$ & 1.227(31) & 0.862(29) & -0.379(18) & 0.468(37) & -0.073(13) & -0.0429(55)\\

\hline

\multirow{2}{*}{a06m220} & $\{4\}$ & 1.256(32) & 0.842(24) & -0.390(17) & 0.434(34) & -0.079(15) & -0.0476(70)\\

 & $\{4^{N\pi}\}$ & 1.271(32) & 0.843(28) & -0.396(19) & 0.426(38) & -0.094(26) & -0.0515(85)\\

\end{tabular}
\end{ruledtabular}
\caption{ Final bare axial charges $g_A^{u,d,s}$. These $g_A^{u-d}$
  results are obtained by doing the ES fit to the difference of the
  connected data with the insertion of the $u$ and $d$ quarks to take
  into account the correlations between the two. The $g_A^{u}$ and
  $g_A^{d}$ are extracted from a fit to the sum of the connected plus
  disconnected correlation functions. Their sum, $g_A^{u} + g_A^{d}$,
  can be compared to the result of a fit to the sum of the correlation
  functions, $g_A^{u+d}$. The errors are calculated using a single
  elimination jackknife method using data binned as described in the text, and each number is the AIC weighted
  average over the various model values. The results from the
  ``standard'' and ``$N \pi$'' ESC fits are given separately.}
\label{tab:gAbare_final}
\end{table*}

\begin{table*}[!p]

\begin{ruledtabular}
\footnotesize
  \begin{tabular}{l l lllllllllll}
  Ensemble & & $g_T^{u-d}$ & $g_T^{u}$ & $g_T^{d}$ & $g_T^{u+d}$ & $g_T^{l}$ & $g_T^{s}$  \\
\hline

\multirow{2}{*}{a15m310} & $\{3\}$ & 1.1231(59) & 0.902(12) & -0.2278(39) & 0.669(11) & \multirow{2}{*}{-0.0075(16)} & \multirow{2}{*}{-0.00204(54)}\\

 & $\{3^{N\pi}\}$ & 1.1096(80) & 0.889(14) & -0.2211(47) & 0.654(14)\\

\hline

\multirow{2}{*}{a12m310} & $\{4\}$ & 1.0608(81) & 0.827(25) & -0.2221(57) & 0.601(34) & \multirow{2}{*}{-0.00517(94)} & \multirow{2}{*}{-0.00188(55)}\\

 & $\{4^{N\pi}\}$ & 1.0509(96) & 0.837(18) & -0.2219(52) & 0.612(25)\\

\hline

\multirow{2}{*}{a12m220} & $\{4\}$ & 1.062(13) & 0.865(21) & -0.2087(92) & 0.657(25) & \multirow{2}{*}{-0.0046(23)} & \multirow{2}{*}{-0.0020(10)}\\

 & $\{4^{N\pi}\}$ & 1.039(28) & 0.846(25) & -0.198(13) & 0.648(36)\\

\hline

\multirow{2}{*}{a09m310} & $\{4\}$ & 1.0288(47) & 0.8210(84) & -0.2086(34) & 0.6099(83) & \multirow{2}{*}{-0.0056(10)} & \multirow{2}{*}{-0.00159(63)}\\

 & $\{4^{N\pi}\}$ & 1.033(19) & 0.822(15) & -0.2080(45) & 0.6030(95)\\

\hline

\multirow{2}{*}{a09m220} & $\{4\}$ & 0.997(13) & 0.791(15) & -0.2015(40) & 0.568(26) & \multirow{2}{*}{-0.0047(19)} & \multirow{2}{*}{-0.00178(69)}\\

 & $\{4^{N\pi}\}$ & 0.982(20) & 0.773(21) & -0.1941(46) & 0.538(39)\\

\hline

\multirow{2}{*}{a09m130} & $\{4\}$ & 1.003(11) & 0.765(25) & -0.2114(96) & 0.528(29) & \multirow{2}{*}{-0.0113(32)} & \multirow{2}{*}{-0.0017(15)}\\

 & $\{4^{N\pi}\}$ & 1.038(86) & 0.741(74) & -0.164(25) & 0.48(11)\\

\hline

\multirow{2}{*}{a06m310} & $\{4\}$ & 0.985(15) & 0.8005(68) & -0.2023(38) & 0.5907(81) & \multirow{2}{*}{-0.0057(14)} & \multirow{2}{*}{-0.00201(74)}\\

 & $\{4^{N\pi}\}$ & 0.962(31) & 0.788(22) & -0.201(12) & 0.581(26)\\

\hline

\multirow{2}{*}{a06m220} & $\{4\}$ & 0.961(18) & 0.757(20) & -0.186(10) & 0.562(21) & \multirow{2}{*}{-0.0058(16)} & \multirow{2}{*}{-0.00155(81)}\\

 & $\{4^{N\pi}\}$ & 0.933(34) & 0.756(27) & -0.182(12) & 0.557(27)\\

\end{tabular}
\end{ruledtabular}
\caption{ Final bare 
  tensor charges $g_T^{u,d,s}$. The rest is the same as in Table~\ref{tab:gAbare_final}.}
\label{tab:gTbare_final}
\end{table*}

\begin{table*}[!p]

\begin{ruledtabular}
\footnotesize
  \begin{tabular}{l l lllllllllll}
  Ensemble & & $g_S^{u-d}$ & $g_S^{u}$ & $g_S^{d}$ & $g_S^{u+d}$ & $g_S^{l}$ & $g_S^{s}$  \\
\hline

\multirow{2}{*}{a15m310} & $\{3\}$ & 0.829(24) & 3.83(11) & 3.068(87) & 6.91(20) & 0.648(77) & 0.701(36)\\

 & $\{3^{N\pi}\}$ & 0.829(31) & 3.94(12) & 3.26(12) & 7.17(23) & 0.761(70) & 0.709(32)\\

\hline

\multirow{2}{*}{a12m310} & $\{4\}$ & 0.906(24) & 4.64(23) & 3.95(33) & 8.54(54) & 1.10(24) & 0.694(80)\\

 & $\{4^{N\pi}\}$ & 0.893(58) & 4.57(15) & 3.73(14) & 8.28(28) & 0.978(72) & 0.671(47)\\

\hline

\multirow{2}{*}{a12m220} & $\{4\}$ & 1.037(88) & 5.70(31) & 4.53(22) & 10.23(50) & 1.15(23) & 0.664(70)\\

 & $\{4^{N\pi}\}$ & 1.13(11) & 6.22(33) & 4.98(28) & 11.33(57) & 1.53(25) & 0.719(64)\\

\hline

\multirow{2}{*}{a09m310} & $\{4\}$ & 1.008(22) & 4.47(14) & 3.77(28) & 8.10(34) & 0.73(18) & 0.499(57)\\

 & $\{4^{N\pi}\}$ & 1.047(33) & 4.73(21) & 3.81(21) & 8.55(43) & 0.88(15) & 0.597(82)\\

\hline

\multirow{2}{*}{a09m220} & $\{4\}$ & 1.008(31) & 5.69(42) & 4.68(43) & 10.34(83) & 1.19(38) & 0.69(13)\\

 & $\{4^{N\pi}\}$ & 1.001(62) & 6.24(41) & 5.18(41) & 11.38(81) & 1.53(30) & 0.87(15)\\

\hline

\multirow{2}{*}{a09m130} & $\{4\}$ & 0.89(10) & 6.37(50) & 5.70(56) & 12.0(1.0) & 1.30(27) & 0.71(14)\\

 & $\{4^{N\pi}\}$ & 0.95(27) & 9.17(81) & 8.74(88) & 17.5(1.5) & 3.25(75) & 1.11(30)\\

\hline

\multirow{2}{*}{a06m310} & $\{4\}$ & 1.180(70) & 5.78(54) & 4.53(43) & 10.47(98) & 0.99(15) & 0.64(10)\\

 & $\{4^{N\pi}\}$ & 1.258(89) & 5.78(49) & 4.59(36) & 10.47(84) & 1.12(16) & 0.69(12)\\

\hline

\multirow{2}{*}{a06m220} & $\{4\}$ & 0.818(88) & 6.15(61) & 6.2(1.2) & 12.8(1.7) & 1.51(31) & 0.520(96)\\

 & $\{4^{N\pi}\}$ & 0.83(11) & 6.19(57) & 6.01(90) & 12.7(1.4) & 1.67(38) & 0.62(20)\\
 
   \end{tabular}
\end{ruledtabular}
\caption{ Final bare 
  scalar charges $g_S^{u,d,s}$. The rest is the same as in Table~\ref{tab:gAbare_final}.}
\label{tab:gSbare_final}

\end{table*}

\section{Chiral-Continuum-Finite-volume (CCFV) extrapolation and error budget}
\label{sec:CCFV}

The renormalized axial, $g_A^{u,d,s}$ and tensor, $g_T^{u,d,s}$ charges 
are extrapolated to the physical point, $a \to 0$, $M_\pi = 135$~MeV, and 
$M_\pi L \to \infty$, using the ansatz 
\begin{align}
 g(a,M_\pi,M_\pi L) =  c_0+c_a a+ c_2 M_\pi^2 + c_3 \frac{M_\pi^2 e^{-M_\pi L}}{\sqrt{M_\pi L}} \,,
 \label{eq:CCFV-1}
\end{align}
that includes the leading corrections, pertinent to our lattice setup, in all three variables $\{a,M_\pi,M_\pi L\}$.

For the scalar charges $g_S^u$ and $g_S^d$, chiral perturbation theory
gives two differences. First, the chiral behavior, $g_S^{u,d} = d_0
+ d_a a + d_1 \mpi + d_2 \mpi^2 + M_\pi^2 \log M_\pi^2 + \ldots$,
starts with a term proportional to
$\mpi$~\cite{Hoferichter:2015hva}. Second, the contribution of $N \pi$
and $N \pi \pi $ excited states from higher order terms (chiral logs,
etc.)  is large as discussed in~\cite{Gupta:2021ahb}.  The difference
on including these in ESC fits is shown in
Fig.~\ref{fig:gAgTgS_ESC}. Thus, one needs to include the higher order terms (chiral logs,
etc.) in both, fits to remove possibly
large ESC and in the chiral part of the CCFV ansatz. Unfortunately, with data
at only three values of $M_\pi$, only the first two terms, $\propto
M_\pi$ and $M_\pi^2$ in addition to lattice spacing dependence, can be
kept without overparameterization. Noting these points, the CCFV ansatz used for the scalar charges
$g_S^{u,d}$ is
\begin{align}
  g_S^{u,d}(a,M_\pi,M_\pi L)  &=d_0+d_a a+ d_2 M_\pi+ d_3 M_\pi^2 \nonumber \\
  &+ d_4 M_\pi \left(1-\frac{2}{M_\pi L}\right) e^{-M_\pi L} \,.
   \label{eq:CCFV-2}
\end{align}
Note that the finite-volume term is also
modified~\cite{Gupta:2021ahb}.  For the strange scalar charge $g_S^s$,
and for $g_{\{A,T\}}^s$, we do not expect a large contribution from
pion loops since the lightest multihadron state is expected to be the
much heavier $\Sigma K$. So we use the ansatz given in
Eq.~\eqref{eq:CCFV-1}.

The final results for the charges, after CCFV (or CC) extrapolations
for the two strategies and the two renormalization methods, are
summarized in Table~\ref{tab:gAST-Z}.  From these we obtain the final
central values and the error budget given in
Table~\ref{tab:final-g}. Adding errors in quadrature, the final
results are given in Table~\ref{tab:Xfinal-g}.  We show, in
Figs.~\ref{fig:CCFV_gAu}---\ref{fig:CCFV_gSs} in
Appendix~\ref{sec:CCFV-fits}, only the CCFV fits but point out that
the finite volume correction is small in all cases as can be inferred
from the values given in Table~\ref{tab:gAST-Z}.  In panels showing
CCFV fits versus $M_\pi^2$, the additional grey bands show fits
keeping only the chiral corrections given in Eqs.~\eqref{eq:CCFV-1}
and~\eqref{eq:CCFV-2}. Overlap between these grey bands and the CCFV
fits (pink bands), implies that the other two systematics,
discretization and finite volume corrections, are small. Overall, no
significant finite-volume corrections are observed for $M_\pi L
\gtrsim 4$.

Some specific observations on these CCFV fits to obtain the charges at the physical point are as follows. 

\subsection{Axial}

CCFV fits in Fig.~\ref{fig:CCFV_gAu} and \ref{fig:CCFV_gAd} for
$g_A^u$ and $g_A^d$, respectively, show similar large positive slope versus $M_\pi^2$,
but with a larger uncertainty in the ``$N\pi$'' analysis.  There also is a significant 
positive slope in $g_A^u$  versus $a$ but 
not in $g_A^d$. These variations decrease $g_A^u$ and make $g_A^d$ more negative
as $a \to 0$ and $M_\pi \to 135$~MeV. They add in $g_A^{u+d}$, making it smaller, and almost cancel in $g_A^{u-d}$. 

The  CCFV estimates for $g_A^u$, summarized in Table~\ref{tab:gAST-Z}, are, within 
errors, independent of the ESC strategy and the renormalization method. 
We therefore, take the weighted average of the four values and the largest error to 
get $g_A^u=0.781(22)$. To 
this we add a systematic uncertainty of 0.011 to account for the difference between the 
CCFV and CC values and quote $g_A^u=0.781(22)(11)_{\rm CC}$ in Table~\ref{tab:final-g}. 

Estimates of $g_A^d$ are consistent between the two renormalization
methods but $\approx 10\%$ more negative with the ``$N\pi$''
analysis. Based on the reasoning given in the extraction of the
isovector axial form factors in~\cite{Park:2021ypf,Jang:2023zts}, the
``$N\pi$'' analysis is suggested, however, given the size of the
errors, we defer choosing. So we take the mean and quote half the
difference, $0.024$, as a systematic uncertainty along with $0.009$
for the difference between the the CCFV and CC values. This gives
$g_A^d=-0.440(29)(9)_{\rm CC}(24)_{N\pi}$.

For $g_A^s$, we choose the ``standard'' analysis as discussed
previously and assign a systematic uncertainty only for the CCFV
versus CC fits to get $g_A^s=0.055(9)(1)_{\rm CC}$. These results for
$g_A^{u,d,s}$, summarized in Table~\ref{tab:final-g}
and~\ref{tab:Xfinal-g}, are consistent with, and update those
published in Ref.~\cite{Lin:2018obj}.

\subsection{Tensor}

The magnitude of ESC in $g_T^u$ is significant, while the errors in
$g_T^d$ and $g_T^s$ are large, especially on the physical pion mass
ensemble $a09m130$, as shown in Fig.~\ref{fig:gAgTgS_ESC} and in
Ref.~\cite{Gupta:2018lvp}. The CCFV fits are shown in
Fig.~\ref{fig:CCFV_gTu}, \ref{fig:CCFV_gTd} and
\ref{fig:CCFV_gTs}. The ``standard'' and ``$N\pi$'' analyses give
consistent estimates but the errors in the ``$N \pi$'' case are
$\approx 50\%$ larger.  For all three charges, $g_T^{u,d,s}$, 
we average the values from
the two renormalization methods and use the difference between CCFV 
and CC fits to assign a systematic uncertainty. 
Since no enhanced contributions from multihadron, ``$N\pi$'',
states is expected in the tensor channel, our choice is the ``standard'' analysis and the CCFV
fits. Nevertheless, to be conservative and account for the $\approx 2 \sigma$ difference 
between the ``standard'' and $N \pi$ values for $g_T^d$, we assign half the difference,  0.012,  
as an additional systematic uncertainty. The final results are
summarized in Tables~\ref{tab:final-g} and~\ref{tab:Xfinal-g}.
\looseness-1

\subsection{Scalar}
\label{sec:scalar}

The behavior of $g_S^u$ and $g_S^d$ versus $a$ is similar and there is
a significant negative slope, i.e., the value increases as $a \to 0$
as shown in Figs.~\ref{fig:CCFV_gSu} and
\ref{fig:CCFV_gSd}. Similarly, the value increases as $M_\pi \to
135$~MeV. There is a significant difference 
between the ``standard'' and ``$N\pi$'' fit values, and the 
extrapolation of the $N\pi$ data is large. The physical pion mass data point $a09m130$,  therefore, plays a crucial role in the
extrapolation.  Results after CCFV extrapolation of the ``standard''
and ``$N\pi$'' analyses data are given in Tables~\ref{tab:gAST-Z}
and~\ref{tab:final-g}. Based on the $\chi$PT analysis presented in
Ref.~\cite{Gupta:2021ahb}, we pick the ``$N\pi$'' analysis for the
final values of $g_S^{u,d}$ given in Table~\ref{tab:Xfinal-g}. Again,
we choose the ``standard'' analysis value for $g_S^s$ as the lowest
multihadron state is expected to be $\Sigma K$.

\subsection{The pion-nucleon sigma term $\sigma_{\pi   N}$ }
\label{sec:sigma_term}

Results for the renormalization group invariant pion-nucleon sigma
term were presented in Ref.~\cite{Gupta:2021ahb} where it was
calculated on each ensemble using the bare quantities $\sigma_{\pi
  N}=m_l^\text{bare}g_S^{u+d,\text{bare}}$ and the data 
extrapolated to the physical point using the N$^2$LO $\chi$PT
expression~\cite{Hoferichter:2015hva}:\looseness-1
\begin{equation}
\sigma_{\pi N} =  (d_2 +d_{2}^{a} a) \mpi^2 + d_3  \mpi^3 + d_4 \mpi^4 + d_{4L} \mpi^4 \log \frac{\mpi^2}{\mN^2} \,.
\label{eq:CPT}
\end{equation}
The $\chi$PT analysis in Ref.~\cite{Gupta:2021ahb} suggested that all
five terms contribute significantly in a CC fit, and finite-volume
corrections are less than $1\MeV$ for the physical pion mass ensemble
$a09m130$.  So we again present results from the CC
extrapolation. With data at only three values of $M_\pi \approx 135,
220, 310$~MeV, we can include at most three chiral terms. The fit with
the leading terms is shown in Fig.~\ref{fig:CCFV_sigma} (top). The
bottom panels show the fit including the fourth chiral term, $\propto
d_{4 L}$, realized by fixing the coefficient $d_3$ to its $\chi$PT
value, $ d_3^\chi$, evaluated with $\mN=0.939\GeV$, $g_A=1.276$, and
$F_\pi=92.3\MeV$~\cite{Gupta:2021ahb}. Averaging these two values
gives our updated results, $\sigma_{\pi N}|_{N\pi} = 61(6)$~MeV for
the ``$N \pi$'' and $\sigma_{\pi N}|_{\rm standard} = 41(6)$~MeV for
the ``standard'' analysis, which overlap with those given in
Ref.~\cite{Gupta:2021ahb}. \looseness-1

In this work, we also do 
the CCFV extrapolation of the
renormalized charges  $g_S^{u+d}$ and
$g_S^{u,d,s}$. Our preferred value for the isoscalar charge,  $g_S^{u+d}$, is obtained 
by doing ESC fits to the sum of the total $u$ and $d$ correlators. The results are $g_S^{u+d}=12.4(1.2)$ for the ``standard''
and $g_S^{u+d}=17.8(1.6)$ for the ``$N\pi$'' analysis. The value   $g_S^{s}=0.37(14)$ is, as explained before, taken from the standard analysis.  
Combining these with the renormalized quark masses $m_{l}=3.427(51)$
and $m_s=93.46(58)$~MeV given in the FLAG
2024 report~\cite{FlavourLatticeAveragingGroupFLAG:2024oxs} for the $N_f=2+1+1$ theory, provides the second 
estimates: $\sigma_{\pi N}|_{N\pi} = 61(6)$~MeV for the $N \pi$
analysis, and $\sigma_{\pi N}|_{\rm standard} = 43(4)$~MeV and
$\sigma_s=35(13)$~MeV from the ``standard'' analysis. Note that if the individual values of   
$g_S^{u}$ and $g_S^{d}$  are taken from Table~\ref{tab:final-g}, then  $\sigma_{\pi N}$ 
is larger by about 2~MeV. 

\begin{figure*}[htb]      
\subfigure{
     \includegraphics[width=0.48\linewidth]{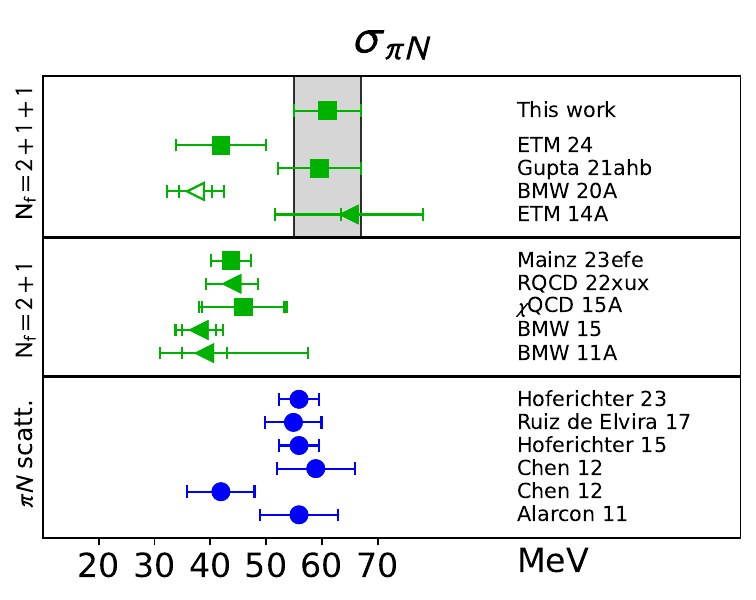}
     \includegraphics[width=0.48\linewidth]{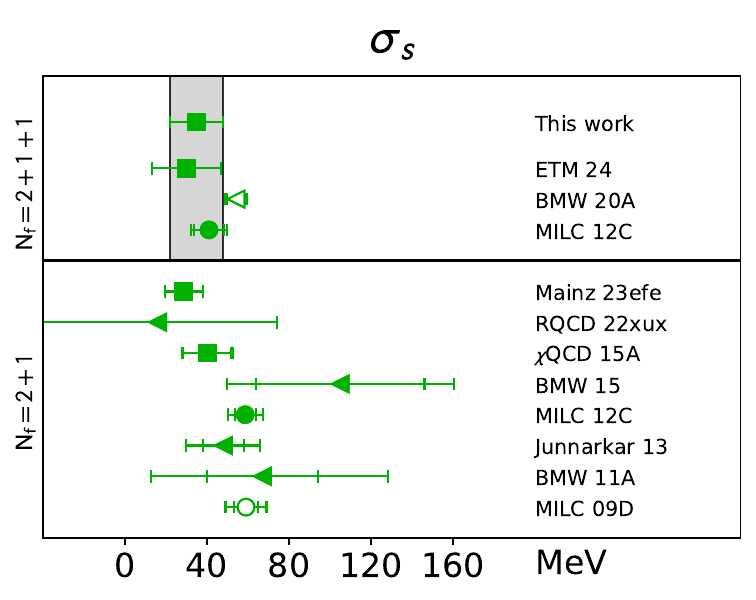}
}
\vspace{-0.1in}
\caption{(Left) Results for $\spiN = { m}_{ud} g_S^{u+d}$ from 2+1- and
  2+1+1-flavor lattice calculations that satisfy the FLAG criteria for
  inclusion in averages.  The BMW 20 1+1+1+1-flavor result is still a
  preprint and listed here under 2+1+1-flavor calculations for
  brevity.  We follow the FLAG conventions and denote direct
  determinations by squares and the results obtained using the Feynman-Helmann method  by
  triangles.  The references for the lattice results are:
  ETM 24~\protect\cite{Alexandrou:2024ozj}, 
  Gupta~21ahb~\cite{Gupta:2021ahb}, BMW~20A~\cite{Borsanyi:2020bpd}
ETM~14A~\cite{Alexandrou:2014sha},
Mainz~23ehe~\cite{Agadjanov:2023efe}
RQCD~22xux~\cite{RQCD:2022xux}
$\chi$QCD~15A~\cite{Yang:2015uis}, 
BMW~15~\cite{Durr:2015dna}, 
BMW 11A~\cite{Borsanyi:2020bpd}. 
The phenomenological estimates (blue filled circles), obtained using $\pi N$ 
scattering data, are from  
  Hoferichter~23~\cite{Hoferichter:2023ptl}, 
  Ruiz de Elvira 17~\cite{RuizdeElvira:2017stg}, 
  Hoferichter 15~\cite{Hoferichter:2015dsa}, 
  Chen 12~\cite{Chen:2012nx}
  Alarcon 11~\cite{Alarcon:2011zs}. 
(Right) Results for $\sigma_s$. The additional references for $\sigma_s$ are
MILC~12C~\cite{McNeile:2012xh},
Junnarkar~13~\cite{Junnarkar:2013ac},
MILC~09D~\cite{Toussaint:2009pz}.
\looseness-1
\label{fig:FLAG-sigma}}
\vspace{-0.1in}
\end{figure*}

Averaging the two estimates for $\sigma_{\pi N}$ and taking the 
larger of the errors, our final values are 
\begin{align}
    \sigma_{\pi N}|_{\rm standard} & = 42(6)~{\rm MeV}  \nonumber \,, \\
    \sigma_{\pi N}|_{N \pi}        & = 61(6)~{\rm MeV}  \nonumber \,, \\
    \sigma_{s}|_{\rm standard}     & = 35(13)~{\rm MeV} \,.
    \label{eq:sigma-final}
\end{align}
These are summarized in Fig.~\ref{fig:FLAG-sigma} along with other lattice 
 determinations with 2+1+1 and 2+1-flavor 
simulations that meet the FLAG criteria~\cite{FlavourLatticeAveragingGroupFLAG:2024oxs}, 
and also the results for $\sigma_{\pi N}$ from phenomenology.\looseness-1

\begin{table*}
\begin{ruledtabular}
\centering
  \begin{tabular}{l|lll|lll}
    & \multicolumn{3}{c|}{``Standard'' analysis for removing ESC} & \multicolumn{2}{c}{``$N \pi$'' analysis}\\
    $q$ & $g_A^q$ & $g_T^q$ & $g_S^q$ & $g_A^q$  & $g_T^q$  &  $g_S^q$  \\

\hline
& \multicolumn{6}{c}{Extrapolation excluding  finite volume corrections (CC fit).  $\rm{Z_1}$ method}\\
\hline
u & 0.803(17)    & 0.784(19)    & 6.35(56)     & 0.804(18)    & 0.776(28)    & 9.28(84)    \\
d & -0.430(17)   & -0.2001(65)  & 6.03(62)     & -0.468(21)   & -0.1903(87)  & 8.79(89)    \\
s & -0.0539(64)  & -0.00164(92) & 0.457(95)    & -0.0725(78)  & -0.00165(92) & 0.73(12)    \\
\hline
& \multicolumn{6}{c}{Extrapolation excluding  finite volume corrections (CC fit).  $\rm{Z_2}$ method}\\
\hline
u & 0.779(18)    & 0.759(19)    & 6.32(57)     & 0.779(18)    & 0.747(29)    & 9.37(87)    \\
d & -0.419(17)   & -0.1939(68)  & 6.03(63)     & -0.458(22)   & -0.1830(91)  & 8.91(91)    \\
s & -0.0543(69)  & -0.0016(10)  & 0.41(10)     & -0.0738(85)  & -0.0016(10)  & 0.68(13)    \\
\hline
& \multicolumn{6}{c}{Extrapolation including  finite volume corrections (CCFV fit).  $\rm{Z_1}$ method}\\
\hline
u & 0.784(22)    & 0.787(25)    & 6.36(56)     & 0.785(22)    & 0.776(37)    & 9.34(85)    \\
d & -0.417(27)   & -0.1957(93)  & 6.05(62)     & -0.465(29)   & -0.172(13)   & 8.81(91)    \\
s & -0.0540(88)  & -0.0016(11)  & 0.38(12)     & -0.0705(95)  & -0.0016(11)  & 0.66(16)    \\
\hline
& \multicolumn{6}{c}{Extrapolation including  finite volume corrections (CCFV fit).  $\rm{Z_2}$ method}\\
\hline
u & 0.776(22)    & 0.778(26)    & 6.32(57)     & 0.778(22)    & 0.768(38)    & 9.38(88)    \\
d & -0.414(27)   & -0.1949(96)  & 6.04(63)     & -0.462(29)   & -0.171(13)   & 8.87(93)    \\
s & -0.0554(94)  & -0.0015(12)  & 0.36(13)     & -0.073(10)   & -0.0015(12)  & 0.66(17)    \\
  \end{tabular}
  \caption{Results for the flavor diagonal charges obtained after
    extrapolation to the physical point with and without finite volume
    corrections (the CCFV and CC fits). Results are presented for the
    two strategies (``standard'' and ``$N\pi$'') for removing ESC and
    the two methods ($\rm{Z_1}$ and $\rm{Z_2}$, defined in
    Appendix~\ref{sec:Zstrategies}) for determining the
    renormalization constants. Note the tiny difference in $g_T^s$
    between the ``standard'' and ``$N\pi$'' analysis since the mixing
    from light flavors depends on $N\pi$.  }
  \label{tab:gAST-Z}
\end{ruledtabular}
\end{table*}

\begin{table*}
\begin{ruledtabular}
\centering
  \begin{tabular}{l|llll|ll}
    & \multicolumn{4}{c|}{This work} & \multicolumn{2}{c}{PNDME 2018}\\
    $q$ & $\phantom{-}g_A^q$ & $\phantom{-}g_T^q$ & $g_S^q|_{N \pi}$ &  $g_S^q|_{\rm St.}$ &$\phantom{-}g_A^q$ \cite{Lin:2018obj} & $\phantom{-}g_T^q$ \cite{Gupta:2018lvp} \\
    \hline
    $u$ & $\phantom{-}0.781(22)(11)_{\rm CC}$       & $\phantom{-}0.782(26)(11)_{\rm CC}$ & $9.36(88)(4)_{\rm CC}$ &  $6.34(57)(1)_{\rm CC}$ & $\phantom{-}0.777(25)(30)$ & $ \phantom{-}0.784(28)(31)$   \\
    $d$ & $-0.440(29)(9)_{\rm CC}(24)_{N\pi}$  & $-0.195(10)(2)_{\rm CC}(12)_{N\pi}$            & $8.84(93)(1)_{\rm CC}$ &  $6.04(63)(1)_{\rm CC}$ & $-0.438(18)(30)$           & $ -0.204(11)(10)$             \\
    $s$ & $-0.055(9)(1)_{\rm CC}$   & $-0.0016(12)(1)_{\rm CC}$           & $0.66(17)(5)_{\rm CC}$ &  $0.37(13)(6)_{\rm CC}$ & $-0.053(8)$                & $ -0.027(16)$                  \\
  \end{tabular}
  \caption{Final results for the flavor diagonal charges obtained by
    combining the values given in Table~\protect\ref{tab:gAST-Z} using
    criteria specified in the text. The subscript CC denotes the
    systematic uncertainty assigned using the difference between the
    CCFV and CC fit values; and $N\pi$ for that between the 
    ``standard'' and ``$N\pi$'' strategies where appropriate. Values
    with the $\rm{Z_1}$ and $\rm{Z_2}$ renormalization methods
    (defined in Appendix~\ref{sec:Zstrategies}) are averaged.  Results
    for $g_A^q$ and $g_T^q$ supersede those published previously in
    Refs.~\protect\cite{Lin:2018obj} and \protect\cite{Gupta:2018lvp}
    and reproduced here.  }
  \label{tab:final-g}
\end{ruledtabular}
\end{table*}

\begin{table}
\begin{ruledtabular}
\centering
  \begin{tabular}{l|lll}
    & \multicolumn{3}{c}{Final results} \\
    $q$ & $\phantom{-}g_A^q$ & $\phantom{-}g_T^q$ & $g_S^q$  \\
    \hline
    $u$ & $\phantom{-}0.781(25)$   & $\phantom{-}0.782(28)$  & $9.39(88)$    \\
    $d$ & $-0.440(39)$             & $-0.195(16)$            & $8.84(93)$    \\
    $s$ & $-0.055(9)$              & $-0.0016(12)$           & $0.37(14)$    \\
  \end{tabular}
  \caption{Final results for the flavor diagonal charges obtained by
    combining, in quadrature, the errors given in
    Table~\protect\ref{tab:final-g}. The results for the axial
    charges, $g_A^{u,d}$, are the average over the ``$N \pi$'' and
    ``standard'' analysis; the tensor, $g_T^{u,d}$, and the strange
    charges, $g_{A,S,T}^s$, are from the ``standard'' analysis; and
    $g_S^{u,d}$ are from the ``$N \pi$'' analysis as explained in the
    text.  }
  \label{tab:Xfinal-g}
\end{ruledtabular}
\end{table}

\subsection{Extracting the isovector charges $g_{A,T,S}^{u-d}$}

\begin{figure}[h]      
  \center
  \includegraphics[width=\linewidth]{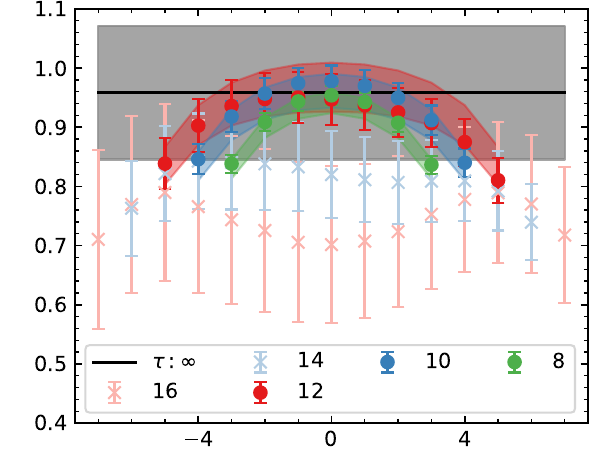}  
  \caption{The ratio plot for $g_S^{u-d,\text{bare}}$ on the a09m130
    ensemble. The $\tau=\infty$ value is taken from the fit to
    $\tau=8, 10$ and 12 data, which show no significant ESC at $t =
    \tau/2$. The $\tau=14$ and 16 data are neglected as, with current
    statistics, they have large errors and deviate from the expected
    behavior of monotonic convergence. }
  \label{fig:gS_u-d_a09m130}
\end{figure}

The disconnected contributions, which have large errors, cancel
exactly in the isovector charges, $g_{A,T,S}^{u-d}$, for the
isospin-symmetric theory. Thus, our overall strategy to get the
isovector charges is to make ES fits to the difference of the
connected contributions, i.e., to $C^{\text{u,conn}}_\Gamma(t;\tau) -
C^{\text{d,conn}}_\Gamma(t;\tau) $.  These fits are much more robust
compared to those for the individual flavor diagonal charges, and benefit from correlated fluctuations that are missed in the construction
$g^{u,bare}-g^{d,bare}$.  For example, extracting  $g_S^{u-d}$ from separate fits to the two 
$C^{\text{conn+disc}}_S(t;\tau) $ gives $O(1)$ errors with current data.  This occurs for the two reasons
illustrated by the $a09m130$ data in Fig.~\ref{fig:gS_u-d_a09m130}.
First, the data for the larger $\tau$ ($\tau = 14 $ and $16$) do not
satisfy the two conditions that must hold for fits to give consistent
results---monotonic convergence from below and symmetry about
$\tau/2$. These features are exhibited by the data from the other
seven ensembles and we have checked that they improve with
increasing statistical precision.  Second, within errors, which
increase rapidly with $\tau$, the data around $\tau/2$ overlap for 
$\tau = 8, 10, 12 $. This, ideally, indicates convergence by the 
plateau criteria, however, for confirmation much higher precision 
$\tau = 14 $ and $16$ data are needed. On the other hand, using 
2-state fits to remove residual ESC not well-constrained.

Our final results for isovector charges from the full set of
clover-on-HISQ calculations (using 13 data points on 11 HISQ
ensembles, CCFV extrapolation, and taking the average of values with $\rm{Z_1}$
and $\rm{Z_2}$ renormalization) have already been published in
Ref.~\cite{Jang:2023zts}. For completeness, Table~\ref{tab:isovector}
summarizes the values from this work using eight ensembles,
``standard'' and ``$N\pi$'' analyses, CCFV and CC extrapolations; and
the same two renormalization methods. The two sets of results are
consistent, with the values obtained here having slightly larger
errors. The only notable feature in both is the difference between the various
estimates for $g_A^{u-d}$ that has already been pointed out in
Ref.~\cite{Jang:2023zts}.

\begin{table*}
\begin{ruledtabular}
\centering
  \begin{tabular}{l|llll|llll}
    & \multicolumn{4}{c|}{``Standard'' analysis for removing ESC} & \multicolumn{4}{c}{``$N \pi$'' analysis}\\
    & This work  & This work &  This work & Ref.~\cite{Jang:2023zts} &   This work  & This work  &   This work & Ref.~\cite{Jang:2023zts} \\
     & ($\rm{Z_1}$)  & ($\rm{Z_2}$)  &  &  &   ($\rm{Z_1}$)  & ($\rm{Z_2}$)  &  &   \\
    \hline

$g_{A}^{u-d}|_{\rm CC}$    & 1.284(29)    & 1.221(30)    &              &              & 1.328(26)    & 1.277(27)    &              &             \\
$g_{A}^{u-d}|_{\rm CCFV}$  & 1.221(39)    & 1.206(40)    &   1.214(40)  & 1.240(27)    & 1.288(32)    & 1.278(33)    &   1.283(33)           & 1.294(49)   \\
$g_{T}^{u-d}|_{\rm CC}$    & 1.007(12)    & 0.969(12)    &              &              & 0.976(27)    & 0.929(28)    &              &             \\
$g_{T}^{u-d}|_{\rm CCFV}$  & 1.010(16)    & 0.998(17)    &    1.004(17)          & 0.989(34)    & 0.976(36)    & 0.969(37)    &  0.973(37)            & 0.991(24)   \\
$g_{S}^{u-d}|_{\rm CC}$    & 0.99(11)     & 0.98(11)     &              &              & 1.17(27)     & 1.17(27)     &              &             \\
$g_{S}^{u-d}|_{\rm CCFV}$  & 0.98(11)     & 0.96(12)     &    0.97(12)          & 1.022(100)   & 1.18(27)     & 1.17(28)     &   1.18(28)           & 1.085(115)  \\

  \end{tabular}
  \caption{ Results for the isovector charges with the two strategies,
    ``standard'' and ``$N\pi$'', used to remove ESC; the two methods,
    $\rm{Z_1}$ and $\rm{Z_2}$, used for renormalization; and the two
     ansatz, CCFV and CC, used to extrapolate data to the physical point. The
    final values from this work are compared with the average of
    $\rm{Z_1}$ and $\rm{Z_2}$ values from Ref.~\cite{Jang:2023zts}
    obtained with CCFV extrapolation of the full set of thirteen
    clover-on-HISQ calculations.  The CCFV extrapolations based on
    eight versus thirteen points account for most of the differences
    between the results.  The ``standard'' analysis strategy is
    labeled ``$3^\ast$" in Ref.~\cite{Jang:2023zts}, while the ``$N
    \pi$'' analysis is similar to ``3-RD".  }
  \label{tab:isovector}
\end{ruledtabular}
\end{table*}

\begin{figure*}[htb]      
\center
\includegraphics[width=0.32\linewidth]{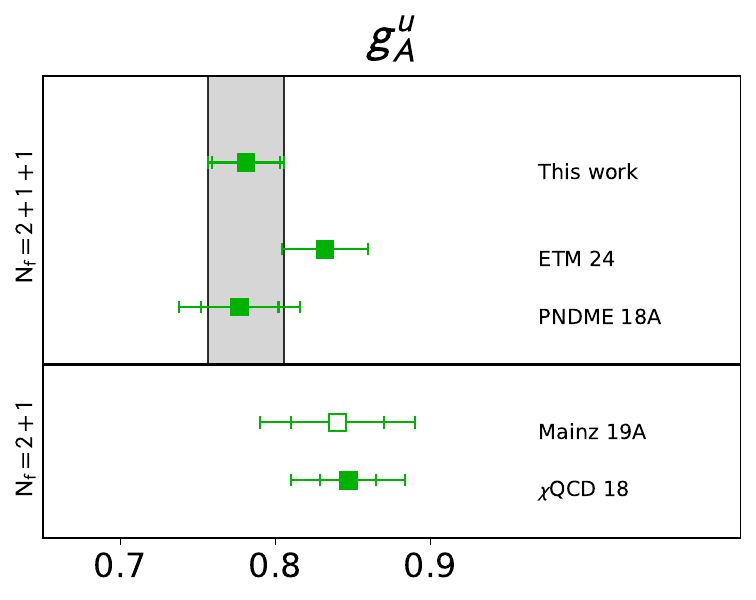}
\includegraphics[width=0.32\linewidth]{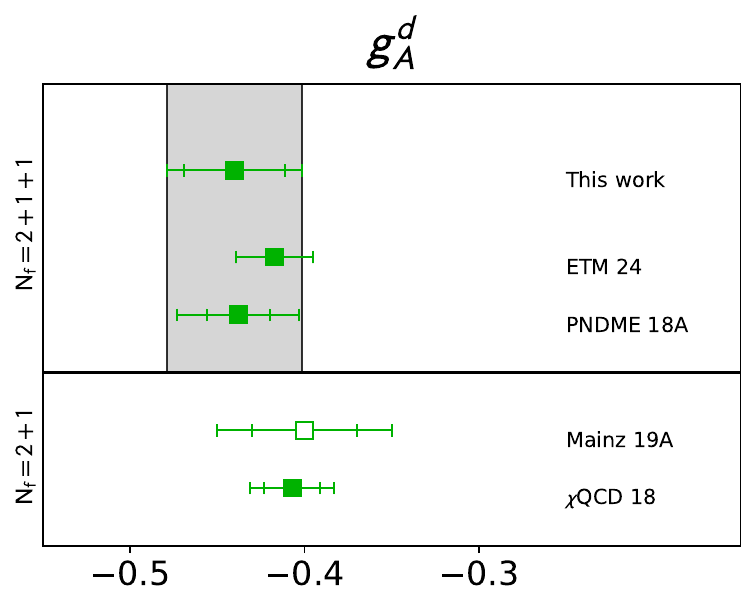}
\includegraphics[width=0.32\linewidth]{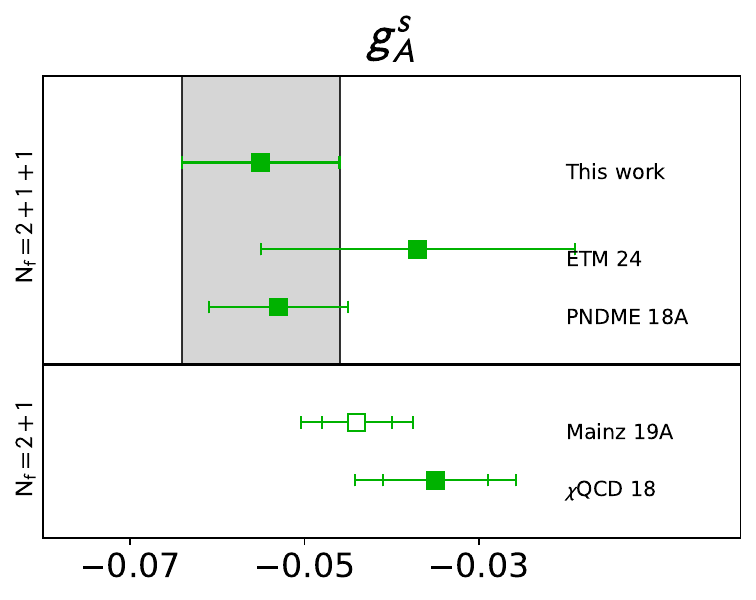}
\includegraphics[width=0.32\linewidth]{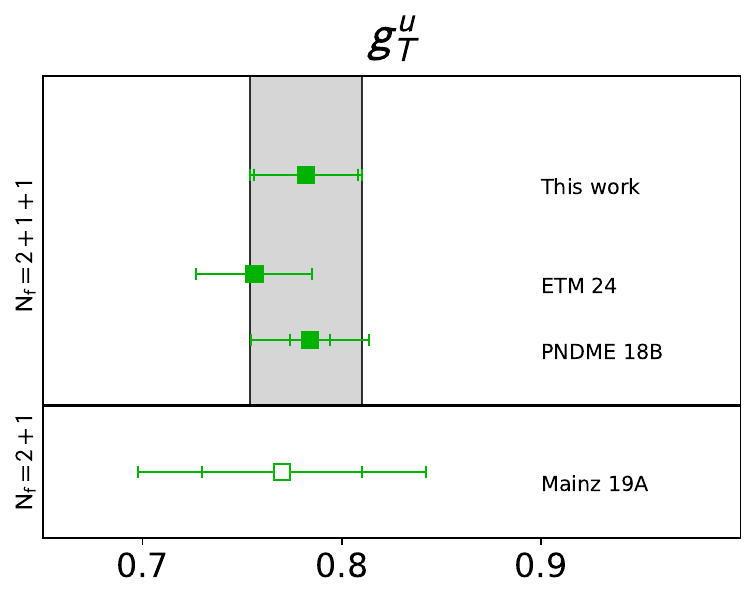}
\includegraphics[width=0.32\linewidth]{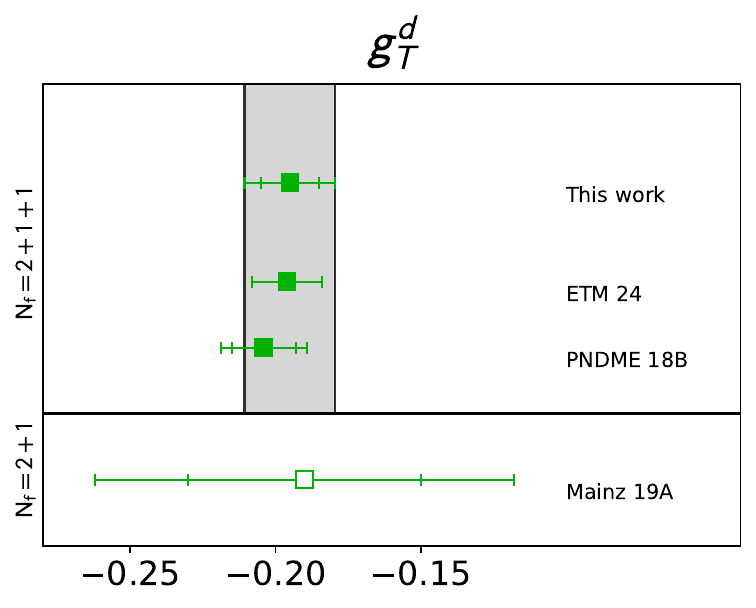}
\includegraphics[width=0.32\linewidth]{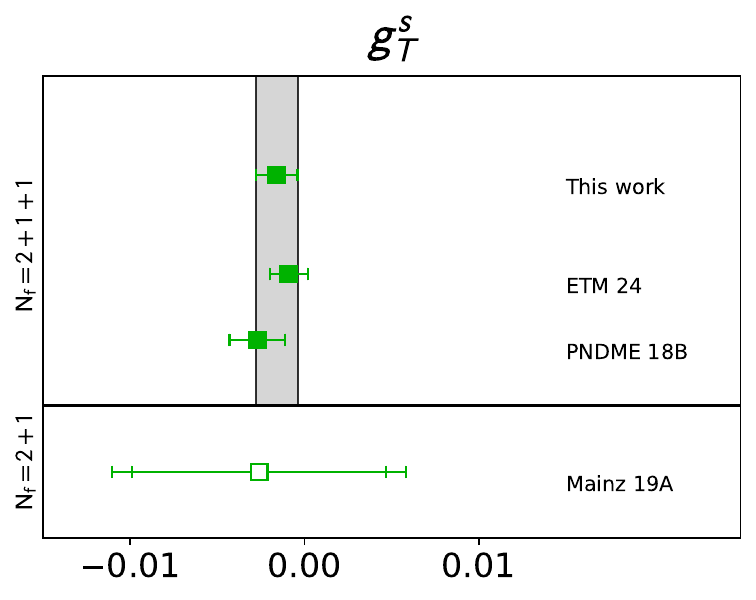}
\vspace{-0.1in}
\caption{ Comparison of axial and tensor charges for the proton with other lattice calculations 
using 2+1+1 and 2+1-flavors of quarks that meet the FLAG criteria. Results for the neutron are given by the $u \leftrightarrow d$ interchange. Unfilled symbols 
indicate that these results have not yet been published or are conference proceedings. The references are 
ETM 24~\protect\cite{Alexandrou:2024ozj}
PNDME 18A~\protect\cite{Lin:2018obj}
PNDME 18B~\protect\cite{Gupta:2018lvp}
Mainz 19A~\protect\cite{Djukanovic:2019gvi}
$\chi$QCD 18~\protect\cite{Liang:2018pis}
}
\looseness-1
\label{fig:FLAG-gAgT}
\vspace{-0.1in}
\end{figure*}

\section{Comparison with Previous Results and Conclusions}
\label{sec:conclusion}

Our final results, which supersede those presented in
Refs.~\cite{Lin:2018obj} and \cite{Gupta:2018lvp}, are summarized in
Tables~\ref{tab:final-g} and ~\ref{tab:Xfinal-g}. The improvements
made in this work are described in the introduction~\ref{sec:into}.
The FLAG 2024 report~\cite{FlavourLatticeAveragingGroupFLAG:2024oxs}
quotes our previous results for $g_{A,T}^{u,d,s}$ (from
Refs.~\cite{Lin:2018obj,Gupta:2018lvp}) as the FLAG values for
2+1+1-flavor simulations. The updated versions of the FLAG summary
figures, including data from other collaborations that meet the FLAG
criteria for averaging, and post FLAG 2024 results from the ETM
collaboration~\cite{Alexandrou:2024ozj}, are shown in
Figs.~\ref{fig:FLAG-gAgT} and~\ref{fig:FLAG-sigma}.

From the axial charges, $g_A^{u,d,s}$ given in
Table~\ref{tab:Xfinal-g}, we calculate the contribution of the
intrinsic spin of quarks to the nucleon spins to be $0.5 \Delta
\Sigma_q = 0.5 \sum_q g_A^q = 0.143(24)$. This is consistent with the
COMPASS analysis $0.13 < 0.5 \Delta \Sigma_q < 0.18$ evaluated at
3~GeV${}^2$~\cite{Adolph:2015saz}.

The lattice determination of tensor charges, $g_T^{u,d,s}$, is far
more accurate than extractions from the transversity distributions of
quarks in semi-inclusive deep-inelastic scattering (SIDIS) experiments
at Jefferson Lab and other facilities
worldwide~\cite{Lin:2017stx,Radici:2018iag,Ye:2016prn}. These charges
also give the intrinsic contributions of the electric dipole moments
of the quarks to the neutron/proton EDM~\cite{Gupta:2018lvp}.

Our results for the scalar charges, $g_S^{u,d,s}$, are new with only
preliminary versions presented in conference
proceedings~\cite{Park:2023tsj,Park:2024vjp}.  Most of the discussion
of scalar charges in literature has been in terms of the sigma terms,
the pion-nucleon, $\sigma_{\pi N}$, and the strangeness content
$\sigma_{s}$~\cite{FLAG:2021npn}. For the pion-nucleon sigma term
$\sigma_{\pi N}$, the main issue, first raised in
Ref.~\cite{Gupta:2021ahb}, is removing the contributions of
multihadron $N \pi$ and $N \pi \pi$ states that are lattice
artifacts. Our selection of the ``$N \pi$'' analysis results, as
discussed in Section~\ref{sec:sigma_term}, rely on the $\chi$PT
analysis and the phenomenological observation of a large scalar
condensate in QCD. Results in Eq.~\eqref{eq:sigma-final} are
consistent with those presented in Ref.~\cite{Gupta:2021ahb}.  The
value $\sigma_{s} = 35(13)$~MeV reflects the large uncertainty in the
calculation of $g_S^s$ and the CCFV extrapolation as shown in
Fig.~\ref{fig:CCFV_gSs}. A comparison with other lattice and
phenomenological results in the FLAG format is shown in
Fig.~\ref{fig:FLAG-sigma}. Our result $\sigma_{\pi N}|_{\rm standard}
= 42(6)~{\rm MeV}$ is consistent with other lattice measurements,
while our preferred value $ \sigma_{\pi N}|_{N \pi} = 61(6)~{\rm MeV}$
is in agreement with phenomenological extraction (dispersive analysis 
of $\pi N$ scattering data).

Note that a larger value of the scalar charges, output of our ``$N
\pi$'' analyses motivated by $\chi$PT, has a significant impact on the
reach of dark matter direct-detection experiments---a $\approx 50\%$
larger $g_{S}^{u,d}$ increases the cross-section by a factor of
$\approx 2.25$ assuming the couplings to dark matter is in the most
promising scalar channel. In fact, all three sets of charges,
$g_{A,S,T}^{u,d,s}$, give the coupling of dark matter or novel
Higgs-like interactions to nucleons in the respective Lorentz
channels.

The most direct comparison of our results is with the latest results
from the ETM collaboration~24~\cite{Alexandrou:2024ozj}, who have
analyzed three physical mass ensembles using 2+1+1-flavors of twisted
mass fermions. We note that these new ETM results supersede all their previously published values as explained in~\cite{Alexandrou:2024ozj}. 

The two sets of results, summarized in Figs.~\ref{fig:FLAG-gAgT} 
and~\ref{fig:FLAG-sigma}, are consistent with two notable ($2\sigma$)
differences: in $g_A^{l,s}$ and $\sigma_{\pi N}$. The reasons are:
\begin{itemize}
    \item Our data show ESC in $g_A^{l,s}$ (see
      Figs.~\ref{fig:gAgTgS_ESC} and~\ref{fig:gAgTgS_s_ESC}) while the
      ETM data (labelled $g_A^{u+d (disc)}$ and $g_A^{s (disc)}$) are
      flat even though they include data at smaller $\tau$, i.e., in
      the range $0.41 < \tau < 1.6$~fm versus $0.9 < \tau < 1.5$~fm in
      our calculation. This accounts for most of the difference in the
      final values of $g_A^{u,d}$ shown in Fig.~\ref{fig:FLAG-gAgT}.
    \item Both calculations find large ESC in $g_S^{u,d,s}$. The
      difference in the estimates of $\sigma_{\pi N}$ comes from the
      choice of ESC fits. We have chosen the ``$N\pi$'' fit for the
      final answer based on our $\chi$PT
      analysis~\cite{Gupta:2021ahb}. Had we chosen our ``standard''
      analysis value, as ETM does, the two results would be consistent.
\end{itemize}
In summary, as already remarked in Sec.~\ref{sec:ESCpatterns}, the largest
uncertainty in these two quantities in our work comes from the ESC
fits and possible contributions from $N\pi$ states. The ETM results
are from a ``standard" analysis.
    
Looking ahead, we are analyzing two physical pion mass HISQ ensembles
at $a \approx 0.087$ and $0.057$~fm with about six times larger statistics (6000 and 4000 configurations, respectively) and better
tuned values of $m_s$ and $m_l$ in the clover action. These
calculations will improve control over excited-state analysis and,
hopefully, the statistics are large enough to provide a data driven
resolution between the ``standard'' and ``$N \pi$'' analyses.

\begin{acknowledgments}
We thank the MILC collaboration for providing the 2+1+1-flavor HISQ
lattices, and acknowledge John Gracey for discussions on the
renormalization of the currents. The calculations used the Chroma
software suite~\cite{Edwards:2004sx}.  This research used resources at
(i) the National Energy Research Scientific Computing Center, a DOE
Office of Science User Facility supported by the Office of Science of
the U.S.\ Department of Energy under Contract No.\ DE-AC02-05CH11231;
(ii) the Oak Ridge Leadership Computing Facility, which is a DOE
Office of Science User Facility supported under Contract
DE-AC05-00OR22725, through the ALCC program project LGT107 and HEP145, and the
INCITE program project HEP133; (iii) the USQCD collaboration, which is
funded by the Office of Science of the U.S.\ Department of Energy; and
(iv) Institutional Computing at Los Alamos National Laboratory.
This work was prepared in part by LLNL under Contract DE-AC52-07NA27344. 
S.~Park acknowledges the support from the ASC COSMON project.
S.~Park acknowledges support from the U.S. DOE Contract
No.\ DE-AC05-06OR23177, under which Jefferson Science Associates, LLC,
manages and operates Jefferson Lab. Also acknowledged is support from
the Exascale Computing Project (17-SC-20-SC), a collaborative effort
of the U.S. DOE Office of Science and the National Nuclear Security
Administration.
T.~Bhattacharya and R.~Gupta were partly
supported by the U.S.\ Department of Energy, Office of Science, Office
of High Energy Physics under Contract No.\
DE-AC52-06NA25396. T.~Bhattacharya, R.~Gupta, S.~Mondal, S.~Park,
and B.~Yoon were partly supported by the LANL LDRD program, and
S.~Park by the Center for Nonlinear Studies.  
\end{acknowledgments}

\appendix

\clearpage

\begin{widetext}
\section{Chiral-continuum-finite-volume (CCFV) extrapolation for
    $g_{A,S,T}^{u,d,s}$}
\label{sec:CCFV-fits}

Figures~\ref{fig:CCFV_gAu}---\ref{fig:CCFV_gSs} in this appendix show
the CCFV fits used to extrapolate the data for the charges on the
eight ensembles to the physical point. In each column, the three panels
show the simultaneous CCFV fit (pink band) plotted versus either $a$
or $M_\pi^2$ or $M_\pi L$, respectively, with the other two variables
set to their physical point values. In the panels plotted versus
$M_\pi^2$, the additional grey bands show the fit with only the chiral
corrections included in Eqs.~\eqref{eq:CCFV-1}
or~\eqref{eq:CCFV-2}. Note that when the discretization and finite
volume corrections are small, the grey bands overlap with the CCFV
fits (pink bands).  The results of these fits are summarized in
Table~\ref{tab:gAST-Z}.

\begin{figure*}[h]      

  \center
  \includegraphics[width=0.23\textwidth]{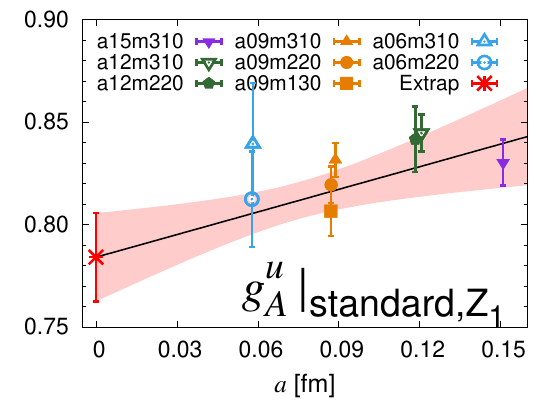}
  \includegraphics[width=0.23\textwidth]{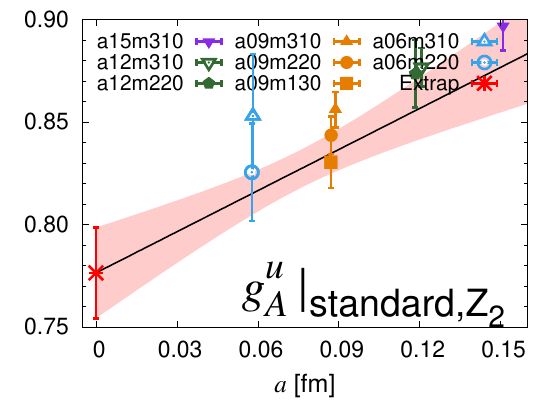}
  \includegraphics[width=0.23\textwidth]{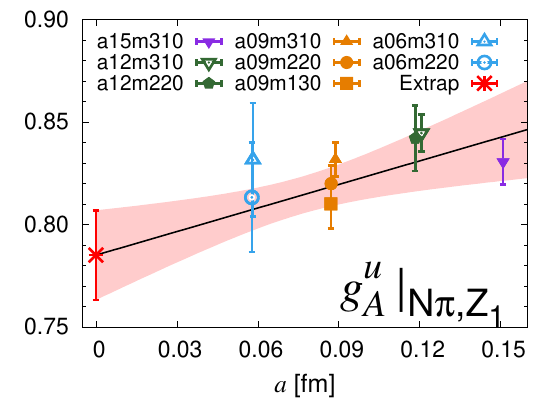}
  \includegraphics[width=0.23\textwidth]{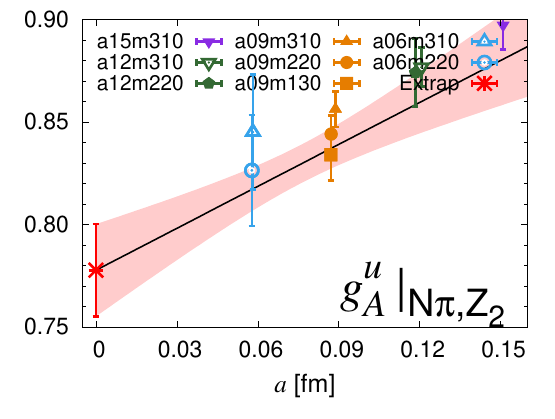}
  
  \includegraphics[width=0.23\textwidth]{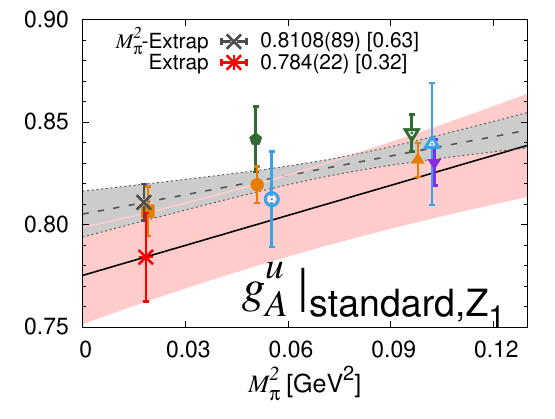}
  \includegraphics[width=0.23\textwidth]{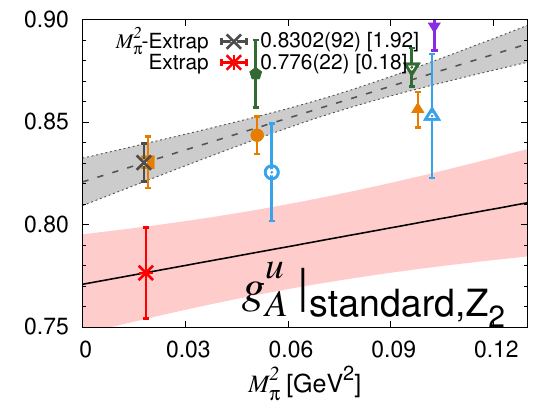}
  \includegraphics[width=0.23\textwidth]{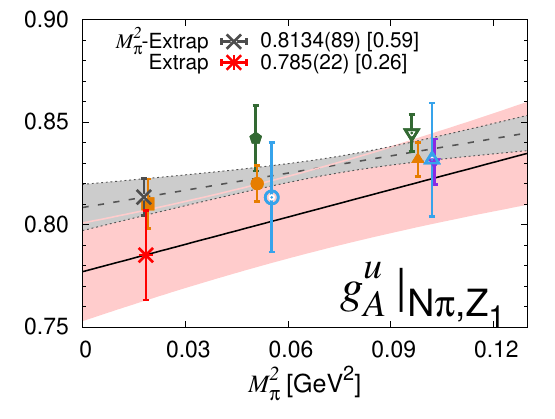}
  \includegraphics[width=0.23\textwidth]{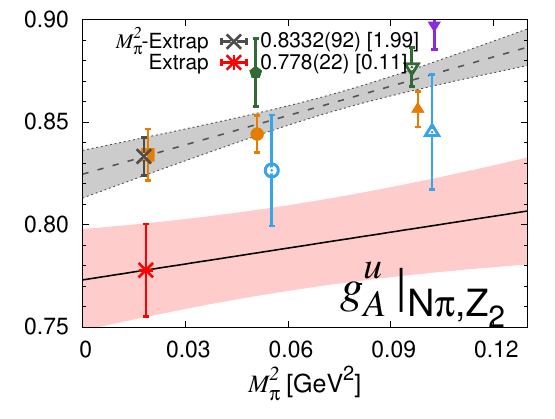}
  
  \includegraphics[width=0.23\textwidth]{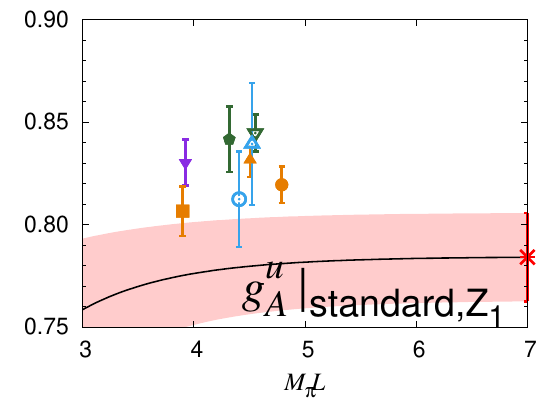}
  \includegraphics[width=0.23\textwidth]{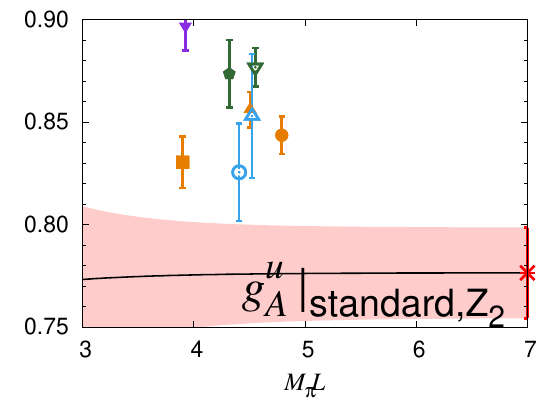}
  \includegraphics[width=0.23\textwidth]{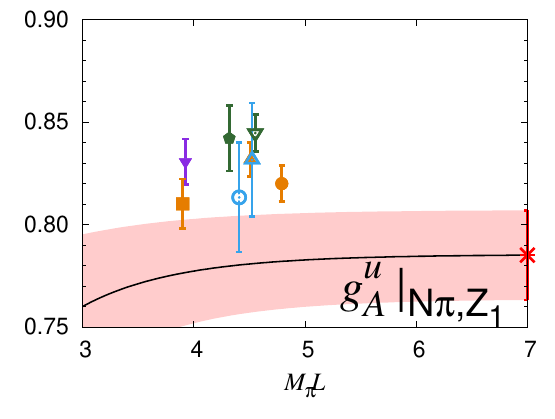}
  \includegraphics[width=0.23\textwidth]{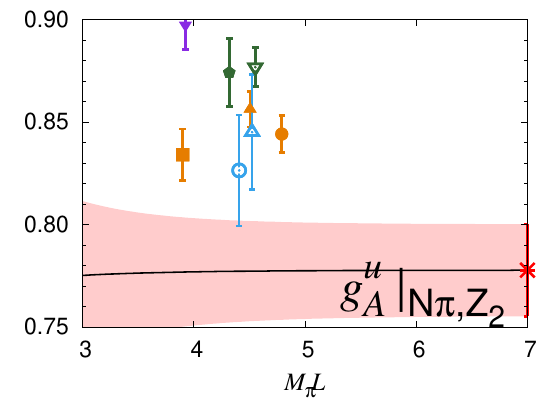}
  \caption{Chiral-continuum-finite-volume (CCFV) 
  extrapolation for $g_A^{u}$ using Eq.~\eqref{eq:CCFV-1}. The 3 panels in each 
  column show the result of the simultaneous
    fit (pink bands) plotted versus $a$, $M_\pi^2$ and $M_\pi L$, respectively, 
    with the other two variables set to their physical point values. The grey 
    band shows a chiral fit, i.e., assuming $c_1 = c_3=0$ in Eq.~\eqref{eq:CCFV-1}. 
    Each row shows, from left to right, the fits for the four different analyses: 
    (i) \{standard,$\rm{Z_1}$\},
    (ii) \{standard,$\rm{Z_2}$\}, (iii) \{$N\pi$,$\rm{Z_1}$\} and
    (iv) \{$N\pi$,$\rm{Z_2}$\}. The red star is the value at the physical point 
    given by the CCFV fit and 
    the black cross is the result of just the chiral fit. The numbers within square parenthesis in the middle panels are the $\chi^2/dof$ of these fits.}
  \label{fig:CCFV_gAu}  
\end{figure*}
\end{widetext}

\begin{figure*}      
  \center
  \includegraphics[width=0.23\textwidth]{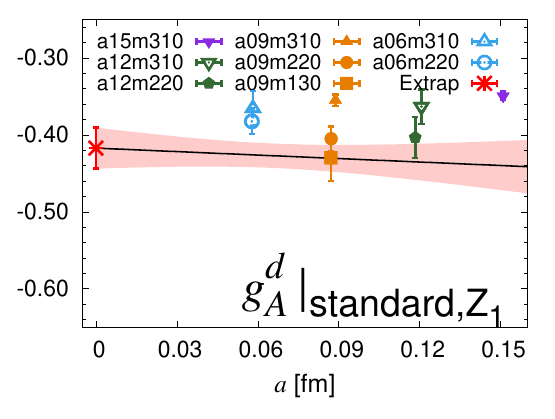}
  \includegraphics[width=0.23\textwidth]{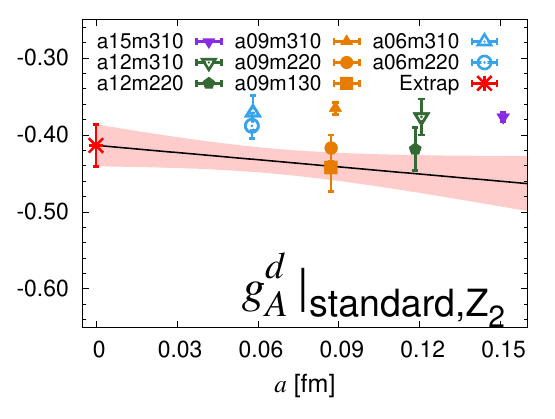}
  \includegraphics[width=0.23\textwidth]{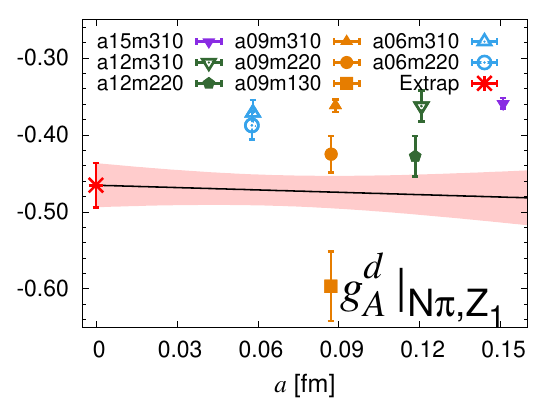}
  \includegraphics[width=0.23\textwidth]{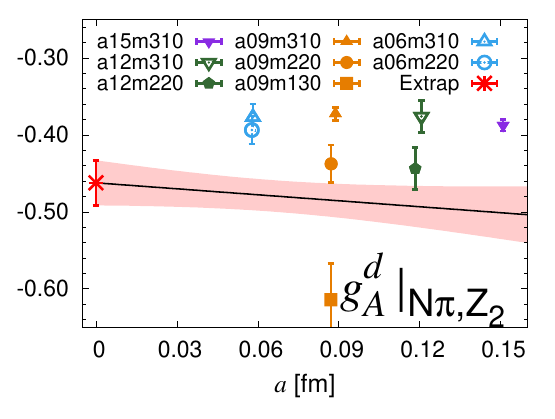}
  
  \includegraphics[width=0.23\textwidth]{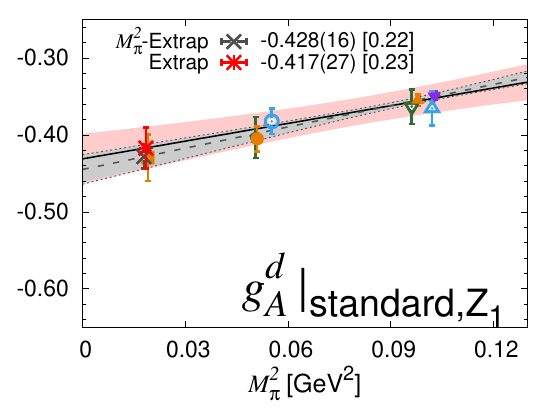}
  \includegraphics[width=0.23\textwidth]{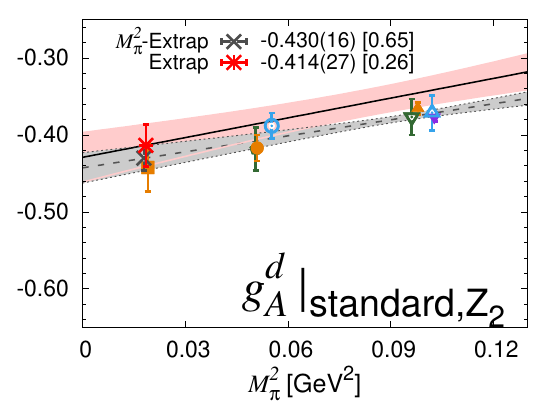}
  \includegraphics[width=0.23\textwidth]{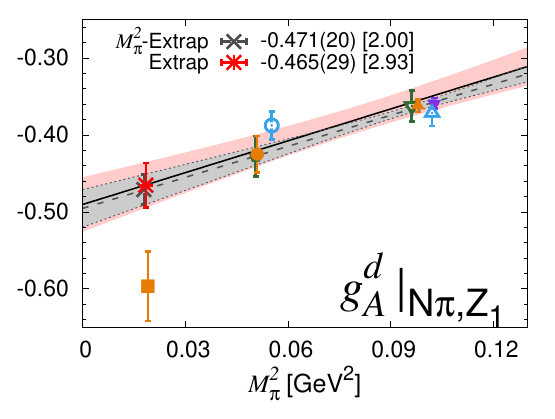}
  \includegraphics[width=0.23\textwidth]{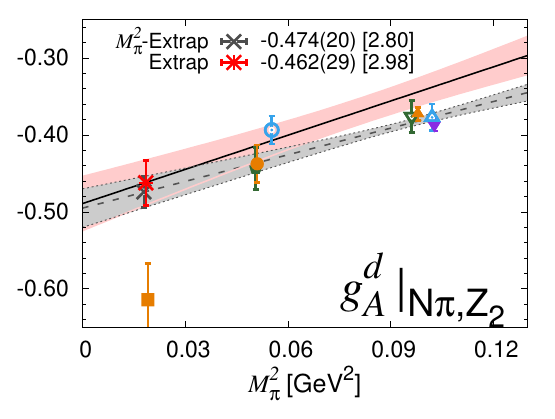}
  
  \includegraphics[width=0.23\textwidth]{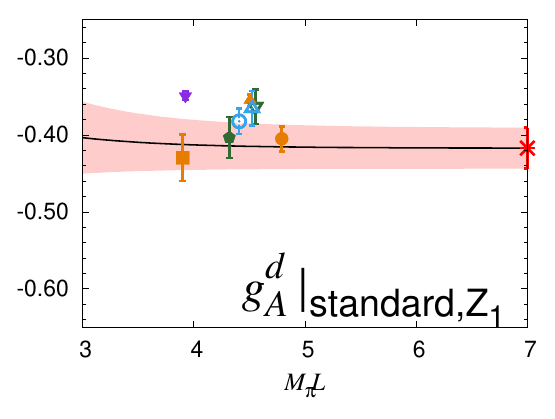}
  \includegraphics[width=0.23\textwidth]{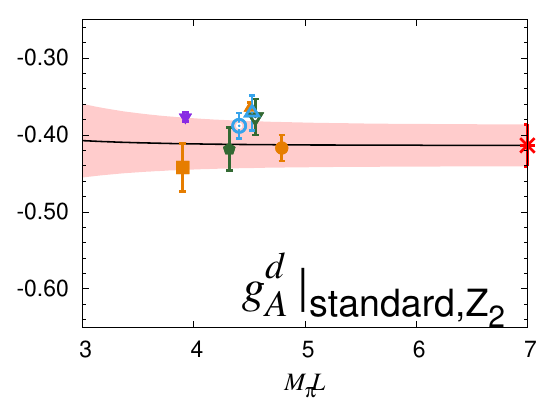}
  \includegraphics[width=0.23\textwidth]{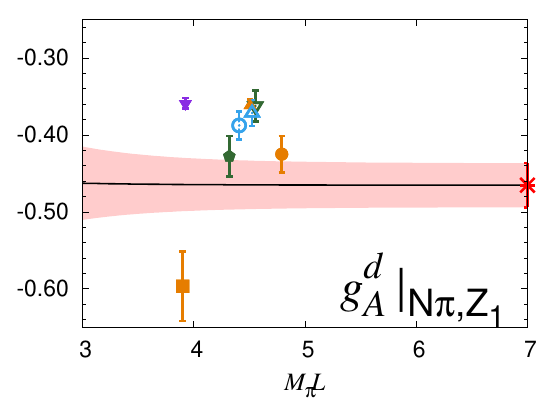}
  \includegraphics[width=0.23\textwidth]{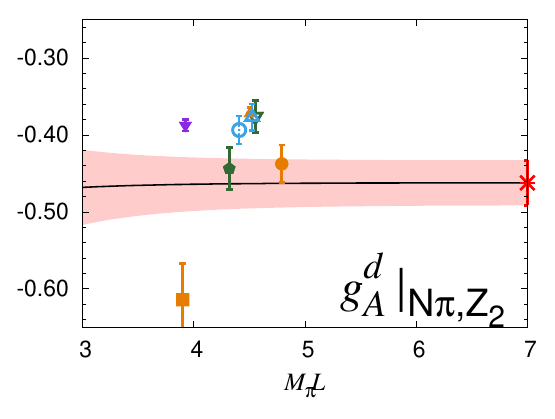}
  \caption{CCFV extrapolation for $g_A^{d}$ using Eq.~\eqref{eq:CCFV-1}. 
  The rest is the same as in Fig.~\ref{fig:CCFV_gAu}.}
  \label{fig:CCFV_gAd}  
\end{figure*}

\begin{figure*}      
  \center
  \includegraphics[width=0.23\textwidth]{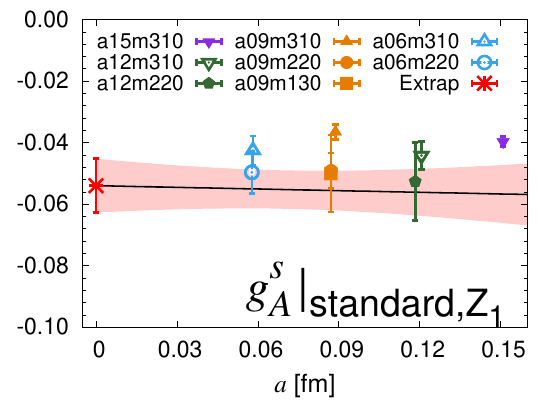}
  \includegraphics[width=0.23\textwidth]{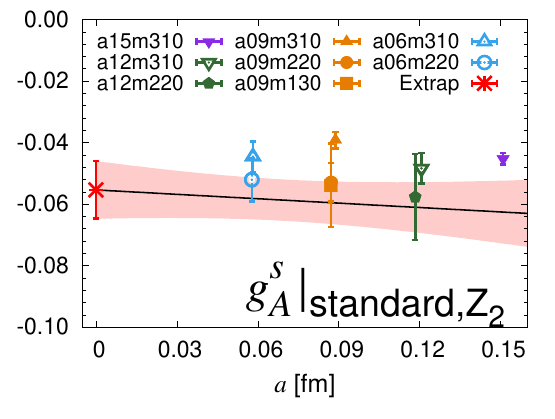}
  \includegraphics[width=0.23\textwidth]{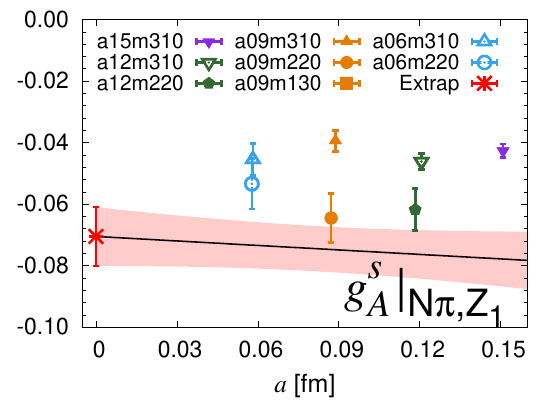}
  \includegraphics[width=0.23\textwidth]{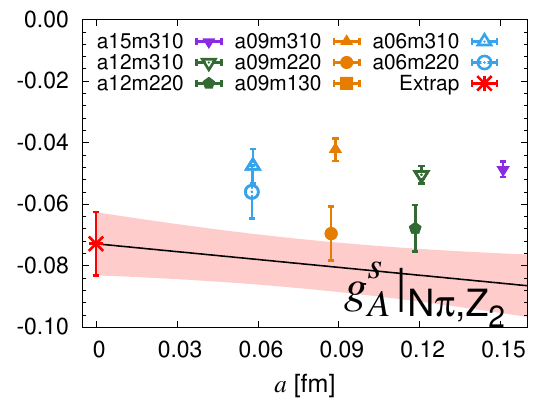}
  
  \includegraphics[width=0.23\textwidth]{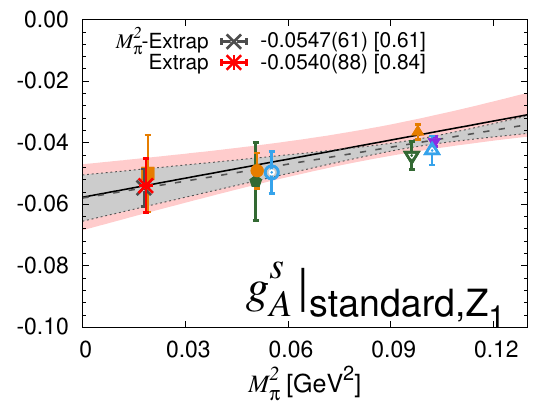}
  \includegraphics[width=0.23\textwidth]{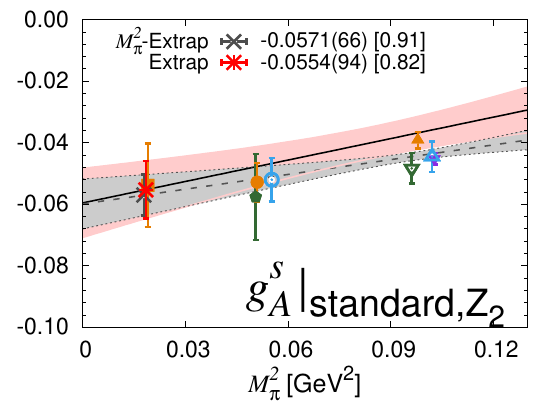}
  \includegraphics[width=0.23\textwidth]{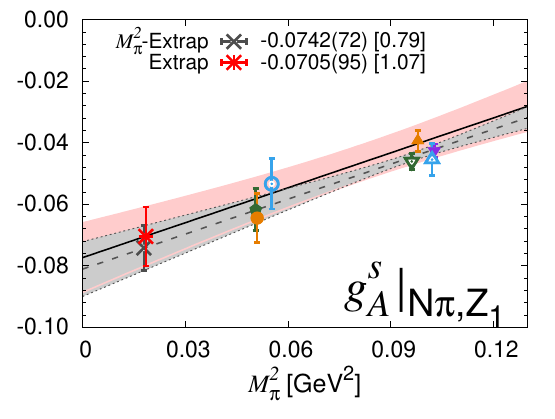}
  \includegraphics[width=0.23\textwidth]{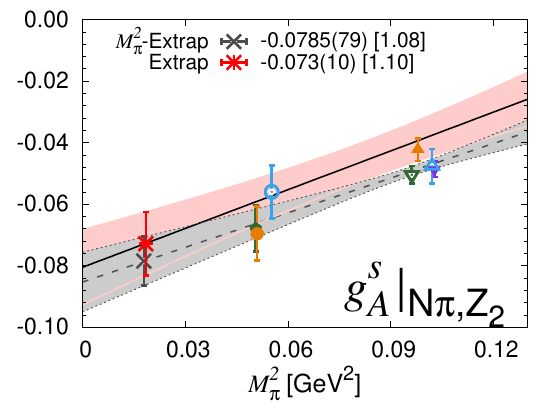}
  
  \includegraphics[width=0.23\textwidth]{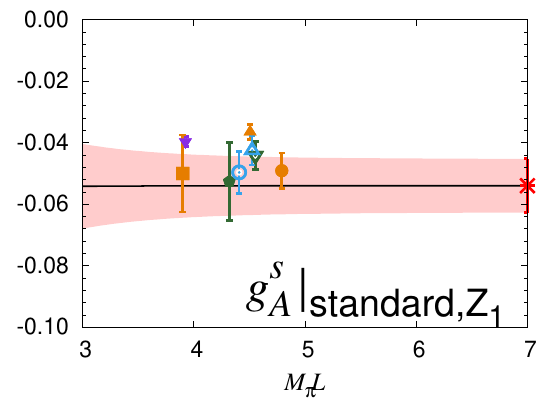}
  \includegraphics[width=0.23\textwidth]{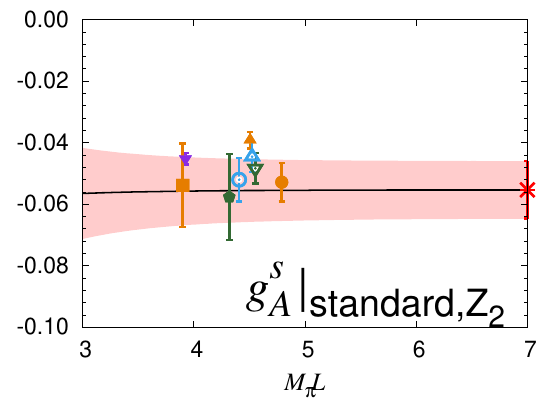}
  \includegraphics[width=0.23\textwidth]{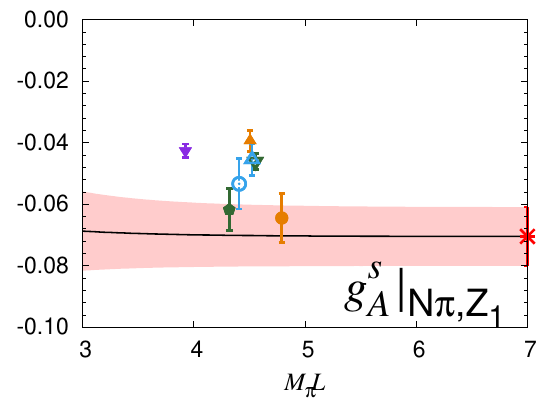}
  \includegraphics[width=0.23\textwidth]{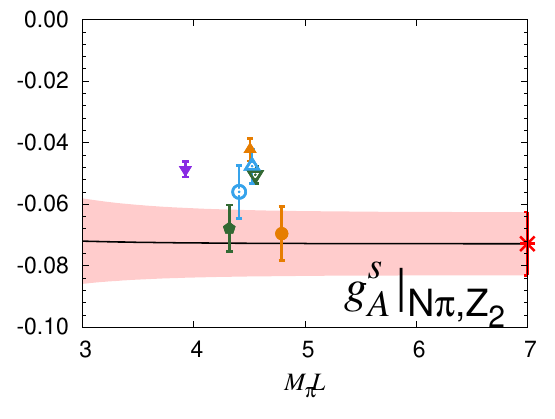}
  \caption{CCFV extrapolation for $g_A^{s}$ using Eq.~\eqref{eq:CCFV-1}. 
  The rest is the same as in Fig.~\ref{fig:CCFV_gAu}.}
  \label{fig:CCFV_gAs}  
\end{figure*}

\begin{figure*}       
  \center
  \includegraphics[width=0.23\textwidth]{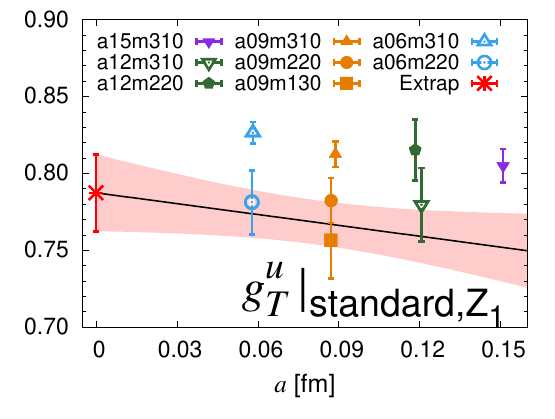}
  \includegraphics[width=0.23\textwidth]{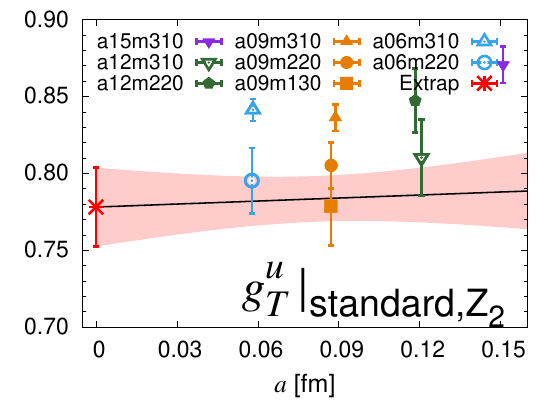}
  \includegraphics[width=0.23\textwidth]{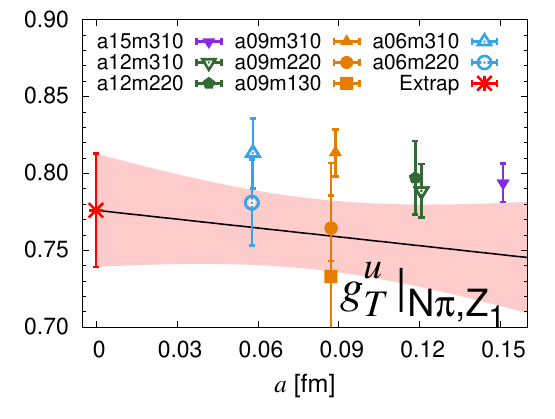}
  \includegraphics[width=0.23\textwidth]{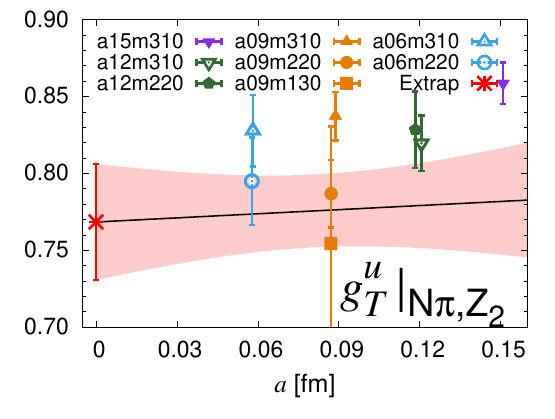}
  
  \includegraphics[width=0.23\textwidth]{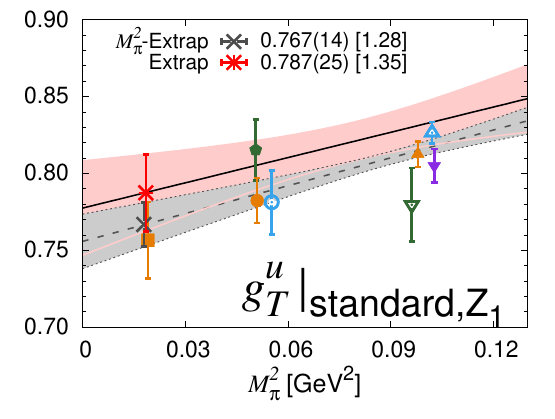}
  \includegraphics[width=0.23\textwidth]{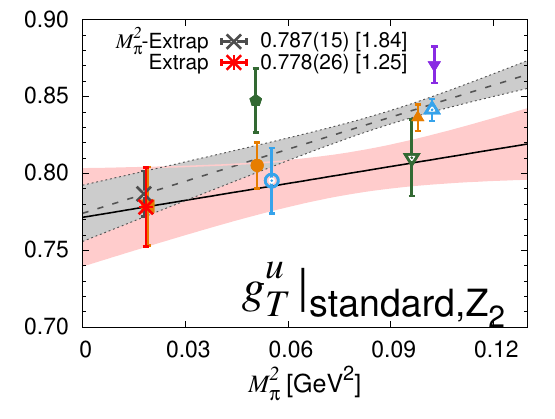}
  \includegraphics[width=0.23\textwidth]{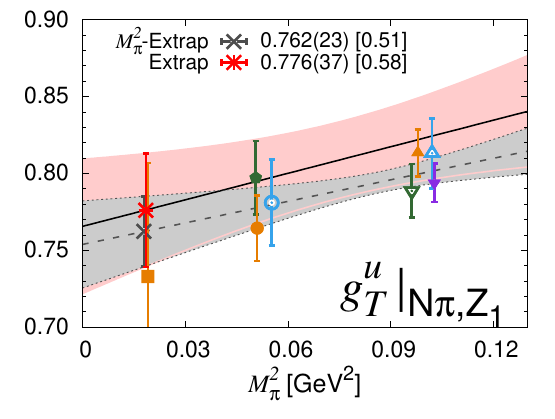}
  \includegraphics[width=0.23\textwidth]{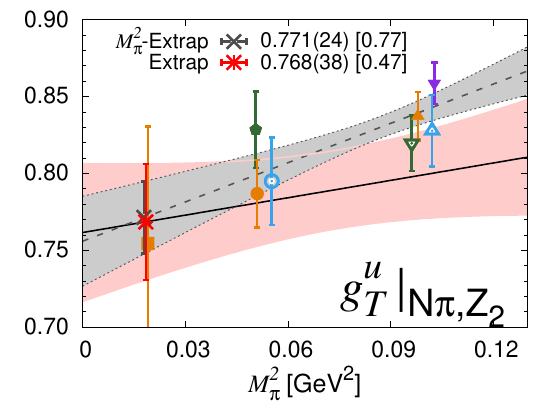}
  
  \includegraphics[width=0.23\textwidth]{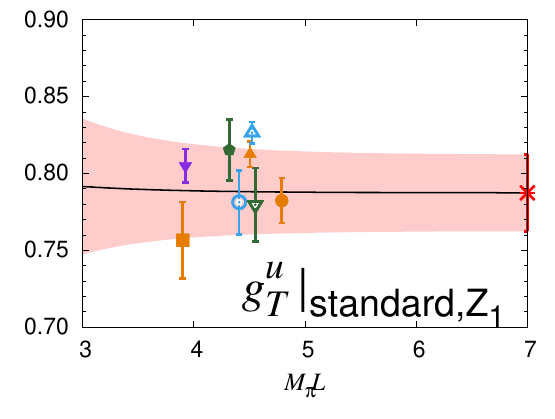}
  \includegraphics[width=0.23\textwidth]{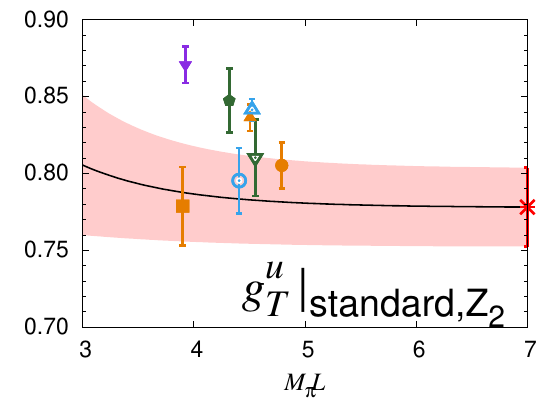}
  \includegraphics[width=0.23\textwidth]{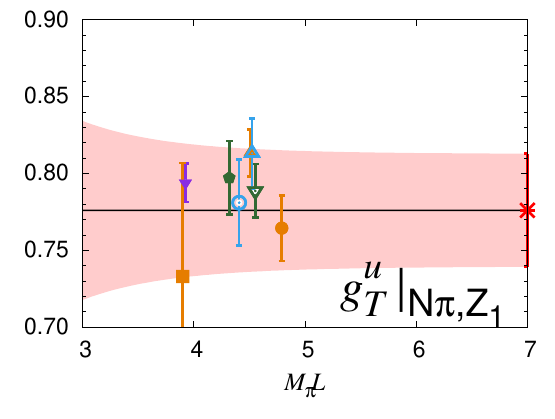}
  \includegraphics[width=0.23\textwidth]{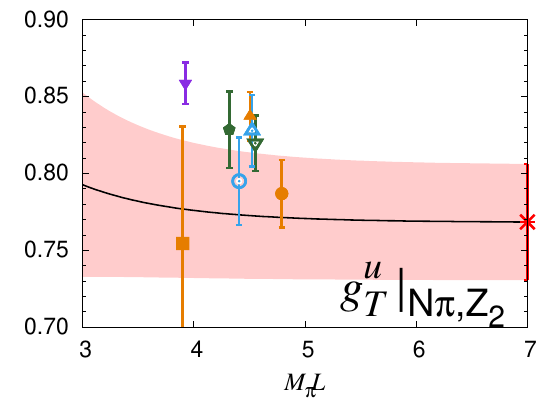}
  \caption{CCFV extrapolation for $g_T^{u}$ using Eq.~\eqref{eq:CCFV-1}. 
  The rest is the same as in Fig.~\ref{fig:CCFV_gAu} }
  \label{fig:CCFV_gTu}  
\end{figure*}

\begin{figure*}       
  \center
  \includegraphics[width=0.23\textwidth]{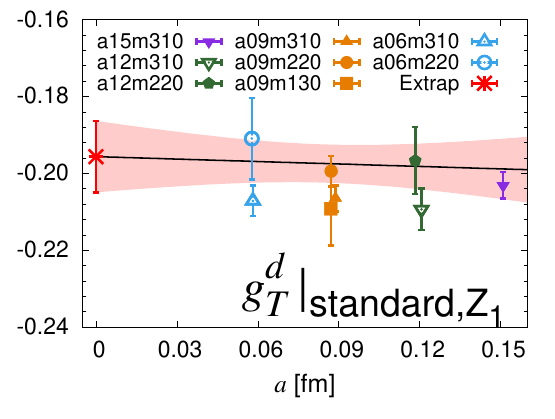}
  \includegraphics[width=0.23\textwidth]{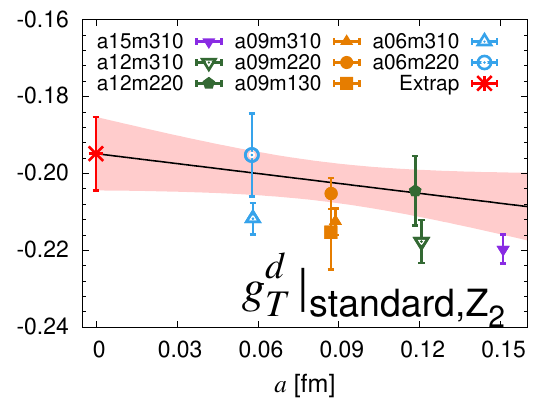}
  \includegraphics[width=0.23\textwidth]{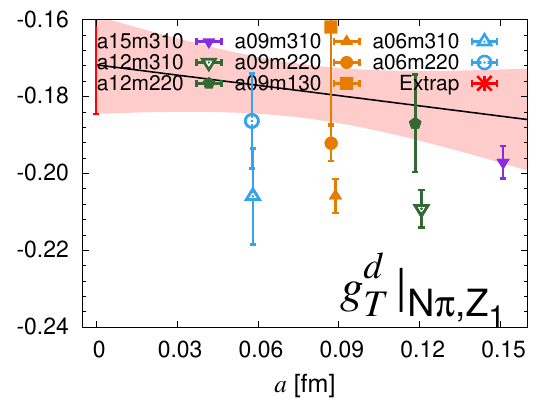}
  \includegraphics[width=0.23\textwidth]{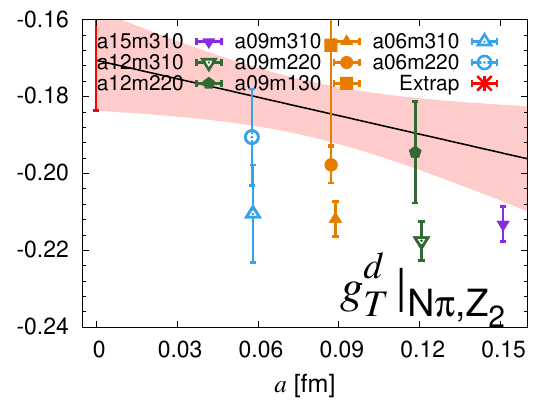}
  
  \includegraphics[width=0.23\textwidth]{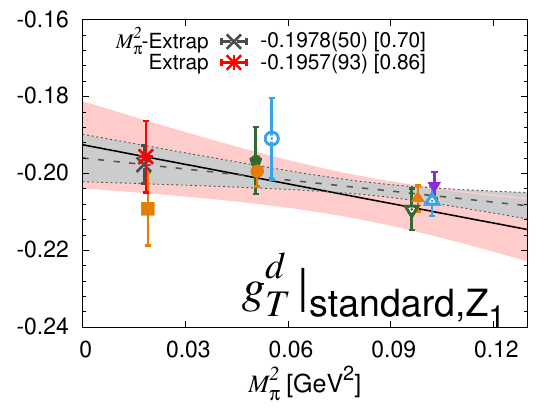}
  \includegraphics[width=0.23\textwidth]{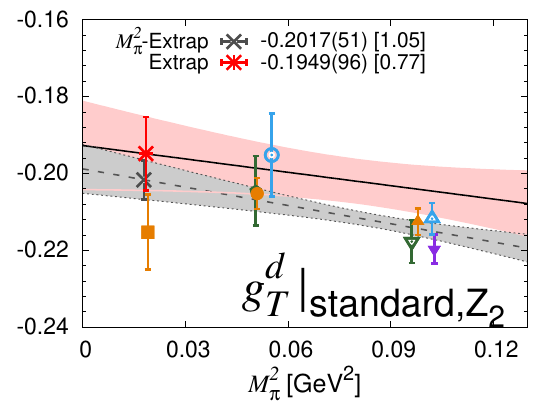}
  \includegraphics[width=0.23\textwidth]{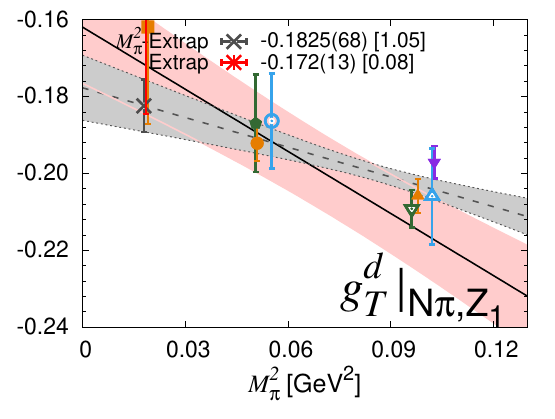}
  \includegraphics[width=0.23\textwidth]{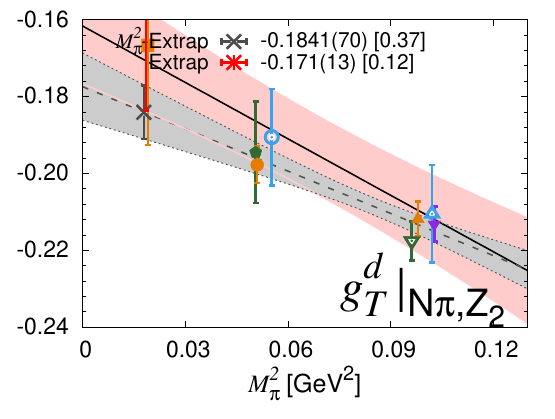}
  
  \includegraphics[width=0.23\textwidth]{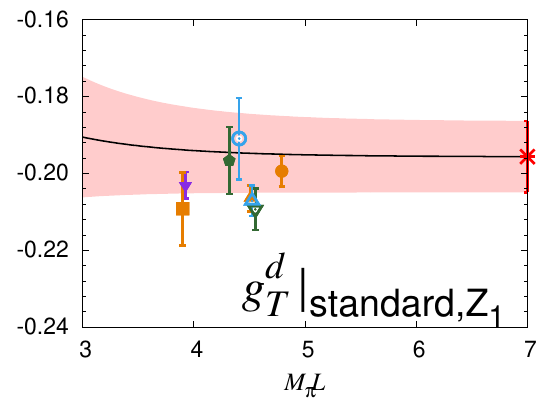}
  \includegraphics[width=0.23\textwidth]{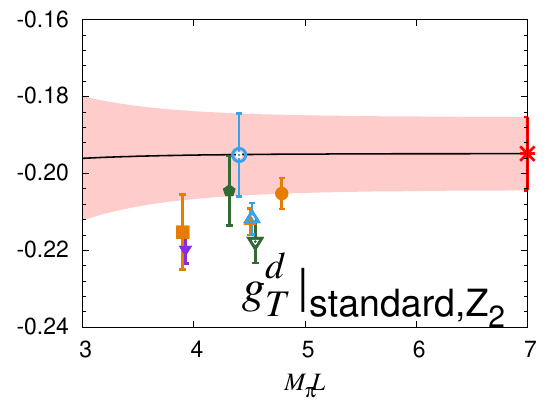}
  \includegraphics[width=0.23\textwidth]{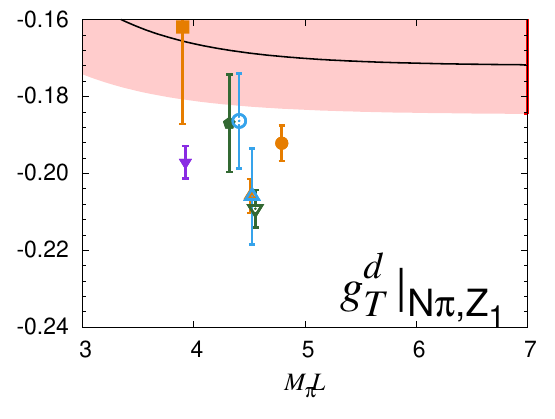}
  \includegraphics[width=0.23\textwidth]{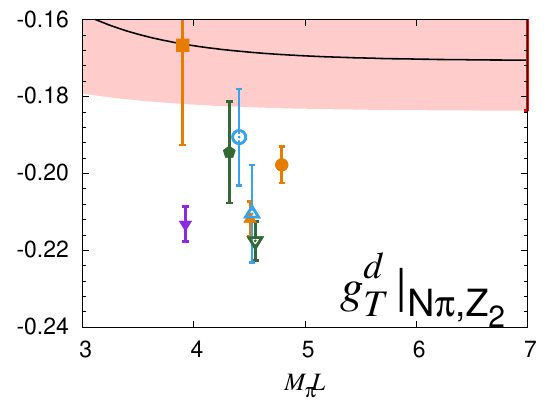}
  \caption{CCFV extrapolation for $g_T^{d}$ using Eq.~\eqref{eq:CCFV-1}. 
    The rest is the same as in Fig.~\ref{fig:CCFV_gAu} }
  \label{fig:CCFV_gTd}  
\end{figure*}

\begin{figure*}      
  \center
  \includegraphics[width=0.3\textwidth]{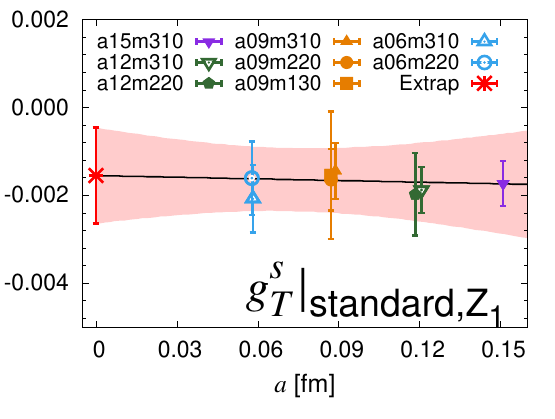}
  \includegraphics[width=0.3\textwidth]{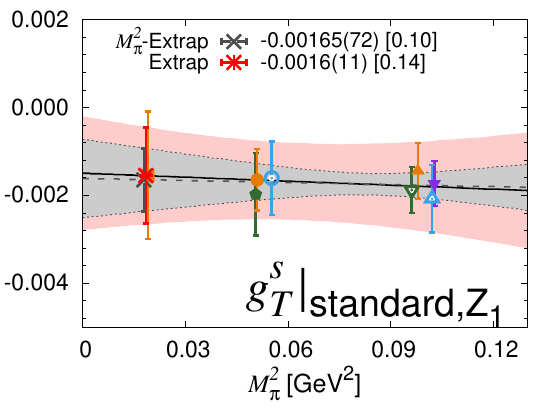}
  \includegraphics[width=0.3\textwidth]{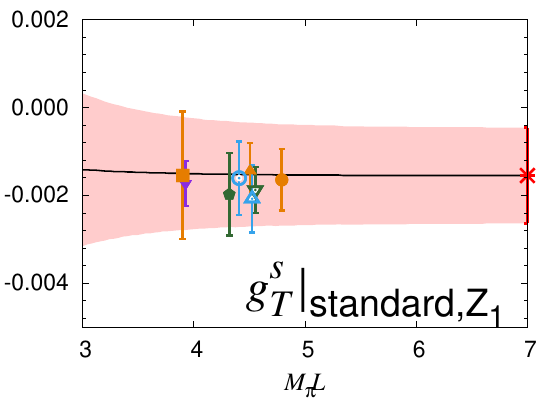}

  \includegraphics[width=0.3\textwidth]{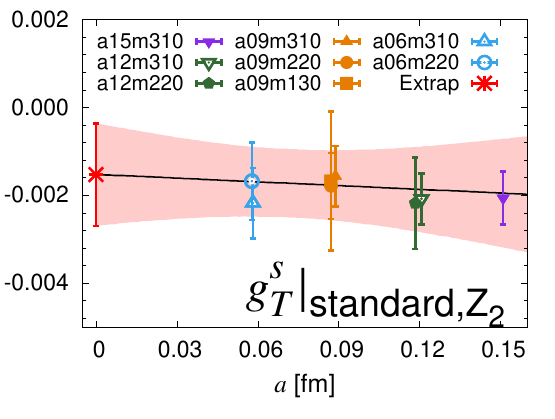}
  \includegraphics[width=0.3\textwidth]{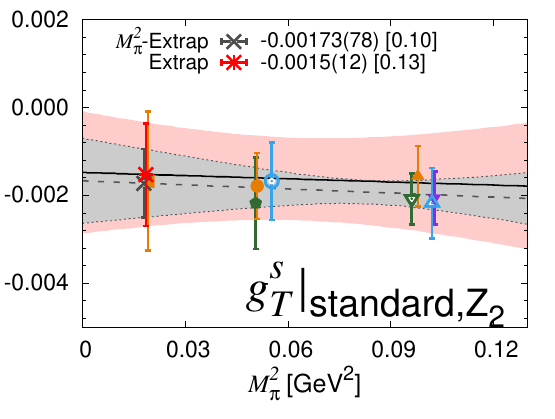}
  \includegraphics[width=0.3\textwidth]{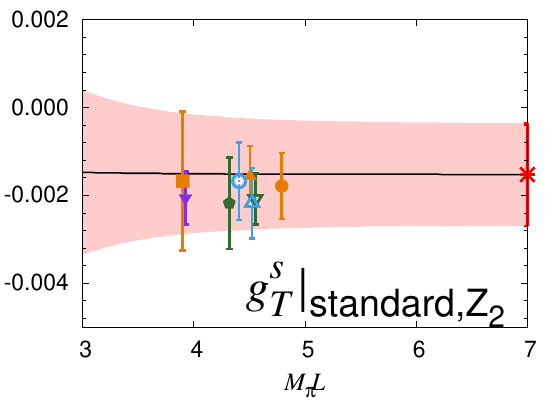}
  \caption{CCFV extrapolation for $g_T^{s}$ using
    Eq.~\eqref{eq:CCFV-1}. The rest is the same as in
    Fig.~\ref{fig:CCFV_gAu}. These data are only analyzed using the
    ``standard" strategy as the likely multihadron lowest excited state,
    $\Sigma K$, is heavier than the $N(1440)$ radial excitation.}
  \label{fig:CCFV_gTs}  
\end{figure*}

\begin{figure*}       
  \center
  \includegraphics[width=0.23\textwidth]{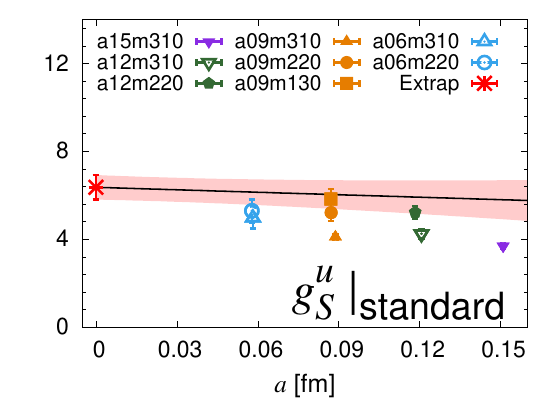}
  \includegraphics[width=0.23\textwidth]{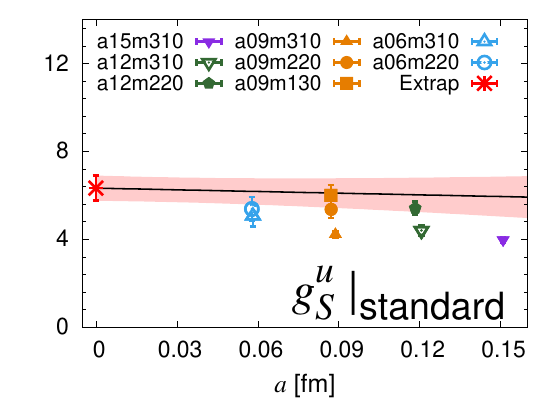}
  \includegraphics[width=0.23\textwidth]{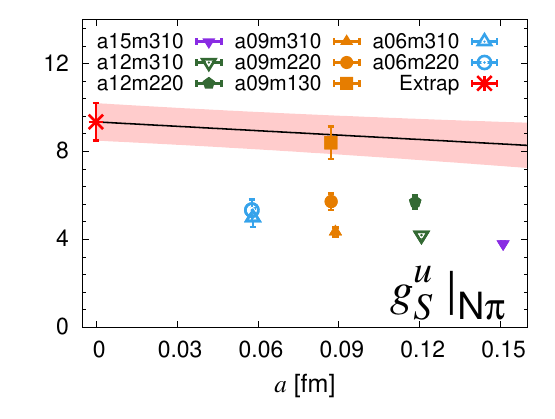}
  \includegraphics[width=0.23\textwidth]{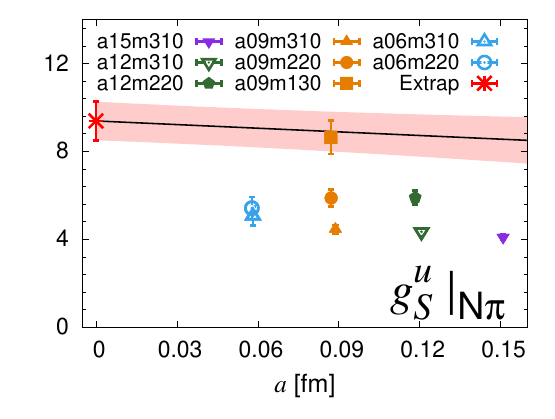}
  
  \includegraphics[width=0.23\textwidth]{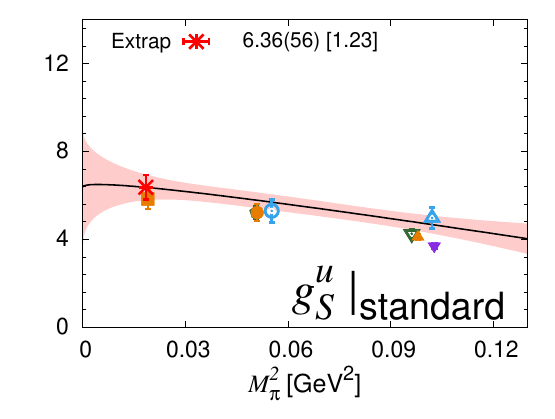}
  \includegraphics[width=0.23\textwidth]{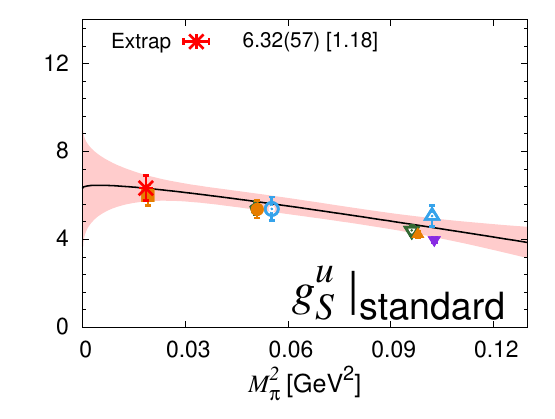}
  \includegraphics[width=0.23\textwidth]{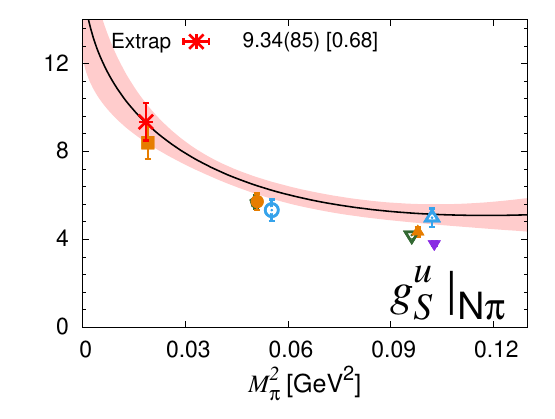}
  \includegraphics[width=0.23\textwidth]{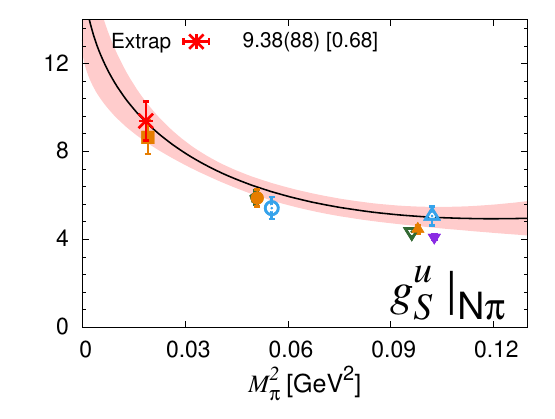}
  
  \includegraphics[width=0.23\textwidth]{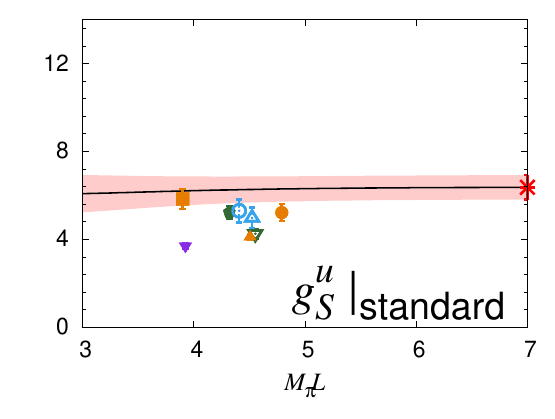}
  \includegraphics[width=0.23\textwidth]{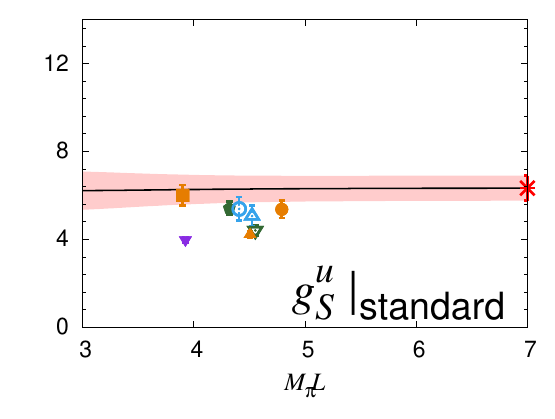}
  \includegraphics[width=0.23\textwidth]{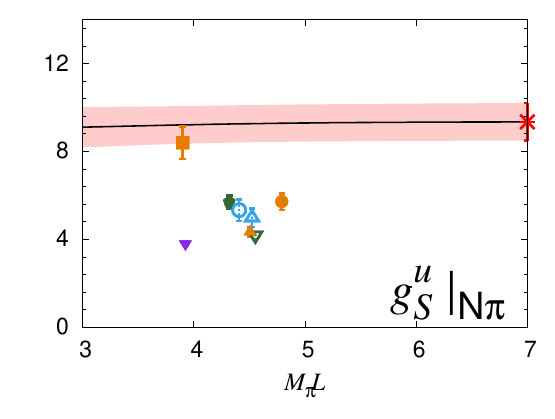}
  \includegraphics[width=0.23\textwidth]{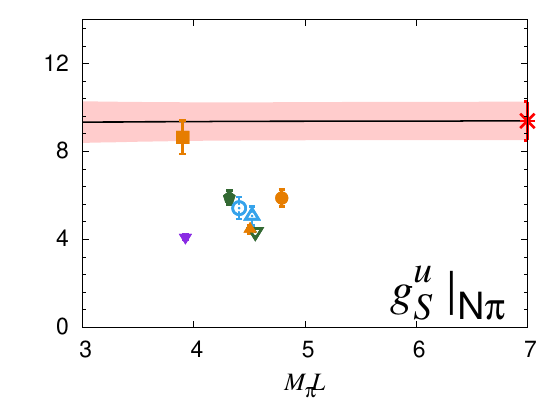}
  \caption{CCFV extrapolation for $g_S^{u}$ using Eq.~\eqref{eq:CCFV-2}. The rest is the same as in Fig.~\ref{fig:CCFV_gAu}. }
  \label{fig:CCFV_gSu}  
\end{figure*}

\begin{figure*}      
  \center
  \includegraphics[width=0.23\textwidth]{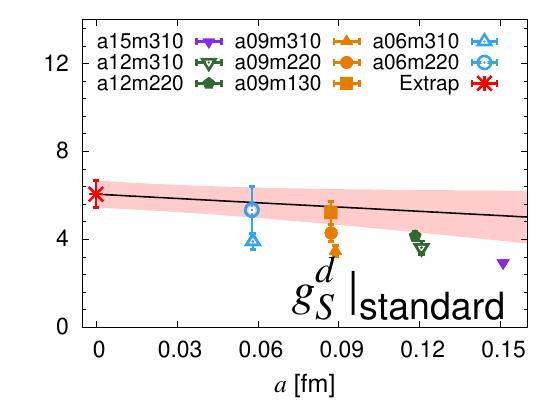}
  \includegraphics[width=0.23\textwidth]{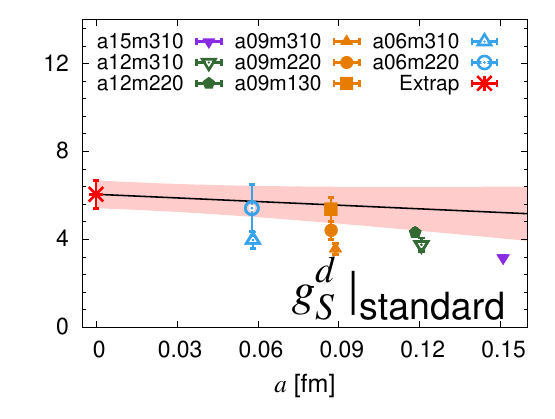}
  \includegraphics[width=0.23\textwidth]{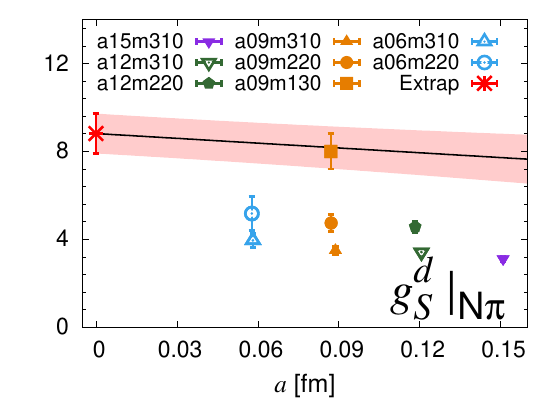}
  \includegraphics[width=0.23\textwidth]{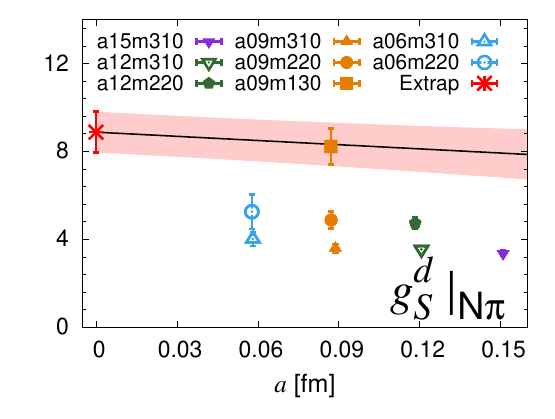}
  
  \includegraphics[width=0.23\textwidth]{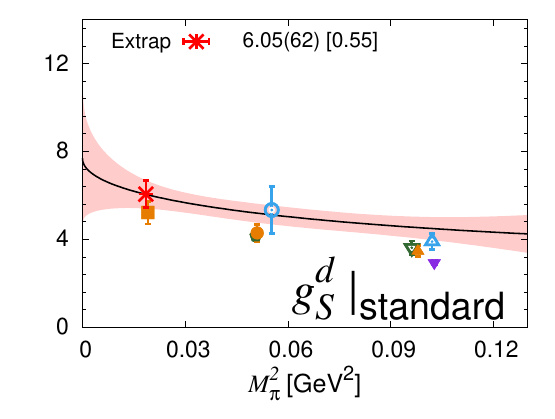}
  \includegraphics[width=0.23\textwidth]{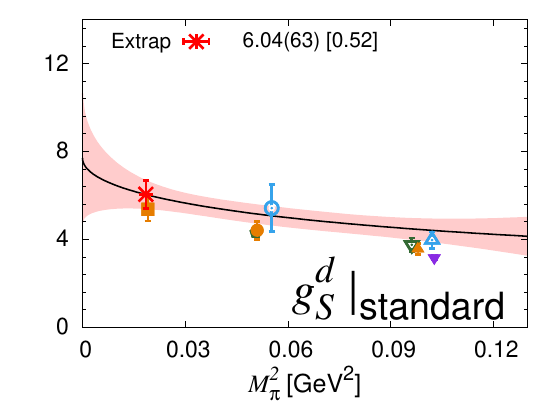}
  \includegraphics[width=0.23\textwidth]{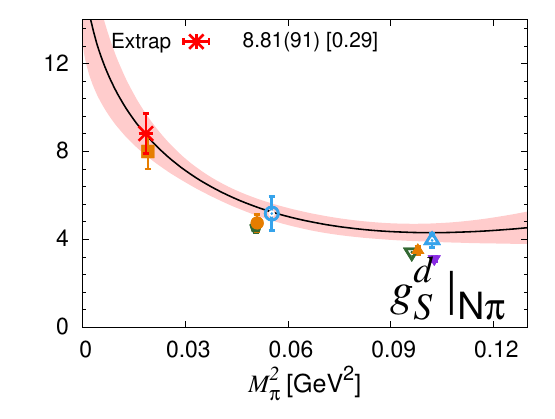}
  \includegraphics[width=0.23\textwidth]{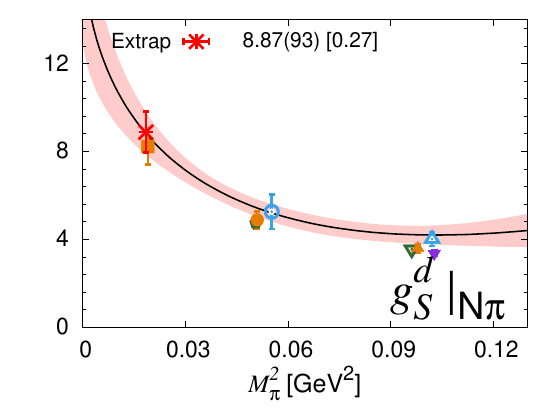}
  
  \includegraphics[width=0.23\textwidth]{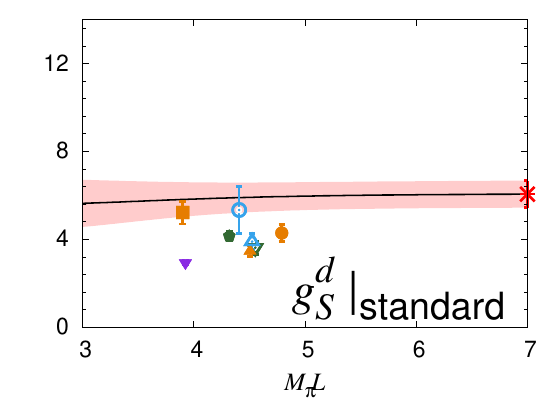}
  \includegraphics[width=0.23\textwidth]{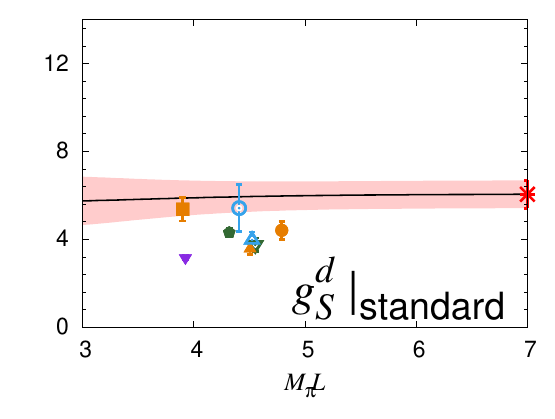}
  \includegraphics[width=0.23\textwidth]{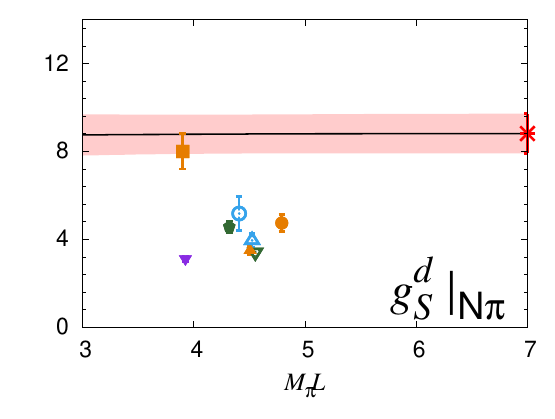}
  \includegraphics[width=0.23\textwidth]{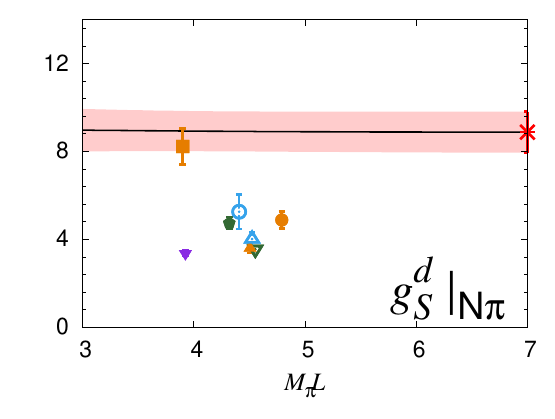}
  \caption{CCFV extrapolation for $g_S^{d}$ using Eq.~\eqref{eq:CCFV-2}. The rest is the same as in Fig.~\ref{fig:CCFV_gAu}. }
  \label{fig:CCFV_gSd}  
\end{figure*}

\begin{figure*}      
  \center
  \includegraphics[width=0.23\textwidth]{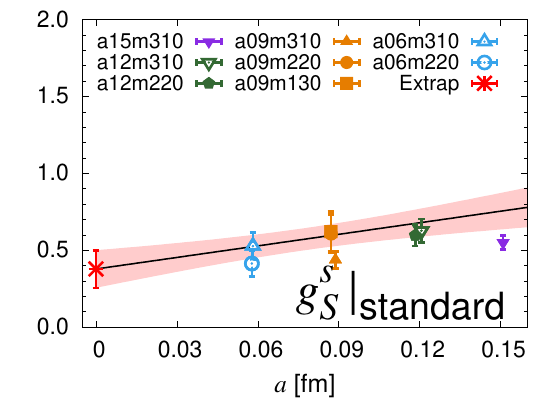}
  \includegraphics[width=0.23\textwidth]{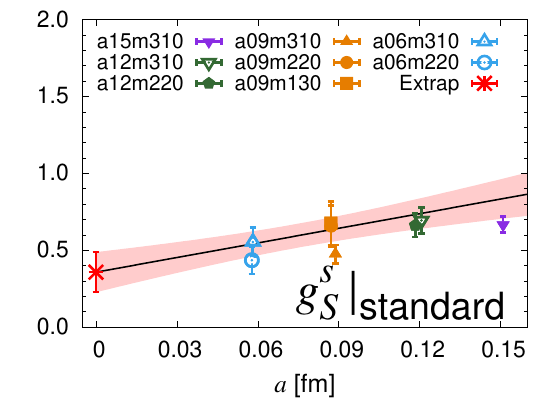}
  \includegraphics[width=0.23\textwidth]{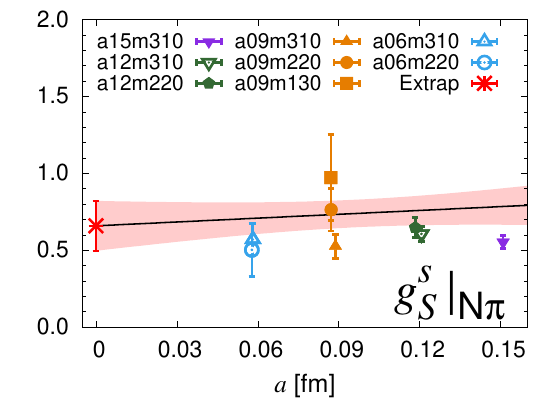}
  \includegraphics[width=0.23\textwidth]{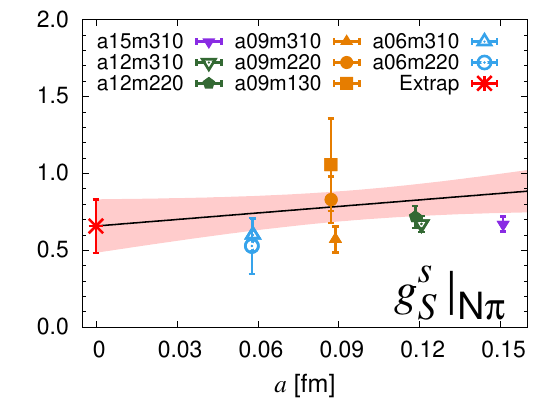}
  
  \includegraphics[width=0.23\textwidth]{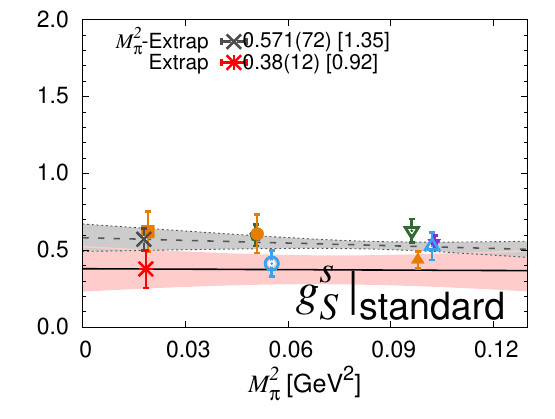}
  \includegraphics[width=0.23\textwidth]{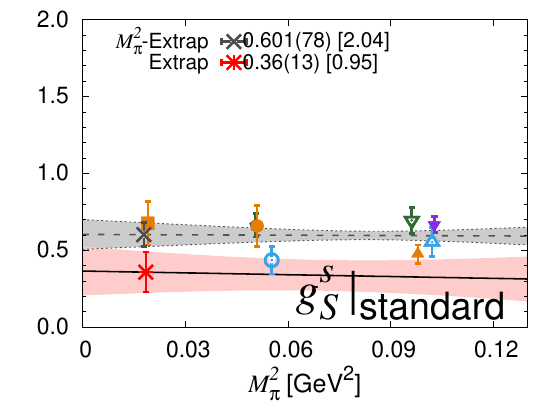}
  \includegraphics[width=0.23\textwidth]{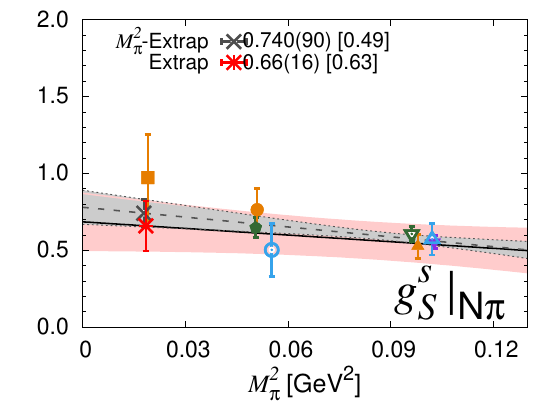}
  \includegraphics[width=0.23\textwidth]{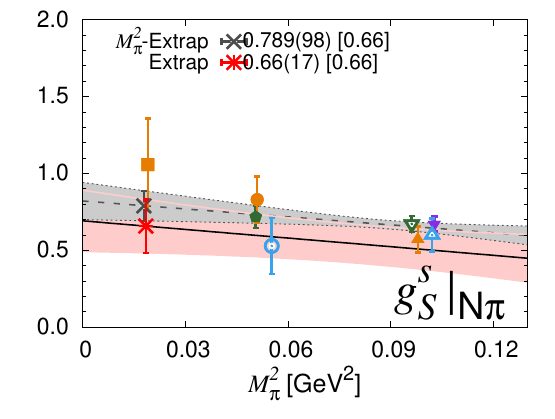}
  
  \includegraphics[width=0.23\textwidth]{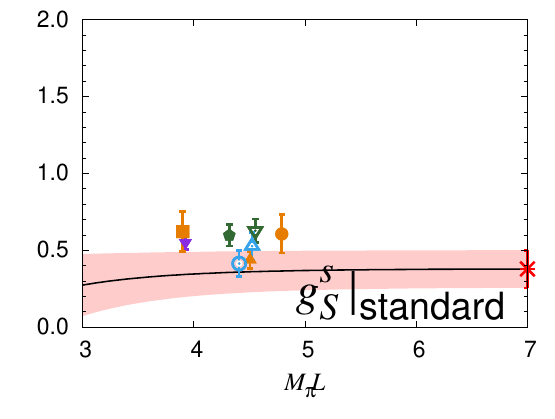}
  \includegraphics[width=0.23\textwidth]{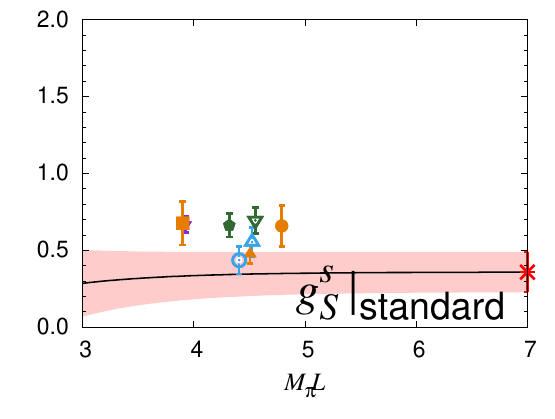}
  \includegraphics[width=0.23\textwidth]{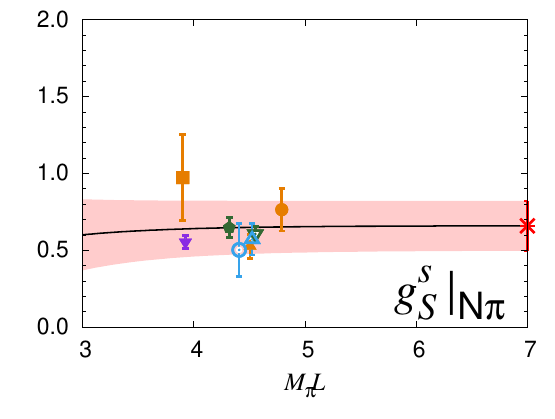}
  \includegraphics[width=0.23\textwidth]{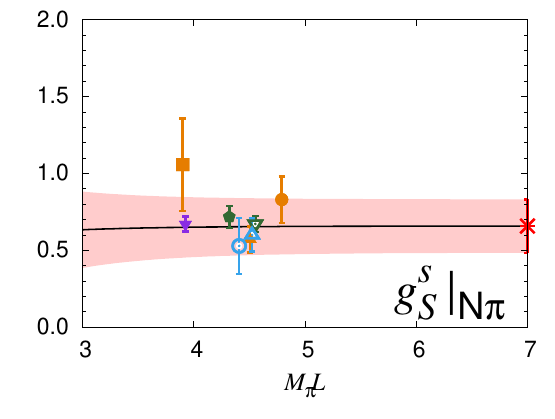}
  \caption{CCFV extrapolation for
    $g_S^{s}$. The rest is the same as in Fig.~\ref{fig:CCFV_gAu} }
  \label{fig:CCFV_gSs}  
\end{figure*}

\begin{figure*}      
  \center
  \includegraphics[width=0.40\textwidth]{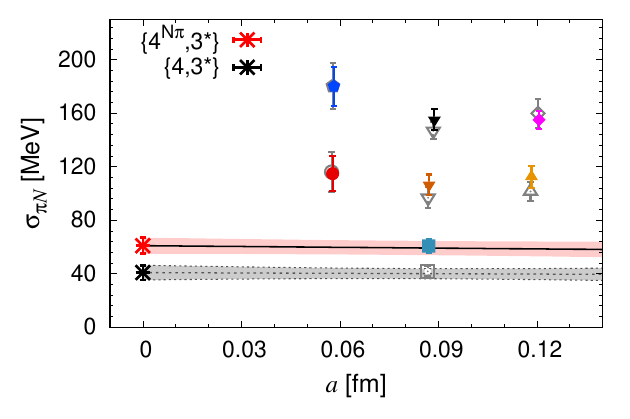}
  \includegraphics[width=0.40\textwidth]{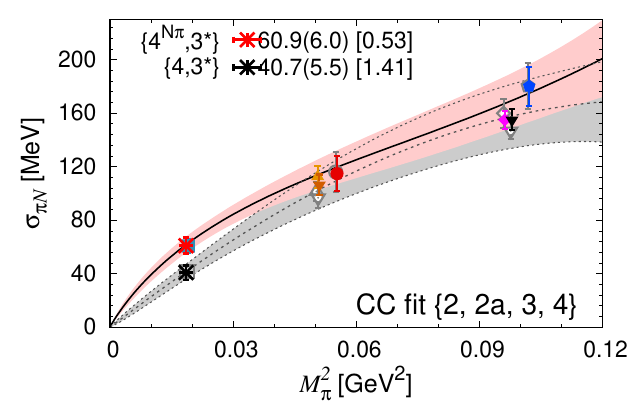}
  \includegraphics[width=0.40\textwidth]{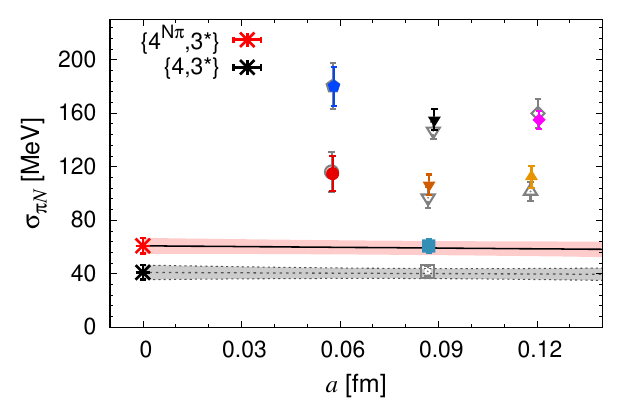}
  \includegraphics[width=0.40\textwidth]{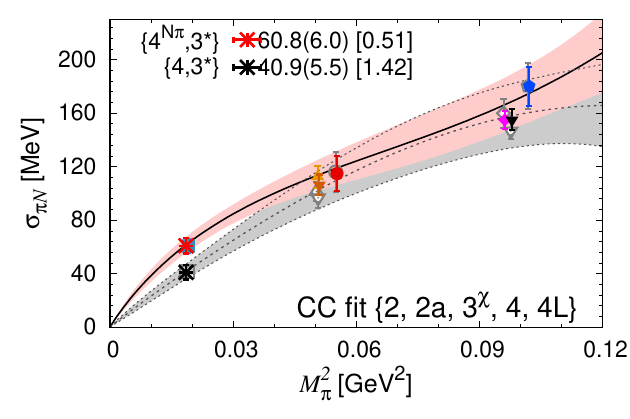}
  \caption{The Chiral-continuum (CC) 
  extrapolation for  $\sigma_{\pi N}=m_{ud}g_S^{u+d}$ using the ansatz 
  given in Eq.~\eqref{eq:CPT}. In the bottom fit, the logarithm term in Eq.~\ref{eq:CPT},  is included by setting  the coefficient $d_3^\chi$ to its   $\chi$PT value taken from Ref.~\cite{Gupta:2021ahb}. Both fits, therefore, have four free  parameters. The color (grey/black) bands and points 
  correspond to the ``$N \pi$" (``standard") analysis. }
  \label{fig:CCFV_sigma}  
\end{figure*}

\clearpage
\section{Nonperturbative renormalization for the 2+1-flavor theory 
using the RI-sMOM scheme}
\label{sec:NPR}

This Appendix presents an overview of the calculation of the
renormalization constants for the flavor diagonal quark bilinear
operators $\CO^f= \bar \psi^f \Gamma \psi^f$ with Dirac matrix
$\Gamma$ and the flavor index $f=\{u,d,s\}$ in the $N_f=3$ theory.
The lattice calculation is done in the RI-sMOM
scheme~\cite{Martinelli:1994ty,Sturm:2009kb} on four ensembles at the 
different values of the lattice spacing used in this study: $a15m310$,
$a12m310$, $a09m310$, and $a06m310$. Since data are obtained at only
one value of $M_\pi \neq 0$ at each $a$, we explicitly assume that the
quark mass dependence can be neglected and these results are a good
approximation to the chiral limit values.  Second, the lattice
calculation is not fully $O(a)$ improved in neither the action nor the
operators. In the latter, we neglect all the lattice spacing dependent 
improvement terms.  Consequently, the results for the
renormalized charges are extrapolated to the continuum limit using an
ansatz starting with a term linear in $a$.

Results for the matching factors between the RI-sMOM and the $\MSb$ scheme, and the running in the $\MSb$
scheme are taken from continuum perturbation
theory~\cite{Kniehl:2020sgo,Baikov:2014qja,Gracey:2022vqr,Gracey:2011fb,Chetyrkin:1999pq,Gracey:2000am}
as specified in Appendix~\ref{sec:MatchingRG}.

The general relation between renormalized, $\CO_R$,  and bare, $\CO$, 
operators including mixing between flavors is: 
\begin{align}
  \CO^f_R &= \sum_{f'} Z_\Gamma ^{ff'}\CO^{f'} \,,
  \label{eq:Z}
\end{align}
with the matrix  $Z_\Gamma^{ff'}$ determined nonperturbatively on the lattice using
the regularization independent (RI) renormalization
scheme~\cite{Martinelli:1994ty} in which the renormalized vertex
function is set to its tree-level value.  The calculation is done with
the gauge fields fixed to the Landau gauge and quark propagators
generated using momentum sources.  The final results for the three
charges, using the two methods, labeled $\rm{Z_1}$ and $\rm{Z_2}$ and 
discussed in Appendix~\ref{sec:Zstrategies}, for partially removing
discretization errors are given in Appendix~\ref{sec:Zfinal}.

\subsection{Flavor mixing in the RI scheme}

\begin{figure}[t]     
  \center
  \subfigure[The connected vertex $\Gamma_\text{conn}(m,\mu^2)$]{ \includegraphics[width=0.4\textwidth]{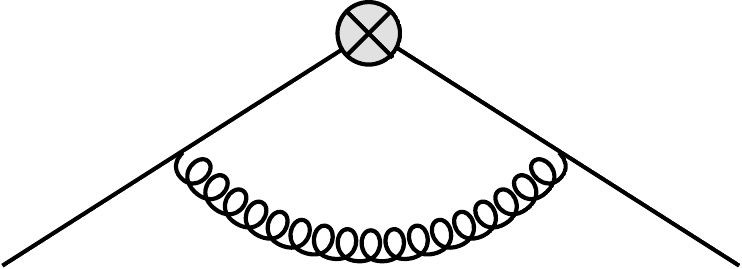}}
  \subfigure[The disconnected vertex $\Gamma_\text{disc}(m_f,m_{f'},\mu^2)$]{ \includegraphics[width=0.4\textwidth]{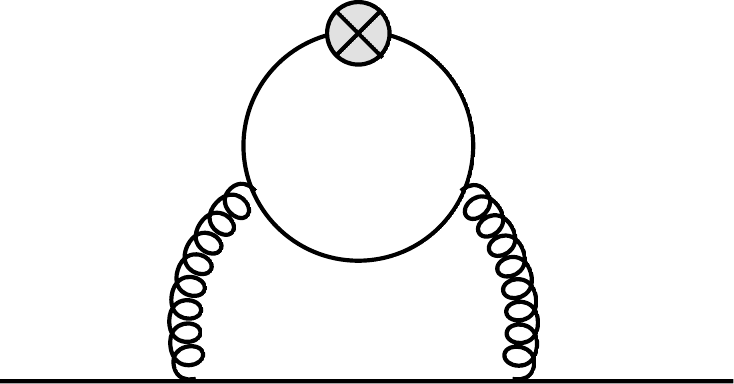}}
  \caption{The amputated vertex diagrams defined in Eq.~\protect\eqref{eq:AVF}}
  \label{fig:diagram}
\end{figure}

 The calculation starts with the amputated vertex function $\Gamma^{ff'}(p_1,p_2)$
defined as,
\begin{align}
  \Gamma^{ff'}(p_1,p_2) &= \langle
  S^f(p_1)\rangle^{-1} \langle \psi^f(p_1) \CO^{f'} \psi^f(p_2) \rangle
  \langle S^f(p_2)\rangle^{-1}
  \label{eq:AVF}
\end{align}
where $\psi^f$ and $S^f$ are the quark field and propagator with
flavor-$f$.  The three-point function $\langle \psi^f(p_1) \CO^{f'}
\psi^f(p_2) \rangle$ gets contributions from both the connected and
the disconnected diagrams shown in Fig.~\ref{fig:diagram}. The
perturbative result for this Green's function $\Gamma^{ff'}(p_1,p_2)$
in the $\MSb$ scheme is given in
Refs.~\cite{Gracey:2020rok,Gracey:2022vqr} in terms of its Lorentz's
components.

Next, we define the projection operators $\mathbb{P}_{\Gamma}$ used in
this calculation to be $\mathds{1}/4, \slashed{q}q^\mu/4q^2,
\gamma_5\slashed{q}q^\mu/4q^2, i\gamma^{[\mu}\gamma^{\nu]}/48$ for the
scalar, vector, axial and tensor operators \cite{Yoon:2016jzj}. The
normalization $\text{Tr}[\gamma_\mu \gamma_\nu ] = \delta_{\mu\nu}
\mathds{1}$ is implied. Then, in the RI-SMOM scheme, the projected
renormalized amputated vertex function is fixed to its tree-level
value
\begin{align}
  Tr[\mathbb{P}_\Gamma \Gamma^{ff'}_R(p_1,p_2)]
   \equiv \frac{Z_\Gamma^{f'f''}}{Z_\psi^f} Tr[\mathbb{P}_\Gamma \Gamma^{ff''}(p_1,p_2)] = \delta^{ff'}\,.
  \label{eq:Gamma}
\end{align}
This relation defines the $3\times 3$ mixing matrix $Z_\Gamma$  for the 3-flavor theory in the RI-sMOM scheme:  
\begin{align}
  Z_\Gamma= \left[\frac{1}{ Z_\psi^f} Tr({\mathbb{P}_\Gamma \Gamma(p_1,p_2)}^T) \right]^{-1} \,.
  \label{eq:ZGamma}
\end{align}
The determination of $Z_\psi^{f}$ is done in a 
separate calculation using two different methods as discussed in Sec.~\ref{sec:Zstrategies}.


The amputated vertex function $\Gamma^{ff'}$ gets contributions from both the connected
and disconnected diagrams shown in Fig.~\ref{fig:diagram}:
\begin{align}
  \Gamma^{ff'}(p_1,p_2)=\Gamma_\text{conn}(m^f,\mu^2)\delta^{ff'} -
  \Gamma_\text{disc}(m^f,m^{f'},\mu^2),
\end{align}
The (-) sign in $\Gamma_\text{disc}$ is due to the anticommuting
nature of the fermion fields contracted into the quark loop. On the lattice
we calculate the projected amputated vertex functions $c_f$
(connected) and $d_{ff'}$ (disconnected) defined as\looseness-1
\begin{align}
  c_\Gamma^f &= \frac{1}{Z_\psi^f}\text{Tr}\left[\mathbb{P}_\Gamma \Gamma_\text{conn}(m^f,\mu^2) \right]
  \label{eq:c-def} \\
  d_\Gamma^{ff'} &=
  \frac{1}{Z_\psi^f}\text{Tr}\left[\mathbb{P}_\Gamma \Gamma_\text{disc}(m_f,m_{f'},\mu^2) \right] \,.
    \label{eq:d-def}
\end{align}

For the $N_f=2+1$ isospin symmetric theory relevant to this work, the
determination of $Z_\Gamma^{ff'}$ require calculating the following 6
quantities on the lattice,
\begin{align}
  c_\Gamma^l,~c_\Gamma^s,~d_\Gamma^{ll},~d_\Gamma^{ls},~d_\Gamma^{sl}\text{, and }d_\Gamma^{ss} 
\end{align}
for each $\Gamma$. Working in the flavor basis $f\in \{u-d,u+d,s\}$, 
the $ Z_\Gamma$, defined in Eq.~\eqref{eq:ZGamma}, is block diagonal:
\begin{align}
    Z_\Gamma&\equiv
  \begin{pmatrix}
    Z_\Gamma^{u-d,u-d} & 0 & 0 \\
    0 & Z_\Gamma^{u+d,u+d} & Z_\Gamma^{u+d,s}\\
    0 & Z_\Gamma^{s,u+d} & Z_\Gamma^{ss} \\
  \end{pmatrix}\\
  &=
  \begin{pmatrix}
    c_\Gamma^l & 0 & 0 \\
    0 & c_\Gamma^l-2d_\Gamma^{ll} & -2d_\Gamma^{sl} \\
    0 & -d_\Gamma^{ls} & c_\Gamma^s-d_\Gamma^{ss}\\
  \end{pmatrix}^{-1}.
  \label{eq:Z_RI}
\end{align}
For the discussion of the matching and running, the more convenient
basis is the isovector (3), octet (8), and singlet (0) one, i.e.,
$f\in \{u-d,u+d-2s,u+d+s\}$. In this case the renormalization matrix in terms of the 6 quantities is\looseness-1

\begin{widetext} 
\begin{eqnarray}
  \label{eqn:diropex}
\hspace{-0.8cm} \left(\begin{array}{ccc}
    Z^{3,3}_{\Gamma} & 0 & 0 \\
    0 & Z^{8,8}_{\Gamma} & Z^{8,0}_{\Gamma}\\0 &  Z^{0,8}_{\Gamma} &  Z^{0,0}_{\Gamma}\end{array}\right) = 
    \left(\begin{array}{ccc}
    c_\Gamma^l & 0 & 0 
    \\
    0 & \frac{1}{3} (c_\Gamma^l+2 (c_\Gamma^s-d_\Gamma^{ll}+d_\Gamma^{ls}+d_\Gamma^{sl}-d_\Gamma^{ss})) & \frac{2}{3} (c_\Gamma^l-c_\Gamma^s-2 d_\Gamma^{ll}+2 d_\Gamma^{ls}-d_\Gamma^{sl}+d_\Gamma^{ss})
    \\
    0 & \frac{1}{3} (c_\Gamma^l-c_\Gamma^s-2 d_\Gamma^{ll}-d_\Gamma^{ls}+2 d_\Gamma^{sl}+d_\Gamma^{ss}) &  \frac{1}{3} (2 c_\Gamma^l+c_\Gamma^s-4 d_\Gamma^{ll}-2 d_\Gamma^{ls}-2 d_\Gamma^{sl}-d_\Gamma^{ss})\end{array}\right)^{-1}  \,. 
\end{eqnarray}
\end{widetext}

\subsection{RI-sMOM scheme}
\begin{figure*}[h]     
  \includegraphics[width=0.323\linewidth]{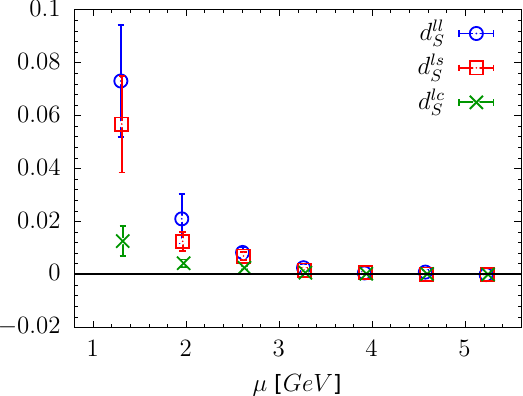}
  \includegraphics[width=0.323\linewidth]{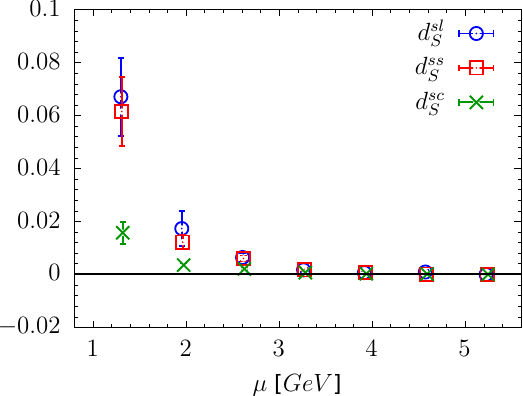}
  \includegraphics[width=0.323\linewidth]{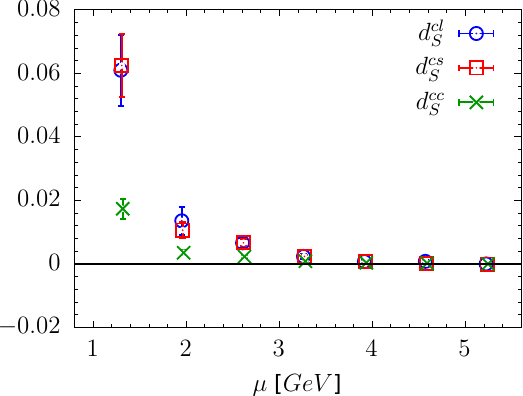}
  \caption{Data for the disconnected projected amputated Green's
    function $d_\Gamma^{ff^\prime}$ for a quark with flavor $f \in
    \{l,s,c\}$ and the disconnected loop with flavor $f^\prime$ for
    the scalar bilinear operator $O_S^{f^\prime}$ in the RI-sMOM
    scheme. The allowed momenta satisfy $\mu^2=p^2={p'}^2=(p-p')^2$.
    Data for the charm quark are also plotted to show the dependence
    on the quark mass even though they are not used in this work. }
  \label{fig:d_lf}
\end{figure*}

\begin{figure*}[t]      
\centering
  \includegraphics[width=0.24\linewidth]{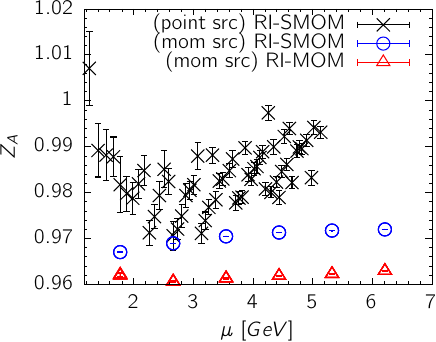}
  \includegraphics[width=0.24\linewidth]{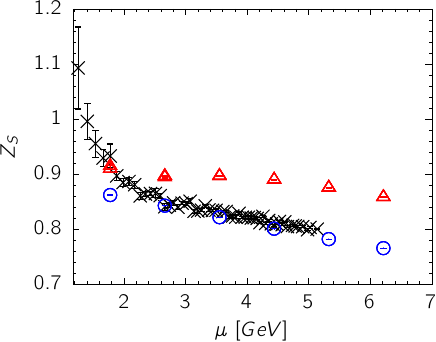}
  \includegraphics[width=0.24\linewidth]{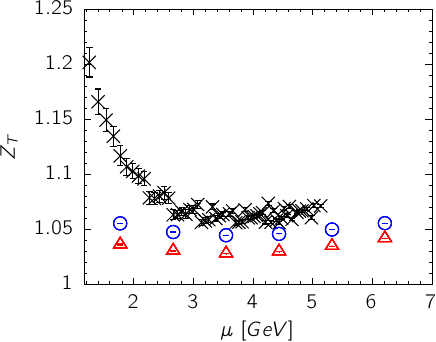}
  \includegraphics[width=0.24\linewidth]{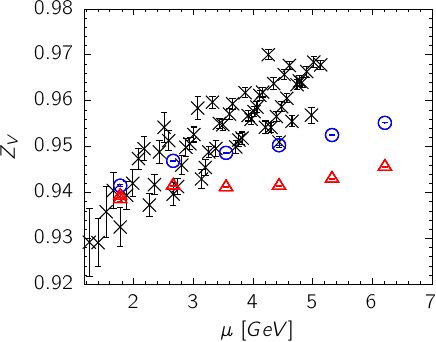}
  \caption{These plots show the improvement in the calculation of the
    four renormalization factors in the RI-sMOM scheme obtained using
    momentum (mom src) versus point source propagators (point
    src). The data are for the a06m310 ensemble.  For comparison, we
    also show the factors obtained in the RI-MOM scheme even though
    they are not used in this work. }
  \label{fig:Z-pt}
\end{figure*}

The lattice calculation is done in the RI-sMOM scheme
\cite{Sturm:2009kb} where the 4-momentum of the external legs
$\{p_1,p_2\}$ satisfies the symmetric momentum condition
$p_1^2=p_2^2=(p_1-p_2)^2=\mu^2$, and $\mu^2$ defines the
renormalization scale. We find that the matrix $Z^\text{RI-sMOM}(\mu)$
is close to diagonal with $c_\Gamma^l\sim O(1)$ and $d_\Gamma^{ff'}$
about a few percent at $\mu\sim2$~GeV.
This is illustrated in Fig.~\ref{fig:d_lf} for the scalar operator
(largest mixing) for 3 of the disconnected projected amputated Green's
functions $d_\Gamma^{lf}$ calculated at various $\mu$. As expected,
the value decreases as the quark mass in the loop is increased from
light to strange to charm. To get a signal for this small mixing, we
use momentum source propagators~\cite{Skullerud:1999gv} in Landau
gauge and take the momenta $\{p_1,p_2\}$ to be proportional to
$(1,1,1,1)$ to minimize $O(4)$-symmetry breaking. A comparison of the
improvement in the extraction of $Z$ using momentum versus point
sources is shown in Fig~\ref{fig:Z-pt}.

\subsection{Matching and RG running}
\label{sec:MatchingRG}

The calculation of the renormalization factors for each operator involves the four steps described below.

\begin{enumerate}
\item To obtain the full $3\times 3$ matrix,
  $Z^\text{RI-sMOM,{latt}}(\mu)$ defined in Eq.~\eqref{eq:Z_RI}, the
  two $c_f$ and four $d_{ff'}$ (Fig.~\ref{fig:d_lf}) are calculated
  using momentum source propagators with the lattice fixed to Landau
  gauge. The simulations are done for a range of lattice momenta
  satisfying $q^2 = p_1^2 = p_2^2 \equiv \mu^2_{\rm latt}$.

\item For each $\mu_{\rm latt}$, a horizontal matching, $\mu_{\rm
  cont} = \mu_{\rm latt}$, to $\MSb$ scheme is performed by
  multiplying $Z^\text{RI-sMOM,{latt}}(\mu)$ with the matching factor
  $C^{\text{RI-sMOM}\to \MSb}(\mu)$ calculated using perturbation
  theory in the continuum:
  \begin{align}
 \hspace{0.20in}   Z^{\MSb}(\mu)=C^{\text{RI-sMOM}\to \MSb}(\mu)Z^\text{RI-sMOM,{latt}}(\mu) \,.
  \end{align}
\item The result for each $\mu$ is run in the continuum to  a common scale, which we choose to be 2~GeV:
  \begin{align}
  \hspace{0.20in}  Z^{\MSb}(2\GeV;\mu)=C^{\MSb}(2\GeV,\mu)Z^{\MSb}(\mu) \,.
  \end{align}
\item Lattice artifacts giving rise to a $\mu$ dependence. These are removed using the ansatz 
  \begin{align}
  \hspace{0.20in}  Z^{\MSb}(2\GeV;\mu)= Z^{\MSb}(2\GeV) + c_1 \mu^2 + c_2 \mu^4 .
    \label{eq:mufit}
  \end{align}
  Note that, in the RI-sMOM calculation, there are, in general, not
  enough data points at small $\mu^2$ to include a $1/\mu^2$ term as
  was done in Refs.~\cite{Bhattacharya:2016zcn,Yoon:2016jzj}. We,
  therefore, fit to points with $\mu^2_{\rm min}$ large enough so that
  the contribution of a $1/\mu^2$ term can be assumed to be
  negligible.
\end{enumerate}

The final results for $Z^{\MSb}_{\{A,S,T\}} (2\GeV)$ are given in
Appendix~\ref{sec:Zfinal}.  We remind the reader of the implicit
assumption that the possible quark mass dependence has been neglected
and the results obtained at $M_\pi \neq 0$ are a good approximation to
the chiral limit values.

In the continuum calculation of $C^{\text{RI-sMOM}\to \MSb}(\mu)$ and
$C^{\MSb}(2\GeV,\mu)$ for the scalar, tensor and pseudoscalar
operators, the singlet and non-singlet factors are the same because
the disconnected loops vanish in the zero mass limit due to
chirality. Thus the $3 \times 3$ matrices for the matching and running
reduce to an overall factor, for example, $ C^{\text{RI-sMOM}\to
  \MSb}(\mu) = Z^{\rm match} \times \mathds{1}$.  The three-loop
result for the scalar and tensor can be obtained, after projection,
from the amputated Green's functions given in
Ref.~\cite{Kniehl:2020sgo}:
\begin{eqnarray}
\hspace{-0.6cm} Z^{match}_{S,\SMOM}&=&1+0.64552 a\ +\  (23.024\, -4.0135 n_f) a^2 
\nonumber\\
&+& \left(2.1844 n_f^2-169.923 n_f+889.74\right) a^3  , \nonumber\\
\hspace{-0.6cm} Z^{match}_{T,\SMOM}&=&1-0.215 a\ + \ (4.103 n_f-43.384)a^2 
\nonumber\\
&-&  \left(7.0636n_f^2-309.828 n_f+1950.76\right) a^3 ,
\end{eqnarray}
where $a(\mu) \equiv \alpha_s(\mu)/4\pi$.  The running factor,
$C^{\MSb}(2\GeV,\mu)$ for the tensor is obtained from the four-loop
results in Ref.~\cite{Gracey:2022vqr}, and for the scalar the
five-loop result in Ref.~\cite{Baikov:2014qja} .

The matching and running factors are unity for both the singlet and
non-singlet vector operator because of the conservation of the vector
current.  The renormalization factor $Z_V$ for the bare local operator
is given by the relation $Z_V g_V = 1$, with $g_V$ obtained from the
forward matrix element as discussed in Appendix~\ref{sec:Zstrategies}.

For the axial operator, only the flavor nonsinglet axial current is
conserved in the chiral limit.  Therefore the matching from RI-sMOM to
$\MSb$ scheme and the RG running factor is unity for the nonsinglet
operator.

The flavor singlet axial current is not conserved due to the $U_A(1)$
anomaly, and the renormalization factors become nontrivial at the
two-loop level. This $1+\CO(\alpha^2)$ correction is taken from
Refs.~\cite{Green:2017keo, Gracey:2020rok}, and the three-loop
anomalous dimension from Ref.~\cite{Larin:1993tq} is used for the RG
running.\footnote{The 4-loop anomalous dimension for the isosinglet
axial current in the continuum in the $\MSb$ scheme has been
calculated in Ref.~\cite{Chen:2021gxv}. We use the 3-loop running for
consistency with the RI-sMOM to $\MSb$ matching that has been
calculated to only 2-loops.\looseness-1}

The off-diagonal terms in the mixing matrix for the singlet axial
current are zero in the chiral limit, leaving only the diagonal
elements. The difference between the singlet and nonsinglet (the
disconnected contribution) has been calculated in the $\MSb$ scheme in
the RI-sMOM setup in Ref.~\cite{Gracey:2020rok}. This is used to get
the following matching factor for the axial current in the $\{3,8,0\}$
basis:
\begin{eqnarray}
\hspace{-0.7cm} C_A^{\text{RI-sMOM}\to \MSb}(\mu)  = \left(\begin{array}{ccc}
    1 & 0  &  0\\
    0 & 1 & 0 \\  
    0 & 0  &1 -26.5\, a^2\, n_f \end{array}\right)
\end{eqnarray}

In the calculation of all the matching and running factors, the number
of flavors used is $n_f=4$. We have checked that the difference in the
final results between using $n_f=3$ and $4$ is below the claimed
precision.

On the lattice, Wilson-clover fermions break chiral symmetry
explicitly and there are non-zero contributions from disconnected
loops for all charges. These give rise to off-diagonal terms as shown
in Eqs.~\eqref{eq:Z_RI} and~\eqref{eqn:diropex}.  The mixing in the
tensor channel is $O(\alpha_s^3)$ in lattice perturbation
theory~\cite{Constantinou:2016ieh} implying $r_T \equiv
Z_T^\text{singlet}/Z_T^\text{non-singlet} = 1+ O(\alpha_s^3) \approx
1$ where $a=\alpha_s/4\pi$ and $n_f$ denotes the number of flavors.
The largest disconnected contribution (off-diagonal elements
$d_\Gamma^{ff'}$) is for the axial. In the $\{u-d,u+d,s\}$ basis, it
is approximately $7\%$ for the coarsest $a \approx 0.15$~fm ensemble
and decreases to below 2\% at the finest $a\approx 0.06$~fm ensemble.

The final factors used to renormalize the charges are given in
Appendix~\ref{sec:Zfinal}.  Note the significant differences between
results with the $\rm{Z_1}$ versus the $\rm{Z_2}$ strategies (defined
in Appendix~\ref{sec:Zstrategies}) used to calculate $Z_\psi$.  There
is a similar pattern of differences between the three diagonal
elements with $Z^{u-d,u-d} \approx Z^{u+d,u+d}$ that differ
significantly from $Z^{s,s}$.  This mass dependence is discussed in
Appendix~\ref{sec:Zstrategies} and our current understanding is, it is
a discretization effect that in the final analysis is, to a large
extent, taken care of in the continuum (CCFV) extrapolation of the
renormalized charges. The mixing is $O(\alpha_s^2)$ in lattice
perturbation theory~\cite{Constantinou:2016ieh}, which would be larger
than what the lattice data for the tensor operator given in
Appendix~\ref{sec:Zfinal} suggests.

In Fig.~\ref{fig:Z_MSb2GeV_a06}, we show the data for $Z_\Gamma$ in
$\MSb$ scheme at $2\GeV$ for the $a06m310$ ensemble as a functions of
$\mu$ along with fits using Eq.~\eqref{eq:mufit}.  The results for the
diagonal parts, for both $\rm{Z_1}$ and $Z_2 $, show a difference that
vanishes in the continuum limit.  The off-diagonal mixing elements
$Z_\Gamma^{u+d,s}$ and $Z_\Gamma^{s,u+d}$ shown in the bottom two rows
are all smaller than $1\%$ and results with the $\rm{Z_1}$ and
$\rm{Z_2}$ strategies essentially overlap.

One advantage of using RI-sMOM scheme for the scalar channel is the
better convergence between 2- and 3-loop running and smaller deviation
from unity in the perturbative series for the matching factor compared
to the RI-MOM scheme \cite{Chetyrkin:1999pq, Sturm:2009kb} as
illustrated in Fig.~\ref{fig:matching}.

To summarize, for our setup, the numerical results given in Appendix~\ref{sec:Zfinal} 
show that the off-diagonal terms and the $r_{\cal O}-1$ factors are small. 

\begin{figure*}      

  \center
  \includegraphics[width=0.24\textwidth]{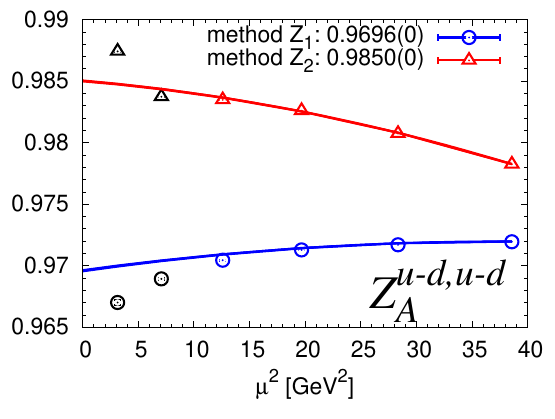}
  \includegraphics[width=0.24\textwidth]{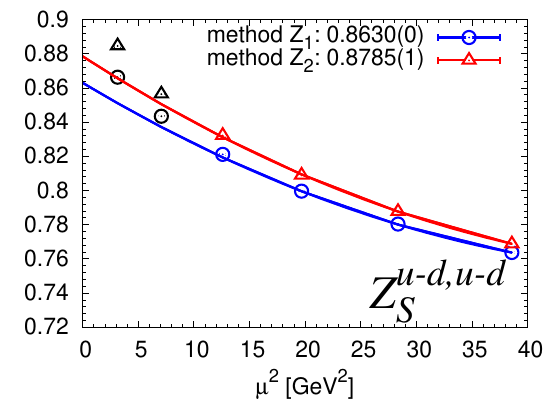}
  \includegraphics[width=0.24\textwidth]{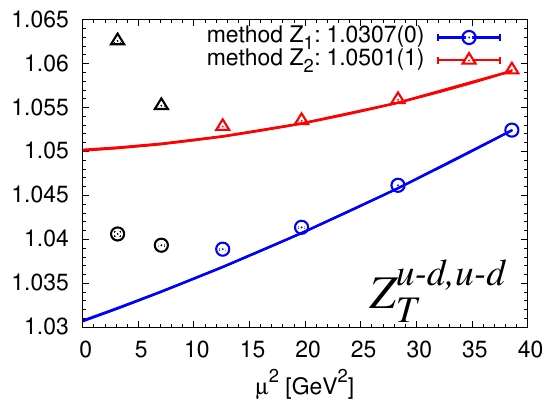}
  \includegraphics[width=0.24\textwidth]{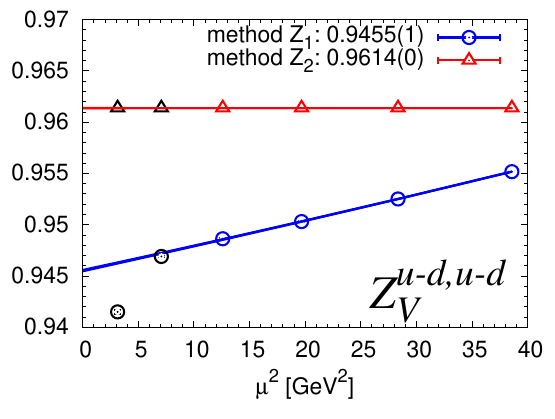}

  \includegraphics[width=0.24\textwidth]{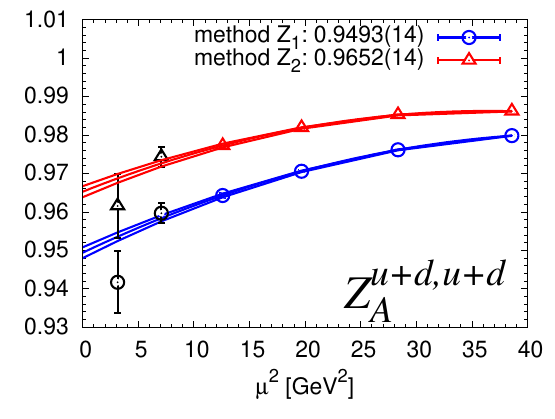}
  \includegraphics[width=0.24\textwidth]{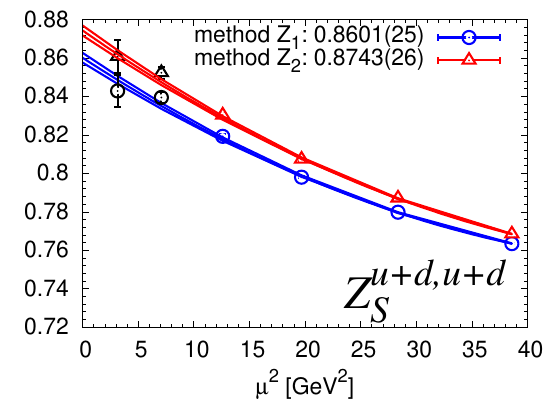}
  \includegraphics[width=0.24\textwidth]{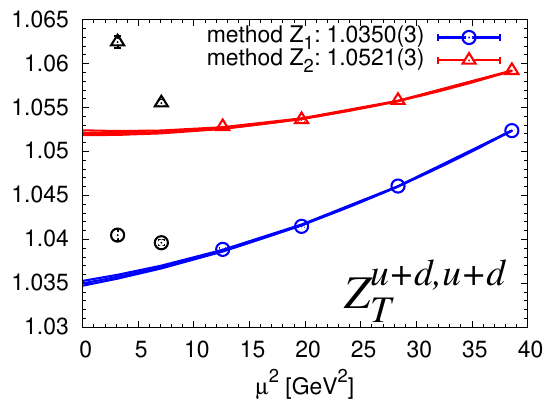}
  \includegraphics[width=0.24\textwidth]{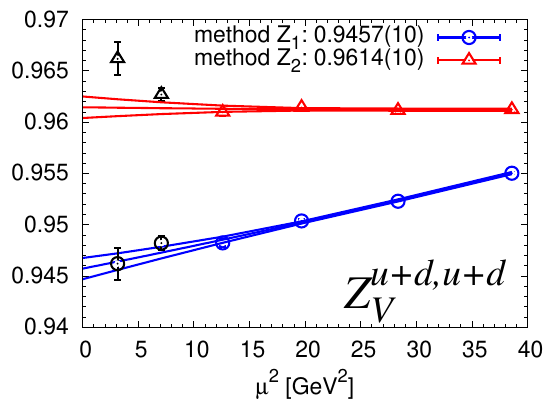}

  \includegraphics[width=0.24\textwidth]{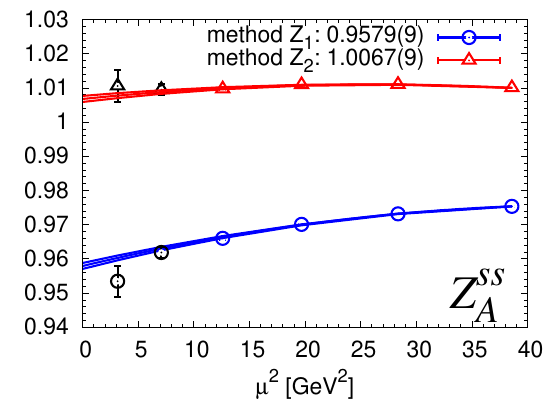}
  \includegraphics[width=0.24\textwidth]{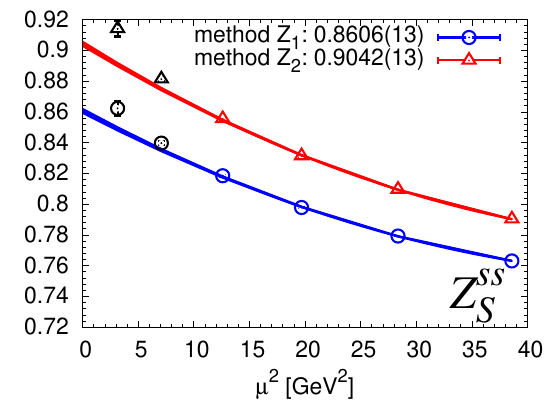}
  \includegraphics[width=0.24\textwidth]{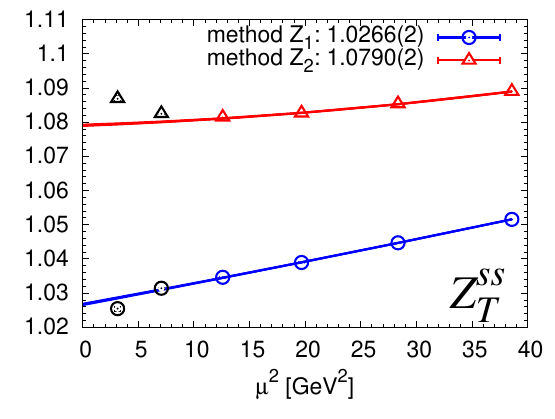}
  \includegraphics[width=0.24\textwidth]{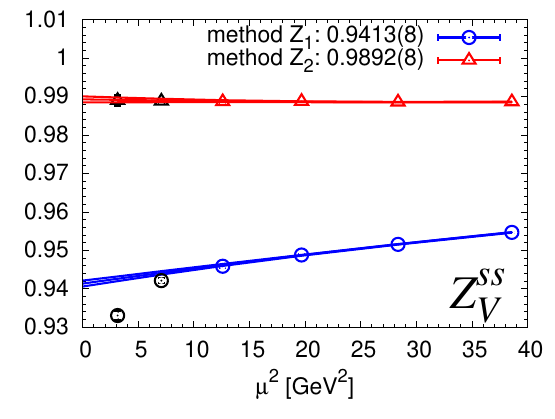}

  \includegraphics[width=0.24\textwidth]{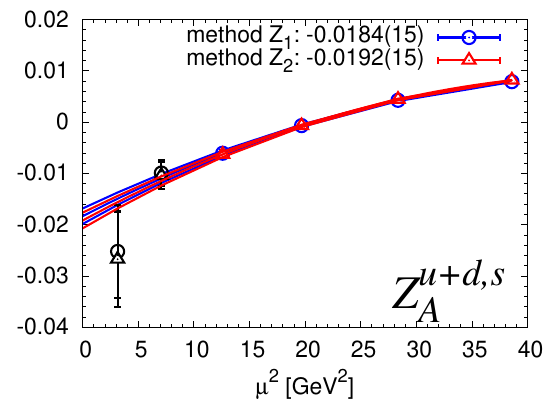}
  \includegraphics[width=0.24\textwidth]{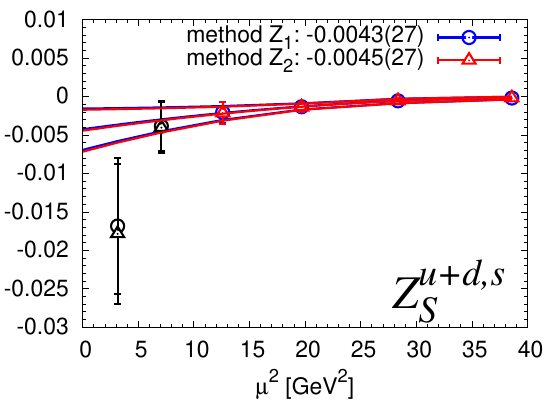}
  \includegraphics[width=0.24\textwidth]{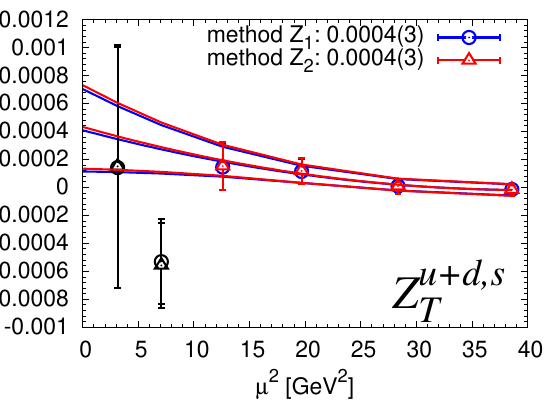}
  \includegraphics[width=0.24\textwidth]{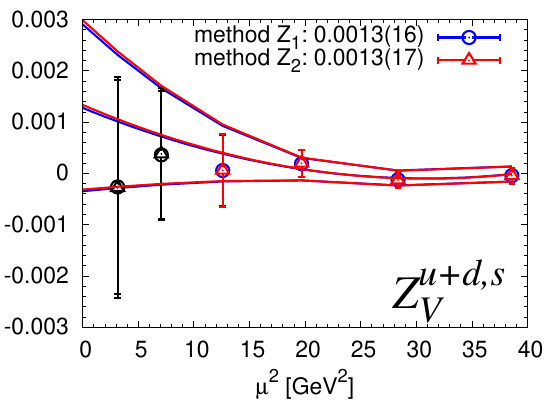}

  \includegraphics[width=0.24\textwidth]{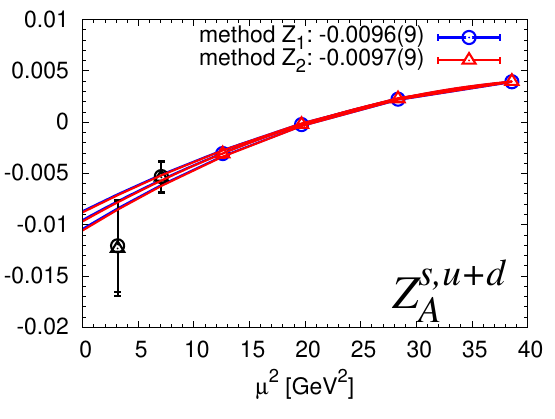}
  \includegraphics[width=0.24\textwidth]{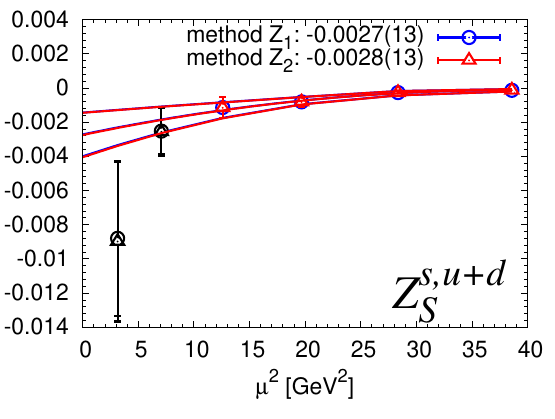}
  \includegraphics[width=0.24\textwidth]{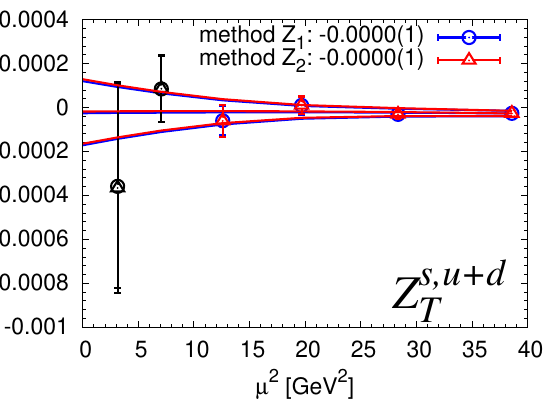}
  \includegraphics[width=0.24\textwidth]{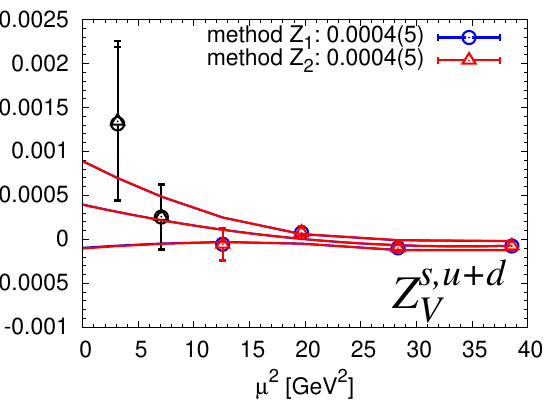}
  \caption{Data for $Z_\Gamma$ in $\MSb$ scheme at $2\GeV$ on the
    a06m310 ensemble along with a quadratic fit versus $\mu^2$ defined
    in Eq.~\eqref{eq:mufit}. Each panel compares results from the two
    different strategies, $\rm{Z_1}$ (blue) and $\rm{Z_2}$ (red). The
    four columns give results for the axial, scalar, tensor and vector
    operators, respectively. Each row represents one of the 5 nonzero
    matrix elements in the $3 \times 3$ matrix defining $Z$ in the
    $\{u-d,u+d,s\}$ basis in Eq.~\eqref{eq:Z_RI}. }
  \label{fig:Z_MSb2GeV_a06}
\end{figure*}

\begin{figure*}[!h]      
  \center
  \includegraphics[width=0.48\textwidth]{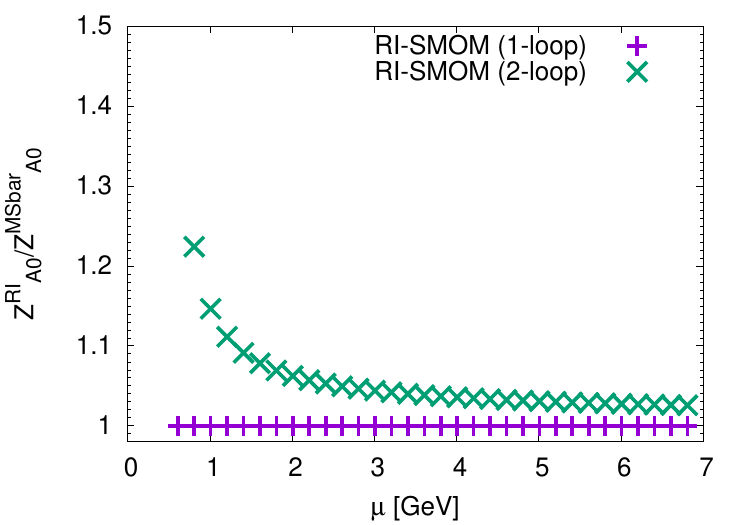}
  \includegraphics[width=0.48\textwidth]{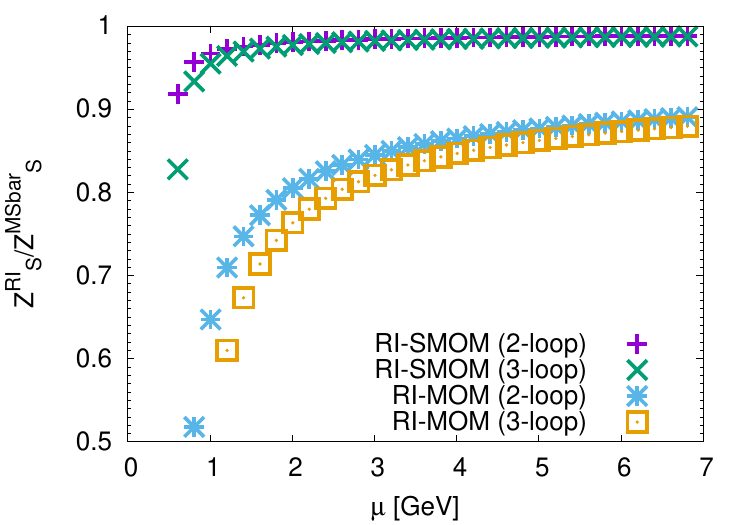}
  \includegraphics[width=0.48\textwidth]{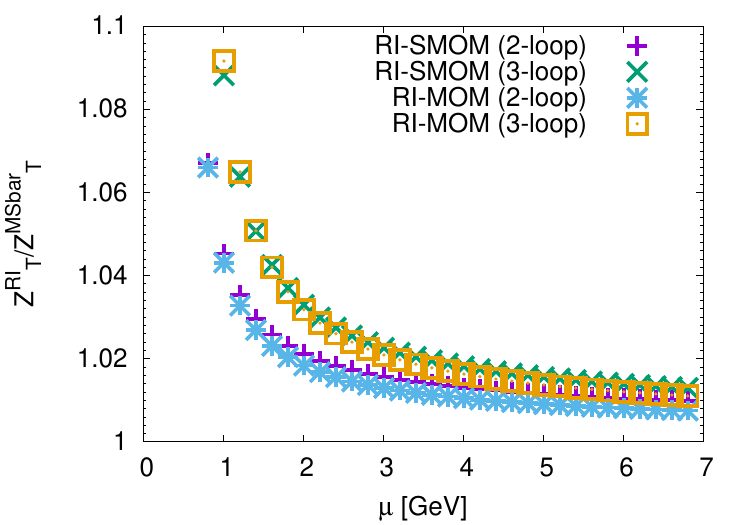}
  \includegraphics[width=0.48\textwidth]{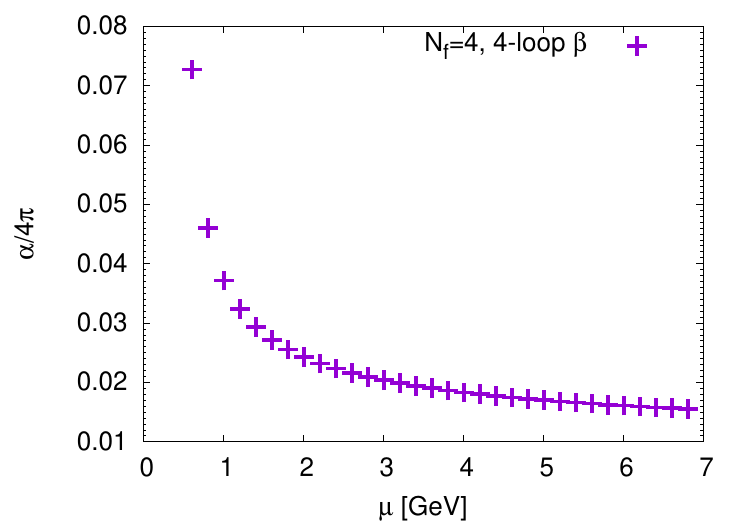}
  \caption{The top left panel shows the matching factor
    $C^{\text{RI-sMOM}\to\MSb}=Z^\text{RI-sMOM}(\mu)/Z^{\MSb}(\mu)$
    for the singlet axial current. It acquires a non-trivial
    dependence on the scale $\mu$ starting at 2-loops.  For scalar and
    tensor operators, we compare the 2- and 3-loop matching factors
    for RI-SMOM $\to \MSb$ and RI-MOM $\to \MSb$.  Bottom Right: The
    4-loop running of $\alpha/4\pi$ versus $\mu$ for the $N_f=4$
    theory used in this study.  }
  \label{fig:matching}
\end{figure*}

\subsection{Two renormalization strategies \texorpdfstring{$\rm{Z_1}$}{Z1} and \texorpdfstring{$\rm{Z_2}$}{Z2} defined by the choice of \texorpdfstring{$Z_\psi^q$}{Zpsi}}
\label{sec:Zstrategies}

To define the $\rm{Z_1}$ and $\rm{Z_2}$ renormalization strategies
used, recall that the renormalization constants for the
\emph{isovector} bilinear operators in the RI scheme are given by
\looseness-1
\begin{align}
  Z_\Gamma(p)=\frac{Z_\psi(p)}{c_\Gamma(p)} 
  \quad {\rm or} \quad 
  Z_\psi(p) = c_\Gamma(p)    Z_\Gamma(p)
  \label{eq:Zpsi}
\end{align}
 where $p$ is the momentum insertion, $Z_\psi$ is the renormalization
 constant for the fermion field, and $c_\Gamma$ is the projected
 amputated connected 3-point function calculated in Landau gauge as
 defined in Eq.~\eqref{eq:c-def}.  $\rm{Z_1}$ and $\rm{Z_2}$ are
 defined by the two ways used to calculate $Z_\psi$.
\begin{itemize}
\item 
$\rm Z_1$: 
  The $Z_\psi$ is calculated from the projected bare quark propagator
\begin{align}
  Z_\psi(p)=\frac{i}{12p^2}\Tr[S_B^{-1}(p)\qslash{p}]
  \label{eq:Z1}
\end{align}
\item $\rm Z_2$ calculated using the vector Ward identity (VWI), $Z_V g_V=1$. Equation~\eqref{eq:Z} then becomes  
  \begin{align}
  Z_\psi^{\rm VWI}(p) = c_V(p)/g_V,
    \label{eq:Z2}
  \end{align}
where $c_V$ is the projected amputated connected 3-point function with
insertion of the local vector operator within the quark state while
the vector charge $g_V$ is obtained from the insertion of the same
operator within any, in principle, hadronic state, i.e., pion,
nucleon, etc. We use the nucleon state.
\end{itemize}

These two strategies were used in
Refs.~\cite{Bhattacharya:2016zcn,Yoon:2016jzj} for \emph{isovector}
bilinear operators of light quarks, i.e., $g_\Gamma Z_\Gamma$ and
$Z_\Gamma/Z_V \times g_\Gamma/g_V$. Later, they were called $\rm Z_1$
and $\rm Z_2$ in Ref.~\cite{Park:2021ypf}.  In
Refs.~\cite{Bhattacharya:2016zcn,Yoon:2016jzj} we showed that they
have different behavior versus $q^2$ as the discretization effects are
different in $Z_\psi$ and $Z_\psi^{\rm VWI}$, which are obtained from
$S_B(p)$ and $c_\Gamma^V(p)$, respectively\footnote{ The $O(a)$
improvement for $S_B(p)$ is studied in
Ref.~\cite{Bhattacharya:2001jk}, and $c_\Gamma^V(p)$ (off-shell
improvement) is studied in Ref.~\cite{Bhattacharya:2001kt}}.  To
within expected discretization errors, we checked
\begin{itemize}
\item $Z_V|_{\rm Z_1}\times g_V^{l,{\rm bare}} = 1$ for $m_l$ as shown
  in Table~\ref{tab:gVZ1}. The deviation from unity decreases as $a
  \to 0$. When this relation is satisfied, renormalization using
  $\rm Z_1$ and $\rm{Z_2}$ should give consistent results.
\item Final continuum limit values of all the nucleon 
charges (and form factors), using $\rm Z_1$ and $\rm Z_2$ agree 
within the quoted errors~\cite{Park:2021ypf}.
\end{itemize}

\begin{figure}[t]      
  \center
  \includegraphics[width=0.48\textwidth]{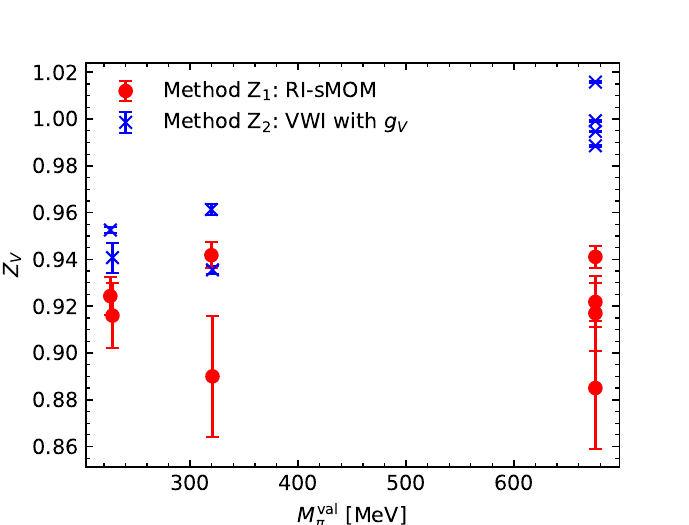}
  \caption{Results for $Z_V$ using the two strategies $\rm{Z_1}$ and $\rm{Z_2}$.}
  \label{fig:ZV_Mpi}
\end{figure}

\begin{figure}[t]      
  \center
  \includegraphics[width=0.48\textwidth]{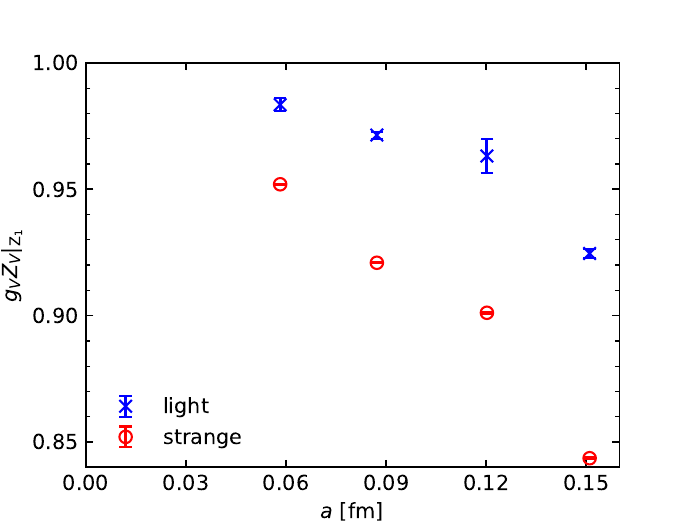}
  \caption{Data for $g_V^f Z_V^f|_{\rm Z_1}$, with the vector
    renormalization factor $Z^f_V|_{\rm Z_1}=1/c^f_V$, for flavor
    $f\in\{l,s\}$. See Eq.~\eqref{eq:Zpsi} for details.  The isovector
    vector charge $g_V^{\{l,s\}}$ is obtained from the forward matrix
    element within the nucleon ground state constructed using light
    and strange valence quark propagators respectively.  The trend in
    the data is consistent with the vector Ward identity (VWI)
    relation $g_V Z_V|_{\rm Z_1}=1$ being satisfied in the continuum
    limit.  }
  \label{fig:gVZV}
\end{figure}

This work extends the calculation to the heavier strange quark. The
data in Tab.~\ref{tab:gVZ1} show that the deviation of the
renormalized $g_V^s$ from unity is significantly larger. While this
deviation for both the light and strange quarks goes to zero
essentially linearly as $a \to 0$ as shown in Figs.~\ref{fig:ZV_Mpi}
and~\ref{fig:gVZV}, it is important to note that the discretization
artifact is large and has a significant mass dependence.

\begin{figure*}      

  \center
  \includegraphics[width=0.23\textwidth]{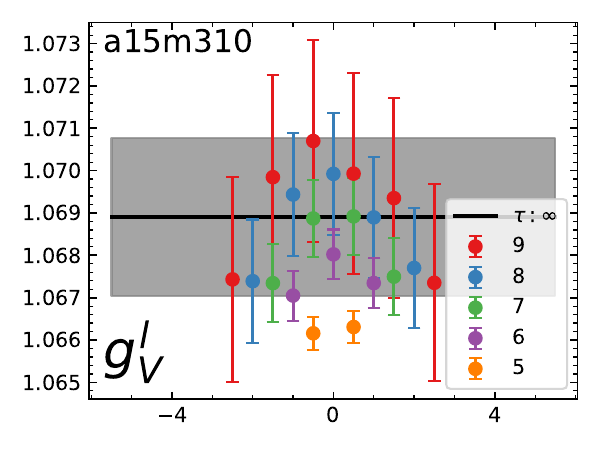}
  \includegraphics[width=0.23\textwidth]{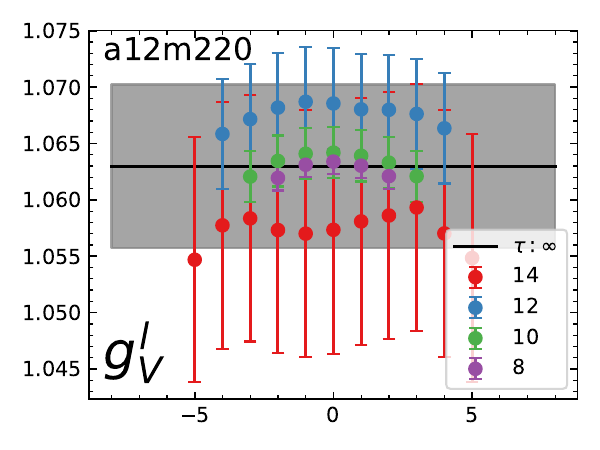}
  \includegraphics[width=0.23\textwidth]{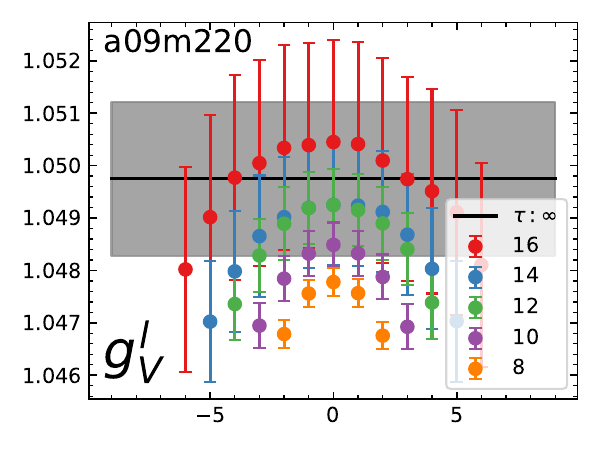}
  \includegraphics[width=0.23\textwidth]{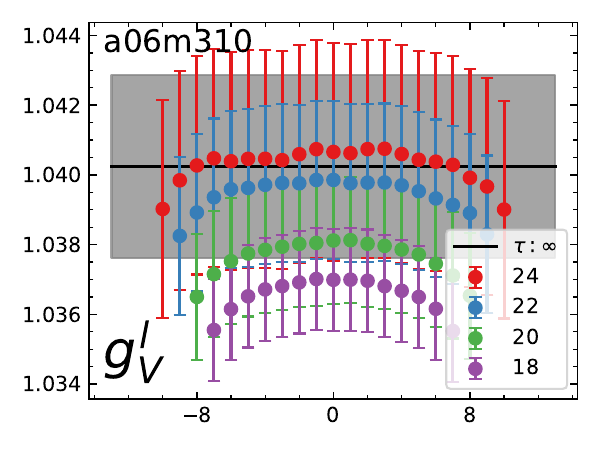}

  \includegraphics[width=0.23\textwidth]{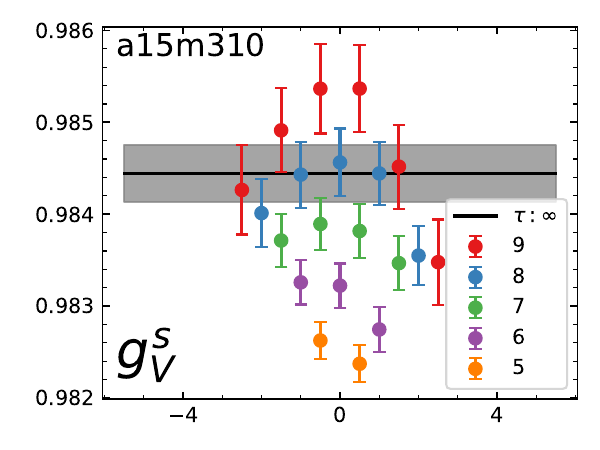}
  \includegraphics[width=0.23\textwidth]{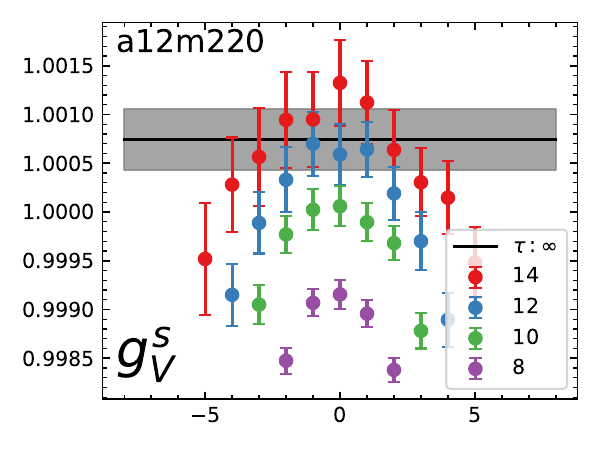}
  \includegraphics[width=0.23\textwidth]{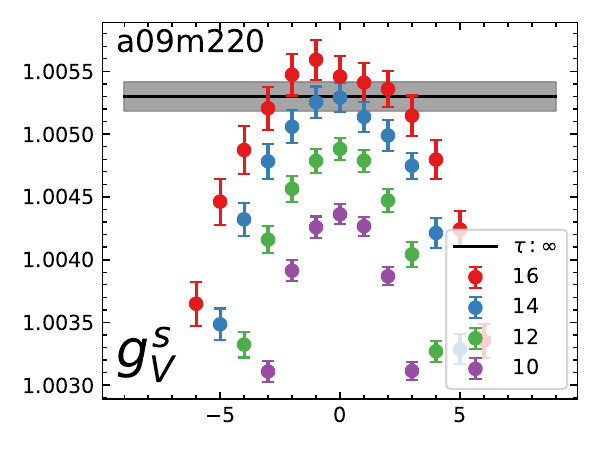}
  \includegraphics[width=0.23\textwidth]{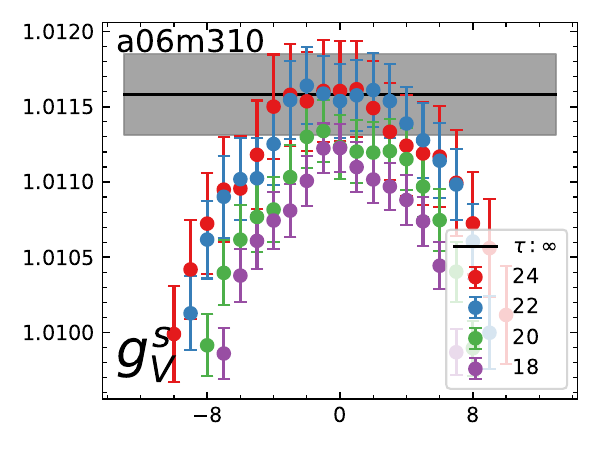}

  \caption{Data for the nucleon vector charge $g_V$ on four HISQ
    ensembles, $a15m310$, $a12m220$, $a09m220$, and $a06m310$, plotted
    from left to right.  The top row is for the light valence quark
    with mass $m_l$ and the bottom for the strange quark with mass
    $m_s$.  The data at the largest three source-sink separations
    $\tau$ agree at the sub-percent level.  For the final result (gray
    band), we take the unweighted average over 4--10 data points
    centered about $t-\tau/2=0$ with the largest two $\tau$.  Residual
    ES effects are neglected.}
  \label{fig:gV_HISQ}
\end{figure*}

We point out that these deviations are not due to incomplete removal
of ESC in the extraction of the isovector vector charge $g_V$.  The
data for the bare vector charge, i.e., the ratio plots versus
$\{\tau,t\}$, are shown in Fig.~\ref{fig:gV_HISQ}.  The ESC are less
than a percent, while the deviations in the data, shown in
Tab.~\ref{tab:gVZ1}, are a 2--8\% effect for even the light quarks.

\begin{table*}[!h]
\center
\begin{ruledtabular}
\begin{tabular}{c| lll |l  ll}
  Ensemble & $g_V^{l,\text{bare}}$ & $Z_V^{l}|_{\rm Z_1}$ & $g_V^{l}|_{\rm Z_1}$ & $g_V^{s,\text{bare}}$ & $Z_V^{s}|_{\rm Z_1}$ & $g_V^{s}|_{\rm Z_1}$   \\ \hline
a15m310 & 1.0689(19) & 0.86503(28) & 0.925(2) & 0.9845(3) & 0.85690(18) & 0.8436(3) \\
a12m220 & 1.0630(73) & 0.90611(6) & 0.963(7) & 1.0007(3) & 0.90048(4) & 0.9011(3) \\
a09m220 & 1.0498(15) & 0.92538(4) & 0.971(1) & 1.0053(1) & 0.91608(3) & 0.9209(1) \\
a06m310 & 1.0402(26) & 0.94550(6) & 0.984(2) & 1.0116(3) & 0.94110(4) & 0.9520(3) \\
\end{tabular}
\end{ruledtabular}
\caption{The bare and the $\rm Z_1$-renormalized vector charge for the
  light and strange quarks on the four ensembles. The non-perturbative
  calculations was done using the RI-sMOM scheme. The factor
  $Z_V^{\{l,s\}}|_{Z_1}$ is obtained using the quadratic extrapolation
  ansatz given in Eq.~\eqref{eq:mufit}. The bare isovector vector
  charge $g_V^{\{l,s\},{\rm bare}}$ is obtained from the forward
  matrix element within the nucleon ground state.}
\label{tab:gVZ1}
\end{table*}

\clearpage
\begin{widetext}
\subsection{Final results for the renormalization factors} 
\label{sec:Zfinal}

The final renormalization factors in the $\MSb$ scheme at 2~GeV,
calculated using the 2+1 theory in the flavor basis $f\in
\{u-d,u+d,s\}$, are given below for the four ensembles, a15m310,
a12m310, a09m310, and a06m310, the three charges, and the two
renormalization methods, $Z_1$ and $Z_2$.  These factors are the
aggregate of the four steps described in Sec.~\ref{sec:MatchingRG} and
used to renormalize the bare charges for all the ensembles at a given
lattice spacing $a$ assuming their dependence on the quark masses is
negligible compared to the other uncertainties.

\begin{align}
 a&\approx0.15\fm & &  \nonumber  \\ 
 Z_A|_{\rm Z_1} &=
\begin{pmatrix}
0.9467(1) & 0 & 0\\ 
0 & 0.8905(21) & -0.0560(27)\\ 
0 & -0.0269(11) & 0.9125(13)\\ 
\end{pmatrix}
 & Z_A|_{\rm Z_2} &=
\begin{pmatrix}
1.0222(3) & 0 & 0\\ 
0 & 0.9627(22) & -0.0647(28)\\ 
0 & -0.0285(12) & 1.0803(14)\\ 
\end{pmatrix}
\\ 
 Z_S|_{\rm Z_1} &=
\begin{pmatrix}
1.0034(2) & 0 & 0\\ 
0 & 0.9642(75) & -0.0323(75)\\ 
0 & -0.0198(39) & 0.9806(38)\\ 
\end{pmatrix}
 & Z_S|_{\rm Z_2} &=
\begin{pmatrix}
1.0792(3) & 0 & 0\\ 
0 & 1.0381(81) & -0.0370(79)\\ 
0 & -0.0208(42) & 1.1554(40)\\ 
\end{pmatrix}
\\ 
 Z_T|_{\rm Z_1} &=
\begin{pmatrix}
0.8925(1) & 0 & 0\\ 
0 & 0.8930(6) & 0.0003(8)\\ 
0 & 0.0001(3) & 0.8764(4)\\ 
\end{pmatrix}
 & Z_T|_{\rm Z_2} &=
\begin{pmatrix}
0.9653(2) & 0 & 0\\ 
0 & 0.9659(6) & 0.0004(8)\\ 
0 & 0.0001(3) & 1.0393(4)\\ 
\end{pmatrix}
\end{align}

\begin{align}
 a&\approx0.12\fm & &  \nonumber  \\ 
 Z_A|_{\rm Z_1} &=
\begin{pmatrix}
0.95705(4) & 0 & 0\\ 
0 & 0.9287(6) & -0.0283(11)\\ 
0 & -0.0141(4) & 0.9386(4)\\ 
\end{pmatrix}
 & Z_A|_{\rm Z_2} &=
\begin{pmatrix}
0.9933(1) & 0 & 0\\ 
0 & 0.9642(6) & -0.0307(11)\\ 
0 & -0.0143(4) & 1.0389(4)\\ 
\end{pmatrix}
\\ 
 Z_S|_{\rm Z_1} &=
\begin{pmatrix}
0.9183(1) & 0 & 0\\ 
0 & 0.9163(43) & -0.0074(63)\\ 
0 & -0.0010(24) & 0.9148(30)\\ 
\end{pmatrix}
 & Z_S|_{\rm Z_2} &=
\begin{pmatrix}
0.94423(4) & 0 & 0\\ 
0 & 0.9493(44) & -0.0080(64)\\ 
0 & -0.0010(25) & 1.0119(31)\\ 
\end{pmatrix}
\\ 
 Z_T|_{\rm Z_1} &=
\begin{pmatrix}
0.94254(5) & 0 & 0\\ 
0 & 0.9424(3) & 0.0003(5)\\ 
0 & -0.0002(2) & 0.9358(2)\\ 
\end{pmatrix}
 & Z_T|_{\rm Z_2} &=
\begin{pmatrix}
0.97970(4) & 0 & 0\\ 
0 & 0.9792(3) & 0.0003(5)\\ 
0 & -0.0002(2) & 1.0380(2)\\ 
\end{pmatrix}
\end{align}

\begin{align}
 a&\approx0.09\fm & &  \nonumber  \\ 
 Z_A|_{\rm Z_1} &=
\begin{pmatrix}
0.96407(4) & 0 & 0\\ 
0 & 0.9356(17) & -0.0286(18)\\ 
0 & -0.0155(9) & 0.9454(9)\\ 
\end{pmatrix}
 & Z_A|_{\rm Z_2} &=
\begin{pmatrix}
0.99252(4) & 0 & 0\\ 
0 & 0.9631(17) & -0.0307(18)\\ 
0 & -0.0159(9) & 1.0251(10)\\ 
\end{pmatrix}
\\ 
 Z_S|_{\rm Z_1} &=
\begin{pmatrix}
0.92167(5) & 0 & 0\\ 
0 & 0.9158(40) & -0.0067(40)\\ 
0 & -0.0025(21) & 0.9141(21)\\ 
\end{pmatrix}
 & Z_S|_{\rm Z_2} &=
\begin{pmatrix}
0.9464(1) & 0 & 0\\ 
0 & 0.9419(42) & -0.0073(41)\\ 
0 & -0.0025(22) & 0.9900(21)\\ 
\end{pmatrix}
\\ 
 Z_T|_{\rm Z_1} &=
\begin{pmatrix}
0.98950(4) & 0 & 0\\ 
0 & 0.9892(6) & -0.0002(7)\\ 
0 & 0.0002(3) & 0.9763(4)\\ 
\end{pmatrix}
 & Z_T|_{\rm Z_2} &=
\begin{pmatrix}
1.01847(4) & 0 & 0\\ 
0 & 1.0182(6) & -0.0002(7)\\ 
0 & 0.0002(3) & 1.0592(4)\\ 
\end{pmatrix}
\end{align}

\begin{align}
 a&\approx0.06\fm & &  \nonumber  \\ 
 Z_A|_{\rm Z_1} &=
\begin{pmatrix}
0.96958(5) & 0 & 0\\ 
0 & 0.9493(14) & -0.0184(15)\\ 
0 & -0.0096(9) & 0.9579(9)\\ 
\end{pmatrix}
 & Z_A|_{\rm Z_2} &=
\begin{pmatrix}
0.98501(4) & 0 & 0\\ 
0 & 0.9652(14) & -0.0192(15)\\ 
0 & -0.0097(9) & 1.0068(9)\\ 
\end{pmatrix}
\\ 
 Z_S|_{\rm Z_1} &=
\begin{pmatrix}
0.86298(5) & 0 & 0\\ 
0 & 0.8601(25) & -0.0043(27)\\ 
0 & -0.0027(13) & 0.8606(13)\\ 
\end{pmatrix}
 & Z_S|_{\rm Z_2} &=
\begin{pmatrix}
0.8785(1) & 0 & 0\\ 
0 & 0.8743(26) & -0.0045(27)\\ 
0 & -0.0028(13) & 0.9042(13)\\ 
\end{pmatrix}
\\ 
 Z_T|_{\rm Z_1} &=
\begin{pmatrix}
1.03069(5) & 0 & 0\\ 
0 & 1.0350(3) & 0.0004(3)\\ 
0 & -0.0000(2) & 1.0266(2)\\ 
\end{pmatrix}
 & Z_T|_{\rm Z_2} &=
\begin{pmatrix}
1.0501(1) & 0 & 0\\ 
0 & 1.0521(3) & 0.0004(3)\\ 
0 & -0.0000(2) & 1.0790(2)\\ 
\end{pmatrix}
\end{align}

\end{widetext}

\bibliography{ref} 

\end{document}